\documentclass[a4paper,12pt]{report}
\pdfoutput=1

\usepackage{titlepic}
\usepackage{mathtools}
\usepackage{amssymb}
\usepackage{physics}
\usepackage{braket}
\usepackage[labelfont=bf]{caption}
\usepackage{subcaption}
\usepackage{nicefrac}
\usepackage[toc,page]{appendix}
\usepackage{float}
\usepackage{bbold}

\DeclareMathOperator{\sign}{sign}
\DeclareMathOperator{\detn}{det}
\DeclareMathOperator{\diag}{diag}

\usepackage{tikz}
\usetikzlibrary{positioning}
\usetikzlibrary{shapes.misc}
\usetikzlibrary{decorations.text}
\usetikzlibrary{calc}

\newdimen\XCoord
\newdimen\YCoord
\newdimen\XotherCoord
\newdimen\YotherCoord
\newcommand*{\ExtractCoordinate}[1]{\path (#1); \pgfgetlastxy{\XCoord}{\YCoord};}
\newcommand*{\ExtractotherCoordinate}[1]{\path (#1); \pgfgetlastxy{\XotherCoord}{\YotherCoord};}

\newcommand\numberthis{\addtocounter{equation}{1}\tag{\theequation}}

\usepackage[top=2.5cm, bottom=2.5cm, left=3.8cm, right=2.5cm]{geometry}
\usepackage{setspace}
\doublespace

\usepackage[backend=bibtex,citestyle=numeric-comp,bibstyle=numeric,doi=false,isbn=false,url=false,sorting=none,minbibnames=5,maxbibnames=5,defernumbers=true,firstinits=true]{biblatex}
\renewbibmacro{in:}{}
\bibliography{Theoretical,Experimental,Manual,Review,Supplementary}

\titlepic{\includegraphics[width=3in]{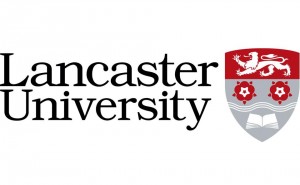}}
\author{Matthew Malcomson \\ PhD Thesis}
\title{Momentum Evolution Numerics of an Impurity in a Quantum Quench}
\date{~\\ Submitted for the degree of Doctor of Philosophy\\
February 15, 2016}
\begin{document}
\maketitle

\chapter*{Declaration}

Except where stated otherwise, this Thesis is a result of the author's original work and has not been
submitted in whole or in part for the award of a higher degree elsewhere.

\chapter*{Acknowledgements}
I would like to thank Evgeni Burovski, my supervisor for introducing me to the power of computers in research, it has
become the catalyst what is now my overriding interest.
Furthermore, I thank Vadim Cheianov, Oleksandr Gamayun, Mikhail Zvonarev and Oleg Lychkovskiy for
interesting discussions that have provided new avenues of research.
\thispagestyle{empty}
\clearpage

\begin{abstract}
  A discussion on the momentum evolution of an impurity interacting via a finite delta potential repulsion with a
  non-interacting fermionic background gas is presented.
  It has recently been shown that the momentum evolution of this system displays two interesting features, namely a non-zero
  thermalised value and a long-lived quantum mechanical oscillation around this plateau named ``quantum
  flutter'' [Mathy, Zvonarev, Demler, \textit{Nat. Phys.} \textbf{2012}].
  We discuss revivals in the momentum of the impurity, which have been seen before but not yet thoroughly investigated.
  Subsequently it is shown the quantum flutter and revivals are caused by disjoint sets of eigenstate transitions, and this
  fact is used to interpret some of their aspects.
  This attribution of momentum features to different eigenstate subsets allows quantitative reproduction of these features
  with much less computational expense than has so far been possible.
  Finally some results on the distribution of the momentum of eigenstates and their relation to the momentum of the impurity
  once the system has been thermalised are presented along with a discussion on the time averaged infinite time value of the
  momentum and its comparison to different eigenstate subsets.
\end{abstract}

\tableofcontents
\listoffigures


\chapter{Introduction}
\label{sec:introduction}
\section{Introduction}
\label{sec:introduction_introduction}

When introduced to a new phenomenon, a natural tendency is to attempt to find the simplest system in which it is
exhibited, and use said system as a playground to explore the phenomenon's features without unnecessary complexity.
An ever-present danger in this methodology is that a simple system may show qualitatively different features to more complex
ones, for example, because features in the complex system are emergent from some complexity not in the smaller system, or
because the limitation on the degrees of freedom in the smaller system prohibit the phenomenon.
Upon encountering this, the investigator must decide whether to add complications piecemeal to their original description,
focus on those features which are shared by the model and observations, or start from scratch with a different model.
It is a rare, but happy, event when the features of said simple model opens up a new and exciting area of research, from
which novel features are found with regularity.
The physics of strongly correlated one-dimensional quantum systems is one such field, revealing numerous phenomena not
present in higher dimensions, both in their mathematical descriptions and their observable nature.
While initially thought of more as a testing ground for methods to apply to the ``real'' 3 dimensional world (see the
introductions of references~\cite{Lenard1964,Calogero1971} for examples), experimental progress creating 1D systems (both
quasi and true) has given these ``toy models'' new-found experimental relevance
\cite{Scheunemann2000,Morsch2001,Moritz2003,Fallani2004,Tolra2004,Paredes2004,Stoferle2004,Kinoshita2004,Kohl2004,Kinoshita2005,Fertig2005,Kinoshita2006,Ryu2007,Hofferberth2007,Olson2007,VanAmerongen2008,Du2008,Palzer2009,Haller2009,Chen2011,Whitlock2011,Trotzky2011,Catani2012,Fukuhara2013,Meinert2013},
that has significantly increased the interest in the field.

One of the most important aspects of one-dimensional physics is the Luttinger liquid universality class.
While many different microscopic models have been used in 1D, for both fermions and bosons, and across a wide range of
interaction types and potentials,
\cite{Tomonaga1950,Luttinger1963,Mattis1965}
\cite{Calogero1969,Calogero1971,Sutherland1971}
\cite{Lieb1963,Lieb1963a,Yang1969}
\cite{Yang1967,Sutherland1968,Sutherland1971}
\cite{Lieb1968}
most one dimensional models share common features at low energy, and from this, a low energy universality class, called a
``Luttinger Liquid'', has been formed, which is in some ways a replacement for the Fermi liquid theory in one
dimension~\cite{Haldane1981,Gogolin1999,Giamarchi2003}.
In one dimension the limitation in dimensionality imposes an inherently collective nature on the excitations of a liquid, as
in order for one particle to move, another must make space.
This breaks the fermionic quasiparticle approximation of a Fermi liquid, and in its place a bosonic quasiparticle of
collective excitations is created.
Non-interacting bosonic quasiparticles are formed by assuming a linear excitation dispersion of the fermionic
particles~\cite{Luttinger1963}, and a small non-linearity can be accounted for by adding interactions between
them~\cite{Haldane1981}.
Initially formulated to describe fermions, the generality of this description has extended to gases of
bosons~\cite{Cazalilla2011}, creating a general description of low energy one-dimensional systems.
Despite the success of this Luttinger liquid theory,
the analogy does not carry over to higher energy excitations, which
allow excitations too far from the approximately linear dispersion to be accounted for.

Given such a widely encompassing description of the low energy physics in one-dimensional systems,
an interest in systems outside the Luttinger liquid paradigm has emerged, focusing on systems at a higher energy, or with
some other feature breaking the linear approximation of the dispersion
relation~\cite{Pustilnik2006,Fiete2007,Zvonarev2007,Imambekov2008,Imambekov2009a,Imambekov2012}.
One method of probing these regimes, is by going back to the microscopic models that have been formed, and finding how they
behave as excitations move them away from the Luttinger liquid paradigm.
It is one of the miracles of one dimension that many of these models can be solved exactly using the ansatz proposed by
Bethe, initially in order to find a solution of the Heisenberg spin-1/2 chain~\cite{Bethe1931}.
This ansatz gives exact eigenstates of the systems, and enables numerical calculation of their energy and
momentum~\cite{Gaudin1983,Giamarchi2003}.

This thesis works on one particular system of this type, modelling the dynamics of a high energy spin impurity quenched in a
spin polarised fermionic gas, presenting a discussion on the features of the impurity's momentum evolution, and the patterns
in the eigenstate transitions that describe individual features.
An impurity in one-dimensional systems has been an active area of research for some
time~\cite{Neto1995,Castella1996,Palzer2009,Lamacraft2009,Girardeau2009,Punk2009,Giraud2009,Giraud2010,Goold2010,Ovchinnikov2010,Johnson2011,Catani2012,Spethmann2012,Recher2012,Windpassinger2013,Astrakharchik2013,Fukuhara2013,Massel2013,Doggen2013,Doggen2014,Massignan2014},
and modelling how a system behaves with a high energy impurity, taking it far away from the Luttinger liquid paradigm
is an interesting variation on a theme that has already provided qualitatively new phenomena~\cite{Mathy2012,Knap2014}.
The interaction between our impurity and background particles is described by a delta function potential, and hence the
system is modelled by the fermionic Yang-Gaudin model~\cite{Yang1967,Gaudin1967,Guan2013} spin polarised but for a single
impurity, as used in references~\cite{McGuire1964,McGuire1990,Edwards1990,Castella1993,Mathy2012,Gamayun2015}.
As a consequence of yet another quirk of a single dimension, this system behaves similarly whether the particles are
free fermions or infinitely repulsive bosons, and hence could also be described with the Lieb-Liniger model~\cite{Lieb1963}
of interacting bosons in the limit of infinite potential and with added terms for an impurity.
This fermionisation of bosons has been directly observed in references~\cite{Kinoshita2004,Kinoshita2005}, and while no
exact realisation of this system has been created as yet, reference~\cite{Palzer2009} has demonstrated many necessary
ingredients, measuring the velocity of a single impurity accelerated with a constant force, using time-of-flight
measurements.

The rest of this thesis is structured in the following way: Chapter~\ref{sec:model_and_method} describes the model used to
probe the system and its solutions, some terminology for different equations, and the difficulties faced when calculating
the impurity's momentum.
The time dependent features of this momentum are presented in Chapter~\ref{sec:observables}, which adds to the existing
literature with a deeper analysis of the momentum revivals in the system.
Further original work is presented in Chapter~\ref{sec:subsets}, which attributes each momentum feature to a subset of
eigenstate transitions, and Chapter~\ref{sec:other} which presents the results of a preliminary look at the structure
inherent in the eigenstates themselves.
Finally, Chapter~\ref{sec:concluding_remarks} summarises the main results of this work and presents some proposed topics of
further work.
A discussion on the technical aspects of the code written for this work is presented in Appendix~\ref{sec:code_details}.

\section{Method and Model}
\label{sec:model_and_method}

\subsection{Model}
\label{sec:m&m_model}
The system we work within has two valid representations: one of hard-core bosons, and one of free fermions.
In the bosonic case, our Hamiltonian, in units of $\hbar = 1$ and $m = m_{\uparrow} = m_{\downarrow} = 0.5$ is
\begin{equation}
  \label{eqn:general_boson_hamiltonian}
  H = \hat{P_{\downarrow}}^{2} +
  \sum_{i=1}^{N}\hat{P_{i}}^{2} +
  g \sum_{i=1}^{N}\delta (x_{i} - x_{\downarrow}) +
  a \sum_{i,j = 1}^{N}\delta (x_{i} - x_{j})
\end{equation}
where $\hat{P_{\downarrow}}$ is the momentum operator of the impurity, the sums are over all particles in the background gas,
$\hat{P_{i}}$ is the momentum operator for a single background gas particle, $g$ defines the interaction strength between the
impurity and the background gas, and $a$ sets the interaction strength between two particles of the background gas.

For the specific case of our system, $a = \infty$, and it is this infinite potential which is the root cause of the
equivalence between representations.
While we are using a delta potential interaction, the correspondence holds for a gas of bosons with any interaction, as long
as the interaction has an impenetrable core, forcing a constant order on the particles~\cite{Girardeau1960}.
Essentially, the correspondence comes from the fact that multiplying a fermionic wavefunction by the unit anti-symmetric function
\begin{equation}
\prod_{j > l} \sign (x_{j} - x_{l})
\end{equation}
produces an eigenfunction of the Hamiltonian that obeys bosonic anti-commutation relations and can be made to satisfy the
same boundary conditions (depending on the parity of the number of particles).
The Pauli exclusion principle, and discontinuities in the anti-symmetric function above do not cause a problem when satisfying
regularity conditions because of the stipulated hard-core interactions between bosons, which require the wavefunction to be
$0$ when two particles share a position.
Both representations share many observables (including the energy of the system), with the determining factor whether the
unit anti-symmetric function above commutes with the corresponding operator.
This argument can be extended to the case of a gas with a single distinguishable impurity, as the impurity puts no extra
constraints on the symmetry of the wavefunction~\cite{Mathy2012sup}.
Recent experiments have managed to record gases in this regime~\cite{Kinoshita2004,Paredes2004,Tolra2004}, called a
Tonks-Girardeau gas, and have even observed the transition between a bosonic state and the Tonks-Girardeau gas with
increasing interaction strength~\cite{Kinoshita2005}.

Of these two representations, we use the fermionic one (called the Yang-Gaudin model) throughout.
This fermionic description gives us some important values, like the Fermi momentum, which is useful in the phenomenological
description of the impurity's momentum, and Fermi time, useful to describe the motion of the impurity in a manner independent
of the system size.
In this case, there is no interaction between pairs of similar particles, only between the single impurity and each particle
in the background gas, so the Hamiltonian is
\begin{equation}
  \label{eqn:our_hamiltonian}
  H = \hat{P_{\downarrow}}^{2} +
  \sum_{i=1}^{N}\hat{P_{i}}^{2} +
  g \sum_{i=1}^{N}\delta (x_{i} - x_{\downarrow}).
\end{equation}
A \emph{dimensionless} interaction strength parameter $\gamma = \nicefrac{g}{2n}$ (where $n=\nicefrac{N}{L}$ is the density of
particles in the system) can be defined to use in place of $g$, which gives a more physically relevant parameter to inspect in
Chapter~\ref{sec:observables}.

This model is integrable, and exactly solvable via the Bethe Ansatz~\cite{McGuire1964,McGuire1965}, we use an alternate
formalism presented in ref~\cite{Edwards1990} which has been previously used to good effect in calculating the spectral
properties of the system~\cite{Castella1993} and the momentum of the impurity and background gas in the
system~\cite{Mathy2012}.

\subsubsection{Bethe Ansatz}
\label{sec:m&m_BA}
Soon after the formulation of quantum mechanics, Hans Bethe discovered a method to find the exact eigenstates and eigenvalues
of the Heisenberg model for one-dimensional chain of spin-1/2 fermions~\cite{Bethe1931}.
As this model only accounts for interactions between neighbouring particles, he noted that when no two down spins were next
to each other the eigenstate must be a linear combination of single down spin wavefunctions.
He used this observation to propose an Ansatz for the form of the eigenstates of the system, and showed that when this
wavefunction satisfies a set of equations now known as the Bethe equations, his Ansatz indeed solves the
Hamiltonian~\cite{Bethe1931}.

Bethe's paper showed that the many-particle problem of the Heisenberg chain reduced to solving how two spin-down
quasiparticles interact when upon neighbouring sites~\cite{Bethe1931}.
This fact allows all interactions to be interpreted as multiple two-body interactions, which has been suggested as a
criterion for integrability~\cite{Sutherland2004}.
Hence the Bethe Ansatz is intrinsically tied with integrability, and indeed almost all integrable systems can be solved in
terms of the Bethe Ansatz~\cite{Batchelor2007}.
While this statement holds, the applicability of the Bethe Ansatz to most integrable systems was not initially seen, and
required slightly different forms to be realised.

The first alternate use came in the 1960's when Bethe's hypothesis was applied to the continuum case of a 1D model of
interacting bosons~\cite{Lieb1963a,Lieb1963,McGuire1964,Yang1967}.
This use, known as the coordinate Bethe Ansatz, is the form used for the current work and is such described in more detail
in Section~\ref{sec:m&m_BA_solution}.
The coordinate Bethe Ansatz draws a parallel between the down spin quasiparticles of Bethe's original work and the physical
bosons in the 1D gas that Lieb and Liniger studied.

After that more applications and generalisations appeared.
The more complicated \emph{nested} Bethe Ansatz was used to account for the additional spin degree of freedom in the
non-polarised fermionic 1D gas~\cite{Flicker1967,Gaudin1967,Yang1967,Sutherland1968}.
For an non-polarised gas, the symmetry between all orderings of particles is broken so there are many different orderings,
the number depending on number of each spin.
Because of this extra degree of complexity a generalised Bethe Ansatz is used and solved using a set of conditions distinct
yet still related to those in Eqn~\eqref{eqn:BA_kvalues}.
Finally an alternate derivation of the Bethe equations called the \emph{algebraic} Bethe Ansatz was found applicable to integrable
systems of quasiparticles above some reference state~\cite{Takhtadzhan1979,Sklyanin1991,Korepin1993,Faddeev1996}.

While we use a fermionic gas system, the nested Bethe Ansatz is not used, as gas is polarised apart from the single impurity.


Since it's inception, the Bethe Ansatz has been found useful in many situations, but as one author notes~\cite{Belliard2008}
``numerous publications have been dedicated to the subject, so that it is becoming difficult to make exhaustive citations''.
Instead we provide the reader with some previously collated references in the introduction of Reference~\cite{Franchini2011}.

\subsubsection{Bethe Ansatz Solution}
\label{sec:m&m_BA_solution}
For a description of how the Bethe Ansatz is used in our system, we first present a work-through of the coordinate Bethe Ansatz for a
two-body example of our system then state the generalisation with a more simple representation used in
references~\cite{Edwards1990,Castella1993,Mathy2012}.
Finally we work through the computationally efficient matrix equations developed in~\cite{Mathy2012} to calculate the
momentum of the impurity.

In essence, the coordinate Bethe Ansatz uses the fact that given a suitable inter-particle potential, the wavefunction of the
system in the asymptotic limit can be described by the wavefunction of free particles when said particles are far enough
apart.
For example, for two particles, at positions $x_{1}, x_{2}$ the wavefunction when they are far enough apart from each other
is
\begin{equation}
\Psi(x_{1}, x_{2})_{asymptotic} = \alpha e^{i(k_{1}x_{1} + k_{2}x_{2})} + \beta e^{i(k_{2}x_{1} + k_{1}x_{2})}
\end{equation}
where $x_{i}, k_{i}$ are the position and momentum respectively of particle $i$, and the energy of this wavefunction is
\begin{equation}
  \label{eqn:BA_energy}
  E = k_{1}^{2} + k_{2}^{2}.
\end{equation}
Within this assumption, any interaction must be accounted for in the coefficients $\alpha$ and $\beta$, which are found using
restrictions imposed by the inter-particle potential and boundary conditions.
While this ansatz is clear in the two-body case, it has also been found correct when generalising to many particles for
multiple two body potentials, as in one-dimension, interactions between multiple particles can be shown to be
non-diffractive.
This non-diffractive nature is the criterion as mentioned above for the Bethe Ansatz where the interaction between multiple
particles can be described as a set of subsequent two-body scattering events.

Reference~\cite{McGuire1964} used a more general form of this ansatz to find the exact eigenstates for the system that we are
using, asserting the asymptotic wavefunction in all configurations where the impurity does not share a position with any
particle in the background gas.
Here we follow that method for a two body system, that is one impurity at position $x_{1}$ and one background fermion at
position $x_{2}$.
In this system, there are two different regions in which the wavefunction must have its asymptotic form, one where
$x_{1} < x_{2}$ and one where $x_{1} > x_{2}$, hence
\begin{equation}
  \Psi(x_{1}, x_{2}) = \Psi_{1}(x_{1}, x_{2}) + \Psi_{2}(x_{1}, x_{2})
\end{equation}
where $\Psi_{1}, \Psi_{2}$ describe the wavefunction in their respective region, and are of the same form as before.
Due to periodic boundary conditions on $x_{2}$, we know the wavefunction when $x_{2} = L$ must be the same as when
$x_{2} = 0$, hence
\begin{gather*}
  \Psi_{1}(x_{1}, L) = \Psi_{2}(x_{1}, 0)
  \implies
  \alpha_{1} e^{i(k_{1}x_{1} + k_{2}L)} + \beta_{1} e^{i(k_{2}x_{1} + k_{1}L)} =
  \alpha_{2} e^{ik_{1}x_{1}} + \beta_{2} e^{ik_{2}x_{1}} \\
  \implies \\
  \alpha_{1}e^{ik_{2}L} = \alpha_{2} \numberthis \label{eqn:x2_pbcs} \\
  \beta_{1}e^{ik_{1}L} = \beta_{2}
\end{gather*}
similarly, applying the same boundary conditions to $x_{1}$ we have the conditions
\begin{gather*}
  \Psi_{1}(0, x_{2}) = \Psi_{2}(L, x_{2}) \\
  \implies \\
  \alpha_{1} = \alpha_{2} e^{ik_{1}L} \numberthis \label{eqn:x1_pbcs} \\
  \beta_{1} = \beta_{2} e^{ik_{1}L}
\end{gather*}
which together imply that
\begin{equation}
  \label{eqn:BA_momentum}
  k_{1} + k_{2} = \frac{2 \pi n}{L} , \qquad n \in \mathbb{N}
\end{equation}
where the value $\frac{2 \pi n}{L}$ is hence the total momentum of the system.

To account for the delta function interaction potential, we assert the condition~\cite{McGuire1964}
\begin{equation}
    \frac{1}{2} \bigg[{(\frac{\partial}{\partial x_{1}} - \frac{\partial}{\partial x_{2}})}_{x_{1} - x_{2} = 0^{+}} -
      {(\frac{\partial }{ \partial x_{1}} - \frac{\partial}{\partial x_{2}})}_{x_{1} - x_{2} = 0^{-}}\bigg] \Psi
    =
    g \Psi
\end{equation}
which, using the split of $\Psi$ depending on the relative positions of each particle, implies the following
\begin{equation}
  \label{eqn:delta_potential_restriction}
  \bigg(\frac{\partial}{\partial x_{1}} - \frac{\partial}{\partial x_{2}}\bigg) \Psi_{1}(x_{1}, x_{2}) -
    \bigg(\frac{\partial}{\partial x_{1}} - \frac{\partial}{\partial x_{2}}\bigg) \Psi_{2}(x_{1}, x_{2})
    =
    2 g \Psi
\end{equation}
where $g$ is the interaction strength from Equation~\eqref{eqn:our_hamiltonian}.
We can combine the requirements found in Equations~\eqref{eqn:x2_pbcs} and \eqref{eqn:x1_pbcs} with the one above
 to form the combined requirement below.
\begin{equation}
  {
    \begin{pmatrix}
      e^{ik_{2}L}\alpha \\  e^{ik_{1}L}\beta
    \end{pmatrix}
  }_{1}
  =
  {
    \begin{pmatrix}
      \alpha \\ \beta
    \end{pmatrix}
  }_{2}
  =
  \begin{pmatrix}
    1 + \nicefrac{g}{i(k_{1} - k_{2})} & \nicefrac{g}{i(k_{1} - k_{2})} \\
    -\nicefrac{g}{i(k_{1} - k_{2})} & 1 - \nicefrac{g}{i(k_{1} - k_{2})}
  \end{pmatrix}
  {
    \begin{pmatrix}
      \alpha \\ \beta
    \end{pmatrix}
  }_{1}
\end{equation}
which requires for self-consistency that
\begin{equation}
  \begin{vmatrix}
    1 + \nicefrac{g}{i(k_{1} - k_{2})} - e^{ik_{2}L} & \nicefrac{g}{i(k_{1} - k_{2})} \\
    - \nicefrac{g}{i(k_{1} - k_{2})} & 1 - \nicefrac{g}{i(k_{1} - k_{2})} - e^{ik_{1}L}
  \end{vmatrix}
  = 0.
\end{equation}
In order to satisfy the above condition, it is sufficient to require that
\begin{equation}
  \label{eqn:BA_kvalues}
    \begin{split}
      \cot(\frac{k_{1}L}{2}) &= \frac{2k_{1}}{g} - const \\
      \cot(\frac{k_{2}L}{2}) &= \frac{2k_{2}}{g} - const
    \end{split}
\end{equation}
where $const$ is some arbitrary value.
Finding an eigenstate of the system is hence reduced to finding $2$ values $k_{1}, k_{2}$ which satisfy the equations
\eqref{eqn:BA_kvalues}, and \eqref{eqn:BA_momentum}, where the energy of the state is given by \eqref{eqn:BA_energy}.
This solution can be generalised to any number of background particles using the assertion that when no background particle
shares a position with the impurity, the wavefunction of the system is a linear combination of free
particles and accounting for interactions in the way described above~\cite{McGuire1964}.
In this more general solution, with a background gas of $N$ particles, the eigenstates of the system are defined by the
$N + 1$ values $k_{1}, k_{2}, ... , k_{N+1}$ satisfying the more generalised versions of Equations~\eqref{eqn:BA_kvalues},
and~\eqref{eqn:BA_momentum} presented alongside the energy of these states below
\begin{equation}
  \label{eqn:Full_BA_eqns}
  \begin{split}
      \sum_{i=1}^{N+1}k_{i} &= \frac{2 \pi n}{L} = Q, \qquad n \in \mathbb{N} \\
      \cot(\frac{L}{2}k_{i}) &= \frac{2k_{i}}{g} - const \\
      E &= \sum_{i=1}^{N + 1} k_{i}^{2}
  \end{split}
\end{equation}
for all $i \in 1,2,..,N+1$, and with $N$ representing the number of particles in the background gas.
These values $k_{1}, k_{2}, ... , k_{N+1}$ are known as the Bethe momenta of the equation and in our case (with a repulsive
potential) they are real.

While this solves the system exactly, it results in a very complicated wavefunction, with many different amplitudes to
calculate.
Reference~\cite{Edwards1990} found the same wavefunctions were reproduced in an easier format, by forming them in the
reference frame of the impurity.
In this alternate reference frame, an ansatz is taken to be
\begin{equation}
  \label{eqn:BA_final_ansatz}
f(y_{2}, ... , y_{N + 1}) = \detn_{N}(\Phi_{j}(y_{l}))
\end{equation}
where $y_{i}$ are the coordinates of each background gas particle in this new frame of reference, and $\Phi_{j}$ are
functions dependent on an individual coordinate.
Coordinates of this wavefunction span from $y_{2}$ onward as the dependence of the function on the position of the impurity
has been factored out when switching reference frame.
This new ansatz can be shown~\cite{Edwards1990} to solve the system when each $\Phi_{j}(y)$ is described as
\begin{equation}
  \label{eqn:phi_constraint}
\Phi_{j}(y) =
\sum_{t=1}^{N+1} a_{j}^{t} e^{ik_{t}y}
\end{equation}
where the $N + 1$ $k_{t}$ values satisfy the conditions in Equation~\eqref{eqn:Full_BA_eqns}, and the $N(N+ 1)$ coefficients
$a_{j}^{t}$ satisfy the equations
\begin{equation}
  \begin{split}
  \sum_{t=1}^{N+1} a_{j}^{t}(1-e^{ik_{t}L}) = 0, \qquad j=1,...,N \\
  \sum_{t=1}^{N+1} a_{j}^{t}[ik_{t}(1-e^{ik_{t}L}) - g] = 0, \qquad j=1,...,N
  \end{split}
\end{equation}
to ensure the wavefunction satisfies restrictions from the periodic boundary conditions and the delta potential in the
Schr\"{o}dinger equation respectively.
This is the form of eigenstates used throughout the current work.

\subsubsection{Equations for Momentum}
\label{sec:m&m_momentum_equations}

This work focuses on the time evolution of the impurity's momentum, which we calculate using the computationally
efficient equations described in~\cite{Mathy2012sup}.
That reference describes in detail both the derivation of equations to find the impurity's momentum in terms of matrix
elements, and a manner to calculate said matrix elements.
Here we state the equations which define the impurity's momentum in order to set the scene for the discussion on
separation of contributions in Chapter~\ref{sec:subsets}.
The expectation value of the impurity's momentum can be found with the equation
\begin{equation}
  \label{eqn:momentum_against_time}
  \expval{P_{\downarrow}(t)} = Q -
  \sum_{f_{Q},f_{Q}^{'}} e^{i t (E_{f} - E_{f^{'}})} \braket{FS|f_{Q}}\matrixel{f_{Q}}{P_{\uparrow}}{f_{Q}^{'}}\braket{f_{Q}^{'}|FS}.
\end{equation}
Here the sum is over all eigenstates $f$ described above, which have been given a subscript of $Q$ to highlight the fact
that they depend on the total momentum of the system.
This total momentum is equal to the initial momentum of the impurity as the system has evolved from an initial state
consisting of the impurity at said momentum and a Fermi sea at $0K$.
The fermionic gas state alone is represented as $\ket{FS}$ in the above equation.
The energy $E_{f}$ of the system is for each different eigenstate, and can be found with the equation given in
\eqref{eqn:Full_BA_eqns} for each eigenstate $f_{Q}$.
In the limit $t \to \infty$, the dependence of Eqn~\eqref{eqn:momentum_against_time} on $E_{f}$ is removed through time
averaging, becoming
\begin{equation}
  \label{eqn:inf_time_theoretical_momentum}
  \expval{P_{\downarrow}(\infty)} = Q - \sum_{f_{Q}}\braket{FS|f_{Q}}\expval{P_{\uparrow}}{f_{Q}}\braket{f_{Q}|FS}
\end{equation}
which finds the infinite time momentum of the impurity by only having to calculate a single sum over eigenstates, instead of
the double one required for Eqn~\eqref{eqn:momentum_against_time}.

Within this equaution there are two non-trivial values to calculate.
The first is the overlap of the eigenstate with the original Fermi sea, the
other is the matrix element of the background gas momentum operator between the
two Bethe eigenstates.
For this work, the overlap values and diagonal matrix elements of the
background gas momentum operator were calculated using a pre-existing
program~\cite{ZhenyaMCBA}, which uses the equations found in
references~\cite{Castella1993,Mathy2012sup} that we describe below.
The code to calculate off-diagonal matrix elements was written by the author,
and combined with the above code into the repo~\cite{MyMCBA}.

In order to calculate either values numerically, a normalisation constant for the eigenstates in
Equation~\eqref{eqn:BA_final_ansatz} must be found.
This is done first finding the dot product of an eigenstate with itself
\begin{equation}
  \label{eqn:self_dot_product}
  \braket{f_{Q}|f_{Q}}
  =
  \frac{Y_{f_{Q}} Y_{f_{Q}}}{N !}
  \int_{0}^{L} \, dx_{1} \, \cdots \, dx_{N}
  \detn_{N}(\overline{\Phi_{j}}(x_{l}))
  \detn_{N}(\Phi_{j}(x_{l}))
\end{equation}
where $Y_{f_{Q}}$ is the normalisation constant of the eigenstate $f_{Q}$.
Using the identity
\begin{multline}
  \label{eqn:magic_identity}
  \frac{1}{N!} \int_{0}^{L} dx_{1} \cdots dx_{N} \,\detn_{N}\big[\psi_{j}(x_{l})\big] \detn_{N}\big[\Phi_{j}(x_{l})\big] \\
  = \quad
  \detn_{N}\bigg[\int_{0}^{L} \,dy \, \psi_{j}(y) \Phi_{l}(y) \bigg]
\end{multline}
valid for any functions $\Phi_{j}, \psi_{j}$, Equation~\eqref{eqn:self_dot_product} can be written
\begin{equation}
  \label{eqn:neat_dot_product}
  \braket{f_{Q}|f_{Q}}
  =
  Y_{f_{Q}} Y_{f_{Q}} \detn_{N} \bigg[ \int_{0}^{L} dx \, \overline{\Phi_{j}}(x) \Phi_{l}(x) \bigg].
\end{equation}
Using the same choice of $a_{j}^{t}$ for Eqn~\eqref{eqn:phi_constraint} as references~\cite{Castella1993,Mathy2012},
$\Phi_{j}(x)$ can be written as
\begin{equation}
  \label{eqn:full_phi}
  \Phi_{j}(x)
  =
  \frac{1}{\sqrt{L}}
  \bigg[
    e^{i\big(k_{j}x + \delta_{j} \big)} -
    \frac{\theta_{j}}{\Theta}
    \sum_{t=1}^{N+1} e^{i\big(k_{t}x + \delta_{t}\big)}
    \bigg]
    .
\end{equation}
Inserting equation~\eqref{eqn:full_phi} into equation~\eqref{eqn:self_dot_product} and solving for $Y_{f_{Q}}$, we get
\begin{equation}
  \label{eqn:final_normalisation}
  \lvert Y_{f_{Q}} \rvert^{-2}
  =
  \frac{1}{\Theta^{2}} \big( \sum_{t=1}^{N+1} \frac{\theta_{t}^{2}}{1 + \theta_{t}^{2}} \big)
  \prod_{t=1}^{N+1} (1 + \theta_{t}^{2}).
\end{equation}
Where the conveniance variables $\theta, \Theta$ are defined below
\begin{gather}
  \label{eqn:theta_variables}
  \frac{L}{2} k_{j}
  =
  n_{j} \pi - \delta_{j} \\
  \theta_{j} = \sqrt{\frac{8}{gL}} \sin(\delta_{j}) \\
  \Theta = \sum_{t=1}^{N+1} \theta_{t}.
\end{gather}
For singular eigenstates, where $\delta_{j} = 0, \qquad j = 1, \ldots, N$, we have the relation
\begin{equation}
  \label{eqn:inf_theta_limit}
  \lim_{c \to -\infty} \frac{\theta}{\Theta}
  =
  \frac{1}{N+1}
\end{equation}
which implies
\begin{equation}
  \label{eqn:final_normalisation_singular}
  Y_{f_{Q}}
  =
  \sqrt{N+1}.
\end{equation}

The equation to calculate the overlaps $\braket{FS|f_{Q}}$ in
Eqn~\eqref{eqn:momentum_against_time} is taken from
reference~\cite{Mathy2012sup}, and restated below
\begin{equation}
  \label{eqn:S52}
  \braket{FS|f_Q}
  =
  Y_{f_Q} \detn_{N} \chi
\end{equation}
where $\chi$ is an $N \times N$ matrix whose elements are defined by
\begin{equation}
  \label{eqn:S53}
  \chi_{j}^{l}
  =
  \frac{\theta_{l}}{\sqrt{a}} \bigg[\frac{1}{u_{j} - \frac{L}{2} k_{l}} -
    \frac{1}{\Theta} \sum_{t=1}^{N + 1} \frac{\theta_{t}}{u_{j} - \frac{L}{2} k_{t}} \bigg],
  \qquad j,l = 1,\ldots,N.
\end{equation}
In the singular case, equation~\eqref{eqn:S52} has an easier representation, using
Equation~\eqref{eqn:final_normalisation_singular} and the alternate equation for the determinant in
Eqn~\eqref{eqn:S52} below.
\begin{equation}
  \label{eqn:final_overlap_singular}
  \detn_{N} \chi =
  \begin{cases}
    \frac{1}{N+1}, & u_{j} = \frac{L}{2} k_{j} \\
    \frac{-1}{N+1}, & u_{j} = \frac{L}{2} k_{j+1} \\
    0, & otherwise
  \end{cases}
\end{equation}

The matrix elements of Equation~\eqref{eqn:momentum_against_time}
$\matrixel{f_{Q}}{P_{\uparrow}}{f_{Q}^{'}}$ are given in
Reference~\cite{Mathy2012sup} as
\begin{equation}
  \label{eqn:S56}
  \matrixel{f_{Q}}{P_{\uparrow}}{f_{Q}^{'}}
  =
  Y_{f_{Q}}Y_{f_{Q}^{'}} \frac{\partial}{\partial \lambda} \bigg(\detn_{N} \big(\mathcal{Y} + \lambda \mathcal{Z} \big) \bigg) \rvert_{\lambda = 0}
\end{equation}
where
\begin{multline}
  \label{eqn:final_Y}
  \mathcal{Y}_{j}^{l}
  =
  \int_{0}^{L} dy \, \overline{\Phi_{j}} (x) \Phi_{l}^{'}(y) \\
  =
  K(k_{l}^{'}, k_{j})
  -
  \frac{\theta_{j}}{\Theta} \sum_{t=1}^{N+1} K(k_{l}^{'}, k_{t}) \\
   -
   \frac{\theta_{l}^{'}}{\Theta^{'}} \sum_{t=1}^{N+1} K(k_{l}^{'}, k_{j})
   +
   \frac{\theta_{j} \theta_{l}^{'}}{\Theta \Theta^{'}} \sum_{t,t^{'}=1}^{N+1} K(k_{t}^{'}, k_{t})
\end{multline}
\begin{multline}
  \label{eqn:final_Z}
  \mathcal{Z}_{j}^{l}
  =
  \int_{0}^{L} dy \, \overline{\Phi_{j}} (x) \partial_{y} \Phi_{l}^{'}(y) \\
  =
  k_{l}^{'}K(k_{l}^{'}, k_{j})
  -
  k_{l}^{'} \frac{\theta_{j}}{\Theta} \sum_{t=1}^{N+1} K(k_{l}^{'}, k_{t}) \\
   -
   \frac{\theta_{l}^{'}}{\Theta^{'}} \sum_{t=1}^{N+1} k_{t}^{'} K(k_{t}^{'}, k_{j})
   +
   \frac{\theta_{j} \theta_{l}^{'}}{\Theta \Theta^{'}} \sum_{t,t^{'}=1}^{N+1} k_{t^{'}}^{'}K(k_{t^{'}}^{'}, k_{t})
\end{multline}
\begin{equation}
  K(k^{'}, k)
  =
  \begin{cases}
    1, & if \quad k^{'} = k \\
  \frac{e^{i(k^{'} - k)L} - 1}{i(k^{'} - k)L} e^{i(\delta^{'} - \delta)}, & otherwise
  \end{cases}
\end{equation}

Note that since $K$ is real
\begin{multline}
  K(k^{'}, k)
  =
  \frac{e^{i(k^{'} - k)L} - 1}{i(k^{'} - k)L} e^{i(\delta^{'} - \delta)} \\
  =
  \frac{e^{2\pi i(n^{'} - n)}e^{-2i(\delta^{'} - \delta)} - 1}{i(k^{'} - k)L} e^{i(\delta^{'} - \delta)}
  =
  \frac{e^{-i(\delta^{'} - \delta)} - e^{i(\delta^{'} - \delta)}}{i(k^{'} - k)L} \\
  =
  \frac{2i \sin(\delta^{'} - \delta)}{i(k^{'} - k)L}
  =
  \frac{2\sin(\delta^{'} - \delta)}{(k^{'} - k)L}
\end{multline}
then so are $\mathcal{Y}$ and $\mathcal{Z}$.

We look at this formula seperately for when $\ket{f_{Q}^{'}} = \ket{f_{Q}}$ and $\ket{f_{Q}^{'}} \neq \ket{f_{Q}}$.
First, for the diagonal matrix elements, we have
\begin{equation}
  \label{eqn:final_diagonal}
  \matrixel{f_{Q}}{P_{\uparrow}}{f_{Q}}
  =
  q
  - \big(
      \sum_{t=1}^{N+1} \frac{\theta_{t}^{2}}{(1 + \theta_{t}^{2})} k_{t}
    \big)
    \big(
      \sum_{t=1}^{N+1} \frac{\theta_{t}^{2}}{(1 + \theta_{t}^{2})}
    \big)^{-1}.
\end{equation}
which, when $f_{Q}$ is singular, becomes
\begin{equation}
  \label{eqn:final_diagonal_singular}
  q \big( 1 - \frac{1}{N+1} \big).
\end{equation}

To find a computationally efficient manner to calculate the off-diagonal case, we take equation~\eqref{eqn:S56} and
manipulate it in two different ways.
First we separate it out into two different determinants
\begin{align}
  \detn(\mathcal{Y} + \lambda \mathcal{Z})
   = \detn \big[ \mathcal{Y}(\mathbb{1} + \mathcal{Y}^{-1} \lambda \mathcal{Z})
    \big] \\
  = \detn(\mathcal{Y}) \detn(\mathbb{1} + \mathcal{Y}^{-1} \lambda \mathcal{Z}) \\
\end{align}
and second we use the identity~\eqref{eqn:sherman_morrison} to transform the more complicated determinant into a trace
\begin{equation}
  \label{eqn:sherman_morrison}
  \ln \big( \detn(X) \big)
  =
  \tr \big( \ln(X) \big)
\end{equation}
\begin{multline}
  \frac{\partial}{\partial \lambda} \detn(\mathcal{Y} + \lambda \mathcal{Z})
  = \frac{\detn(\mathcal{Y} + \lambda \mathcal{Z})}{\detn(\mathcal{Y} + \lambda \mathcal{Z})}
  \frac{\partial}{\partial \lambda} \detn(\mathcal{Y} + \lambda \mathcal{Z}) \\
  = \detn(\mathcal{Y} + \lambda \mathcal{Z})
  \frac{\partial}{\partial \lambda} ln \big( \detn(\mathcal{Y})  \detn(\mathbb{1} + \mathcal{Y}^{-1} \lambda \mathcal{Z}) \big)
\end{multline}
\begin{multline}
  \label{eqn:use_sherman_morrison}
  \frac{\partial}{\partial \lambda} \ln \big( \detn(\mathcal{Y})  \detn(\mathbb{1} + \mathcal{Y}^{-1} \lambda \mathcal{Z}) \big) \\
  =
  \frac{\partial}{\partial \lambda} \ln \big( \det(\mathcal{Y}) \big) \rvert_{\lambda = 0}
  + \frac{\partial}{\partial \lambda} \ln \big( \det(\mathbb{1} + \mathcal{Y}^{-1} \lambda \mathcal{Z}) \big) \rvert_{\lambda = 0} \\
  =
  0
  +
  \frac{\partial}{\partial \lambda} \tr \big( \ln(\mathbb{1} + \mathcal{Y}^{-1} \lambda \mathcal{Z}) \big) \rvert_{\lambda = 0} \\
  =
  \tr \frac{\partial}{\partial \lambda} \big( \ln(\mathbb{1} + \mathcal{Y}^{-1} \lambda \mathcal{Z}) \big) \rvert_{\lambda = 0}
\end{multline}
Next the logarithmic expansion is taken from the last form of equation~\eqref{eqn:use_sherman_morrison}, and we again use the
position the derivative is taken at to simplify the form
\begin{multline}
  \ln (\mathbb{1} + \mathcal{Y}^{-1} \lambda \mathcal{Z})
  =
  \lambda \mathcal{Y}^{-1} \mathcal{Z} + \frac{\big(\lambda \mathcal{Y}^{-1} \mathcal{Z}\big)^{2}}{2} + \ldots \\
  \implies
  \frac{\partial}{\partial \lambda} \big( \ln (\mathbb{1} + \mathcal{Y}^{-1} \lambda \mathcal{Z}) \big) \rvert_{\lambda = 0}
  =
  \mathcal{Y}^{-1} \mathcal{Z} \\
  \implies
  \frac{\partial}{\partial \lambda} \ln \big( \detn(\mathcal{Y} + \lambda \mathcal{Z}) \big) \rvert_{\lambda = 0}
  =
  \tr(\mathcal{Y}^{-1} \mathcal{Z}) \\
  \implies
  \frac{\partial}{\partial \lambda} \detn(\mathcal{Y} + \lambda \mathcal{Z}) \rvert_{\lambda = 0}
  =
  \tr(\mathcal{Y}^{-1} \mathcal{Z}) \detn(\mathcal{Y} + \lambda \mathcal{Z})
\end{multline}
where we have taken advantage of the fact that the derivative is taken at $\lambda = 0$.

In order to solve the above equation we split the definition of $\mathcal{Y}^{-1}$ via singular value decomposition (SVD)
\begin{equation}
  \tr(\mathcal{Y}^{-1}\mathcal{Z})
  =
  \tr \big((\mathcal{U} \Sigma_{Y}\mathcal{V}^{*})^{-1} \mathcal{Z}\big)
  =
  \tr (\mathcal{V}^{*^{-1}} \Sigma_{Y}^{-1} \mathcal{U}^{-1} \mathcal{Z})
\end{equation}
which, using the fact that $\mathcal{Y}$ is real, and hence $\mathcal{V}$ and $\mathcal{U}$ are both unitary and real, can be
represented as
\begin{equation}
  \tr \big( \Sigma_{Y}^{-1} (\mathcal{U}^{T} \mathcal{Z} \mathcal{V}) \big).
\end{equation}

Using the fact $\Sigma$ is diagonal, and that $\detn (\Sigma_{Y} ) = \detn (\mathcal{Y})$, we can write
\begin{multline}
  \tr \big( \Sigma_{Y}^{-1} (\mathcal{U}^{T} \mathcal{Z} \mathcal{V}) \big) \detn(\Sigma_{Y})
  =
  \diag (\Sigma_{Y}^{-1}) \cdot \diag (\mathcal{U}^{T} \mathcal{Z} \mathcal{V}) \cdot \detn(\Sigma_{Y}) \\
  =
  \bigg[
    \frac{1}{\Sigma_{2}},
    \frac{1}{\Sigma_{3}},
    \cdots,
    \frac{1}{\Sigma_{N+1}}
    \bigg]
  \diag (\mathcal{U}^{T} \mathcal{Z} \mathcal{V})
  \prod_{n=2}^{N+1} \Sigma_{n} \\
  =
  \bigg[
    \frac{\prod_{n=2}^{N+1} \Sigma_{n}}{\Sigma_{2}},
    \frac{\prod_{n=2}^{N+1} \Sigma_{n}}{\Sigma_{3}},
    \cdots,
    \frac{\prod_{n=2}^{N+1} \Sigma_{n}}{\Sigma_{N+1}}
    \bigg]
  \diag (\mathcal{U}^{T} \mathcal{Z} \mathcal{V}) \\
  =
  \bigg[
    \prod_{n \neq 2}^{N+1} \Sigma_{n},
    \prod_{n \neq 3}^{N+1} \Sigma_{n},
    \cdots,
    \prod_{n=2}^{N} \Sigma_{n}
    \bigg]
  \diag (\mathcal{U}^{T} \mathcal{Z} \mathcal{V}).
\end{multline}
Finally, using the fact that the definition of $\mathcal{Y}$ is the same as the
matrix used in the dot product of eigenstates~\eqref{eqn:neat_dot_product} we
know that $\detn(\mathcal{Y}) = 0$ for off-diagonal states.
This means that one value of $\Sigma_{n}$ must be zero.
Without loss of generality we can set this to be the element $N$, so we have the computationally efficient representation
\begin{equation}
  \label{eqn:final_off_diagonal}
  \matrixel{f_{Q}}{P_{\uparrow}}{f_{Q}^{'}}
  =
  Y_{f_{Q}} Y_{f_{Q}^{'}} \frac{\partial}{\partial \lambda} \bigg(\detn_{N} \big(\mathcal{Y} + \lambda \mathcal{Z} \big) \bigg) \rvert_{\lambda = 0}
  =
  Y_{f_{Q}} Y_{f_{Q}^{'}}  \prod_{n=2}^{N}\Sigma_{n} \cdot (\mathcal{U}^{T}\mathcal{Z}\mathcal{V})_{NN}
\end{equation}
where $(\mathcal{U}^{T}\mathcal{Z}\mathcal{V})_{NN}$ is the final element of the $N \times N$ matrix $\mathcal{U}^{T}\mathcal{Z}\mathcal{V}$.
This equation is only valid for off-diagonal elements, so the diagonal elements must be calculated with
Equation~\eqref{eqn:final_diagonal}.
For the special case of $c = - \infty$ calculating the matrix elements requires accounting for the singularities in the
$\mathcal{Y}$ and $\mathcal{Z}$ matrices.
This is done by using equation~\eqref{eqn:inf_theta_limit} for the singular roots in equations~\eqref{eqn:final_Z} and
\eqref{eqn:final_Y}.

The equations~(\ref{eqn:final_diagonal},\ref{eqn:final_normalisation},\ref{eqn:S52}), and their special case equivalents for
singular Bethe roots~(\ref{eqn:final_diagonal_singular},\ref{eqn:final_overlap_singular},\ref{eqn:final_normalisation_singular}) were
already encoded into reference~\cite{ZhenyaMCBA}.
This work required implementing equation~\eqref{eqn:final_off_diagonal} in a distributed manner, which allowed calculating the full momentum against
time evolution of the impurity via Equation~\eqref{eqn:momentum_against_time}.

As there are an infinite number of eigenstates on the RHS of Eqn~\eqref{eqn:momentum_against_time}, some subset must be taken
for a numerical calculation of the momentum.
Given that this will inevitably introduce some error in the momentum calculated, we need some way to ensure the subset of
states we are using reproduces the actual value of Eqn~\eqref{eqn:momentum_against_time} close enough for quantitative results.
A quantitative bound on the error in momentum has been derived in reference~\cite{Mathy2012sup}, which depends on a bound in
the absolute value of the matrix element $\bar{P} = \sup(\abs{\matrixel{f_{Q}}{P_\downarrow}{f_{Q}^{'}}})$, and the saturation of
\begin{equation}
  \label{eqn:saturation_value}
  \varsigma = \sum_{i=1}^{N} \abs{\braket{FS|f_{Q,i}}}^{2}.
\end{equation}
This value $\varsigma$ must approach $1$ as $N \to \infty$ due to the completeness of the Bethe
eigenstates~\cite{Mathy2012sup}.
It is noteworthy that this bound on the error,
\begin{equation}
  \label{eqn:error_bound}
  \sqrt{\bar{P}^{2}(2\varsigma(1 - \varsigma) + 2(1 - \varsigma)^{2})}
\end{equation}
is independent of time, which allows us to plot the impurity's momentum for large time with the same accuracy as any other
point.
We will use this attribute heavily when inspecting the momentum revivals of the system in Section~\ref{sec:revivals}, which
can happen on a time scale of $t \approx 140t_{F}$.

Throughout the text we talk of the saturation of the sum rule $\varsigma$ instead of the bound on the error in the momentum.
This is done to keep the relation between the number of states counted, and the value of $\abs{\braket{FS|f_{Q}}}^{2}$ for
those states clear.
We wish to maintain the connection between these values, as while the time dependent momentum evolution has been studied
before~\cite{Mathy2012,Knap2014}, the isolation of eigenstate pairs responsible for each feature of the momentum in
Chapter~\ref{sec:subsets} is wholly novel work and can be better understood in these terms.

\subsection{Method}
\label{sec:m&m_method}

Despite these pre-existing solutions and methods, the evaluation of Eqn~\eqref{eqn:momentum_against_time}, is still difficult,
as when the system is highly excited, a reliable calculation has to account for the contribution of a large number of
eigenstates~\cite{Mathy2012sup}.

In order to calculate states and choose which states to use, we use a program written in the Python programming
language~\cite{Python} with the Scipy and Numpy~\cite{Scipy,Numpy} external libraries.
The program uses a stochastic sampling algorithm to find and choose a smaller subset than in~\cite{Mathy2012}, that will
still reliably reproduce observables, the discovered states are then accumulated with the greatest $\abs{\braket{FS|f_{Q}}}$
first.
This program has been used in work before~\cite{Burovski2014}, is freely available online~\cite{ZhenyaMCBA}, and provides not
only the overlap value for every state used, but the diagonal matrix elements of the momentum operator.
To this program, we add the functionality to calculate the off-diagonal matrix elements of the momentum operator
$\matrixel{f_{Q}}{P_{\uparrow}}{f_{Q}^{'}}$ for $f_{Q} \ne f_{Q}^{'}$ (see Equation~\eqref{eqn:final_off_diagonal}, and hence
the impurity's time-dependent momentum evolution.
This additional code is also freely available, an overview of its structure is given in Appendix~\ref{sec:code_details}, and
the source code can be found at~\cite{MyMCBA}.

\chapter{Observables of the System}
\label{sec:observables}

%
%
%
%


\section{Introduction}
\label{sec:mom_features}

This chapter details the difference in $\expval{P_{\downarrow}(t)}$ with differing parameters of the system.
The three physical parameters we can change within the restrictions of our model are the system size, the initial momentum
of the impurity, and the dimensionless interaction strength parameter $\gamma$.
Note the only dependency of $\expval{P_{\downarrow}(t)}$ on the density of the background gas $n = \nicefrac{N}{L}$ or the
interaction strength $g$ is via the dependency on $\gamma$ and not on the values themselves.
While the momentum against time of the impurity has been discussed in other work~\cite{Mathy2012,Mathy2012sup,Knap2014}, we
look further into the revivals of the impurity's momentum that come from finite size effects in Section~\ref{sec:revivals}.
We then present comparisons between $\expval{P_{\downarrow}(\infty)}$ as calculated from
Eqn~\eqref{eqn:inf_time_theoretical_momentum} and the momentum plateau obtained when plotting the full evolution of
$\expval{P_{\downarrow}(t)}$ in Section~\ref{sec:inf_time}.
We also discuss the variation of $\expval{P_{\downarrow}(t)}$ with $\varsigma$ (see Section~\ref{sec:m&m_method}), which will
provide grounding for the discussion on separating contributions provided in Chapter~\ref{sec:subsets}.
Where this chapter overlaps with~\cite{Mathy2012,Knap2014}, there is consistent agreement, corroborating their results and
increasing the confidence that both our programs give the correct numerical value for the solution of this model.


\section{Justification}
\label{sec:justification}
As mentioned in Section~\ref{sec:m&m_model}, when calculating $\expval{P_{\downarrow}(t)}$, a large number of states must be
accounted for in the sum of Eqn~\eqref{eqn:momentum_against_time}.

\begin{figure}[ht!]
  \centering
  \includegraphics[width=\linewidth]{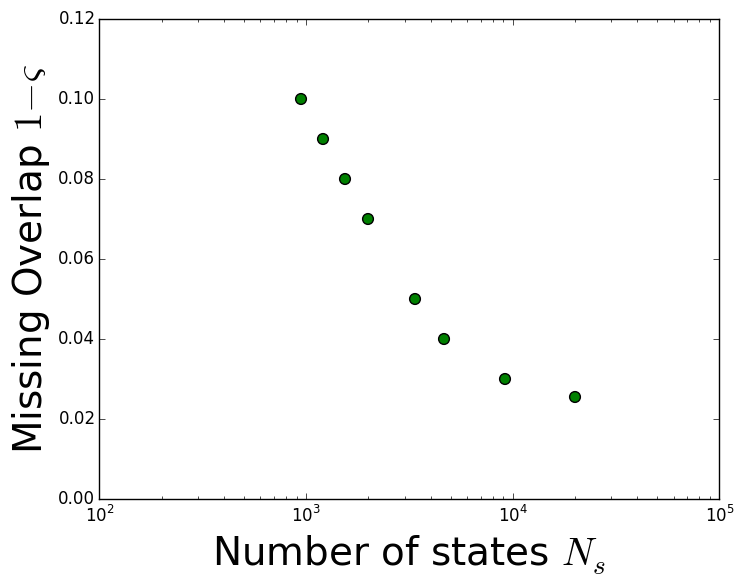}
  \caption{\label{fig:overlap_vs_numstates}
    The number of states required for a range of saturation values for $\varsigma$.
    We show the semi-log plot of how $\varsigma$ changes with the number of states.
    The progression is linear until a $\varsigma \approx 0.96$, at which point many more states are required to provide
    further accuracy.
  }
\end{figure}

\begin{figure}[ht!]
  \centering
  \includegraphics[width=\linewidth]{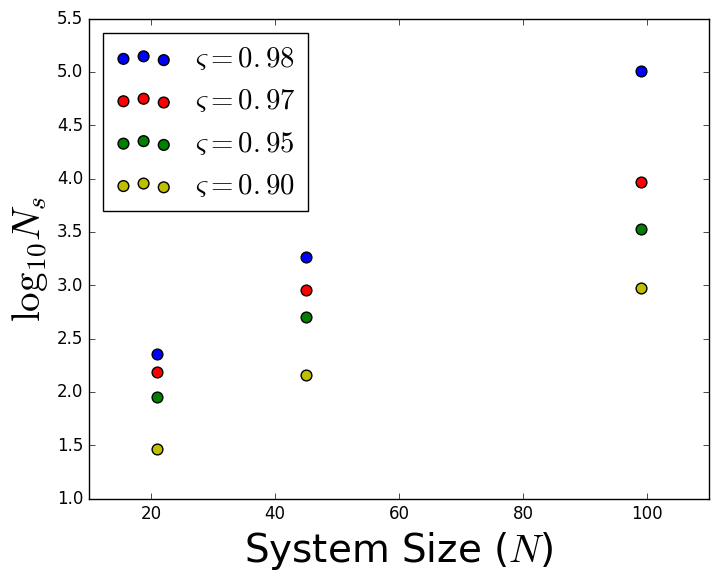}
  \caption{\label{fig:states_overlap_systemsize}
    The number of states required for a range of saturation values for $\varsigma$.
    We show the semi-log plot of how the number of states required for a given $\varsigma$ changes with increasing system
    size, the progression is not linear, so this is not an exponential relation, but the plot does show a large increase in
    the number of states required as larger systems are used.
  }
\end{figure}

Reaching a high saturation of $\varsigma$ is easier said than done, as while $1 - \varsigma$ decreases linearly with the
log of the number of states (see Fig~\ref{fig:overlap_vs_numstates}), this relation only happens until $\varsigma
\approx 0.96$, and the number of states required for a given $\varsigma$ strongly increases with system size, as seen in
Fig~\ref{fig:states_overlap_systemsize}.

Because of the computational restrictions on the number of states used and the system size (see
Appendix~\ref{sec:code_details}),
calculating the impurity's momentum against time is prohibitively expensive for a number of states $ N_{s} > 20000 $ and a
system size of $ N = 99 $, which gives an overlap of $\varsigma \approx 0.97$ which is not a large enough $\varsigma$ for
confidence in our results from the saturation itself.
This is a problem, as for those systems where we can reach $\varsigma = 0.99$ some of the more interesting features of the
system are hidden by finite size effects (see Sections~\ref{sec:revivals}~\ref{sec:inf_time})

\begin{figure}[H]
  \centering
  \begin{subfigure}{0.9\linewidth}
    \centering
    \includegraphics[width=\linewidth]{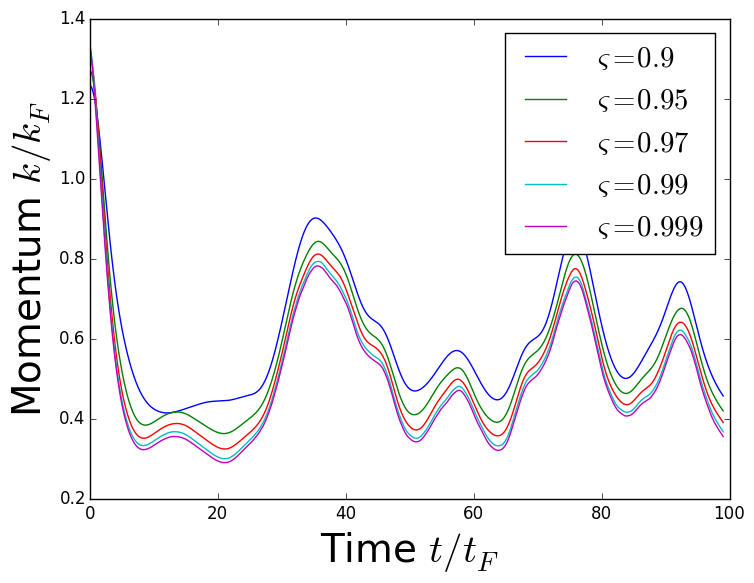}
    \caption{}
    \label{fig:downwards_shift}
  \end{subfigure}
  \begin{subfigure}{0.9\linewidth}
    \centering
    \includegraphics[width=\linewidth]{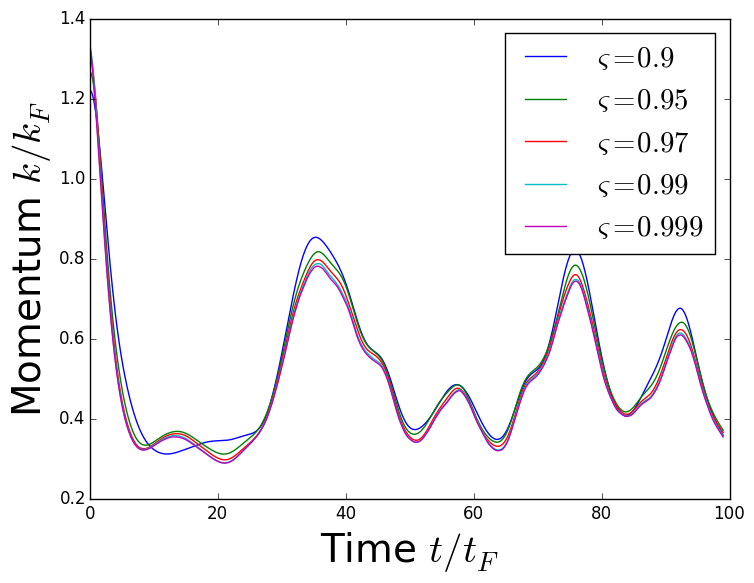}
    \caption{}
    \label{fig:normalised_overlap}
  \end{subfigure}
  \caption{
    \subref{fig:downwards_shift}) As $\varsigma$ is increased past $0.9$ to $0.999$, the main observable change in
    $\expval{P_{\downarrow}(t)}$ is a downwards shift over the entire time range.
    \subref{fig:normalised_overlap}) This downwards shift can be normalised out by the value of $\varsigma$ to provide
    approximate results when the number of states required for a satisfactory $\varsigma$ value is too high.
  }
  \label{fig:overlap_shift}
\end{figure}

Fortunately, we find evidence in these smaller systems that a missing sum rule contribution this small does not change the
general shape of the momentum evolution, but rather introduces some minor variances, and a total downwards shift in the
momentum of the impurity as seen in Fig~\ref{fig:downwards_shift}.
While the normalisation in this manner has no rigorous mathematical backing, it is a useful approximation for values of
the plateau in a wider range of parameters than otherwise available.
Fig~\ref{fig:changing_overlap} provides more details on how a change in $\varsigma$ affects $\expval{P_{\downarrow}(t)}$.
As our main focus in this chapter is on the momentum revivals, which are the feature least investigated in reference
\cite{Mathy2012}, the fact that these are stable once $\varsigma$ has passed $\varsigma \approx 0.9$ puts good confidence behind our
results.
As shown in Fig~\ref{fig:normalised_overlap}, normalising $\expval{P_{\downarrow}(t)}$ by the sum rule saturation we have
reached goes some way to accounting for the difference in the plateau of $\expval{P_{\downarrow}(t)}$.
This normalisation is done with the equation
\begin{equation}
  \label{eqn:renormalisation}
  \widetilde{\expval{P_\downarrow(t)}} = Q -
  \frac{\sum_{f_{Q},f_{Q}^{'}} e^{i t (E_{f} - E_{f^{'}})} \braket{FS|f_{Q}}\matrixel{f_{Q}}{P_\downarrow}{f_{Q}^{'}}\braket{f_{Q}^{'}|FS}}
       {\sum_{f_{Q}} \abs{\braket{FS|f_{Q}}}^{2}}
\end{equation}
which would provide the total and correct $\expval{P_{\downarrow}(t)}$ if the set of states accounted for provided a fully representative evolution.
This extends the range of system sizes we can investigate to 99 particles, beyond what has been seen previously
\cite{Mathy2012,Knap2014}.
The additional range lets us view the evolution of the system for a much longer time without the intrusion of
finite size effects (discussed in detail in later chapters).
However, the normalisation is not sufficiently effective to allow the approach to be used for investigating systems that
contain more than about $99$ particles.
The dependence of the properties of the momentum on $\varsigma$ is further explored in Sections~\ref{sec:inf_time} and \ref{sec:flutter}.

\section{Overall Momentum}
\label{sec:overall_momentum}

\begin{figure}[ht!]
  \centering
  \includegraphics[width=\linewidth, keepaspectratio]{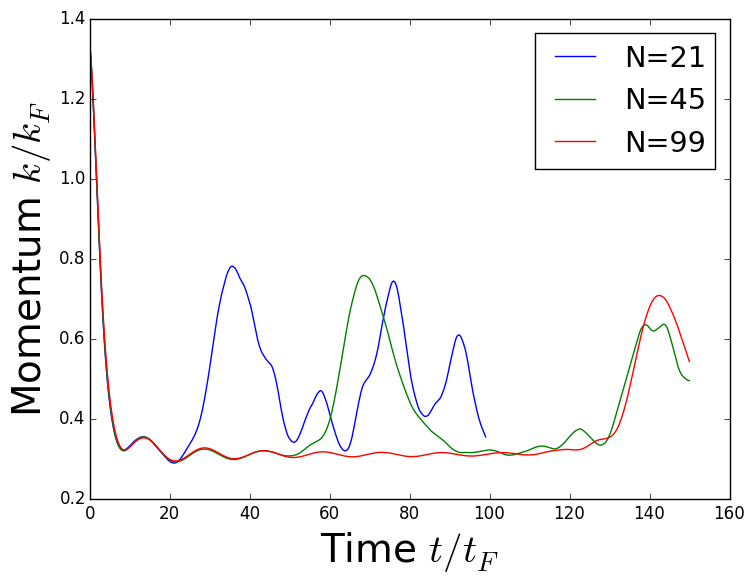}
  \caption{
    Impurity momentum evolution for multiple system sizes.
    Plots showing $\widetilde{\expval{P_{\downarrow}(t)}}$ over systems of $21$, $45$, and $99$ particles with constant
    $\gamma = 3$ and constant initial momentum of $Q = \frac{4}{3}k_{F}$.
    Until the revival in the impurity's momentum, the evolution of the impurity is identical for all system sizes, the
    revivals increase in period with a linear progression on the system size (see Fig~\ref{fig:revivals_vs_syssize}), and
    they are the only finite size effect apparent here.
    The consistency of the flutter and plateau is in agreement with~\protect\cite{Mathy2012}.
    In this plot we ensure $\varsigma$ is consistent for $N=21$ and $N=45$, however we were unable to match the $\varsigma$ for
    $N=99$, so we plot all data once appropriately normalised.
    When not normalised by the value of $\varsigma$, the only noticeable difference is a total shift downwards in the entire
    plot for $N=99$.
  }
  \label{fig:changing_syssize}
\end{figure}

\begin{figure}[ht!]
  \centering
  \begin{subfigure}{0.9\linewidth}
    \includegraphics[width=\linewidth]{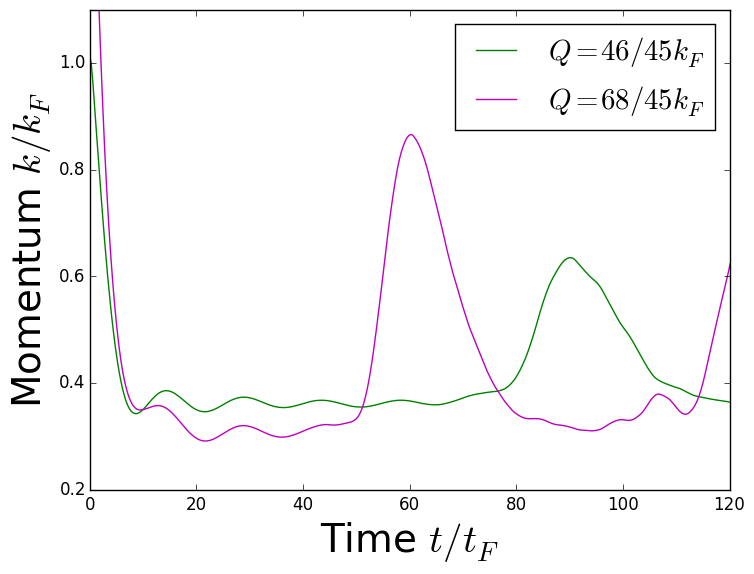}
    \caption{}
    \label{fig:change_initmom_high}
  \end{subfigure}
  \begin{subfigure}{0.9\linewidth}
    \includegraphics[width=\linewidth]{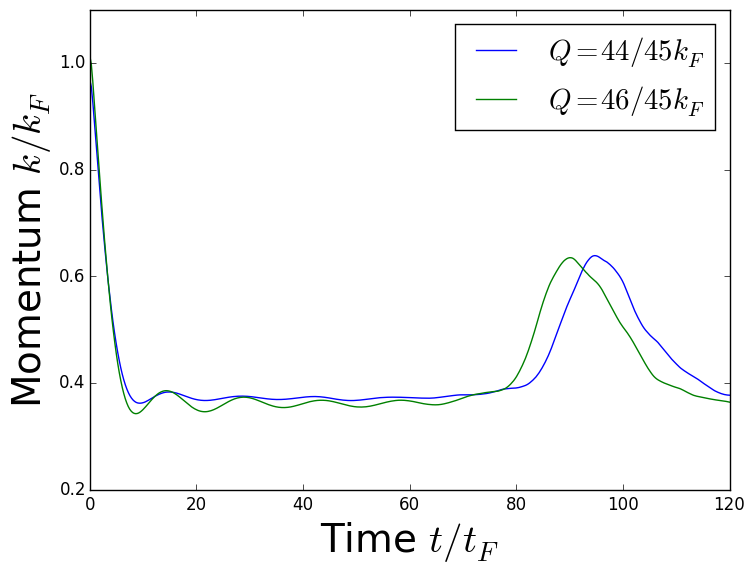}
    \caption{}
    \label{fig:change_initmom_crossover}
  \end{subfigure}
  \caption{
    $\expval{P_{\downarrow}(t)}$ over multiple values of $Q$, with a fixed system size $N = 45$ and interaction strength $\gamma = 3$.
    \subref{fig:change_initmom_high}) While $Q > k_{F}$, increasing $Q$ decreases both the plateau and time to the momentum
    revival (see Sections~\ref{sec:revivals} and \ref{sec:inf_time}).
    \subref{fig:change_initmom_crossover}) As $Q$ decreases past $k_F$, the flutter goes away, which is a central feature of
    reference~\protect\cite{Mathy2012}.\\
  }
  \label{fig:changing_initmom}
\end{figure}

\begin{figure}[ht!]
  \centering
  \includegraphics[width=\linewidth]{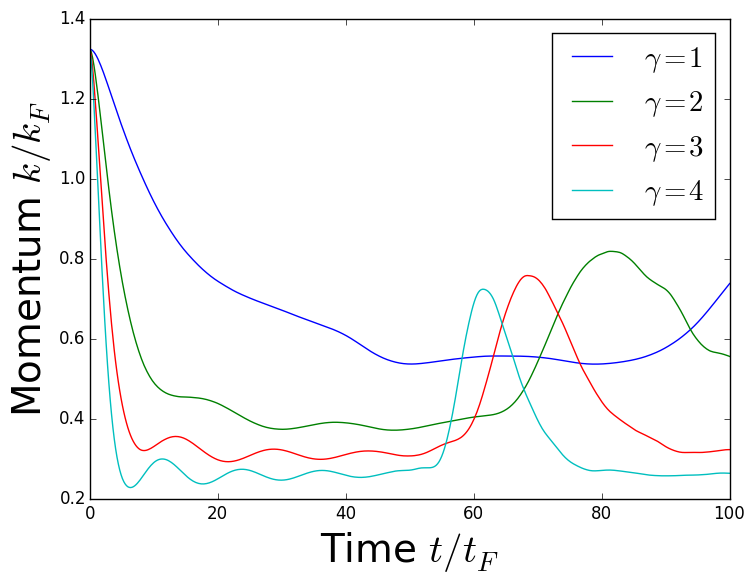}
  \caption{
    Change in $\expval{P_{\downarrow}(t)}$ with $\gamma$ for fixed system size $N = 45$ and $Q = \frac{4}{3}k_{F}$.
    As $\gamma$ increases, the revival period decreases, the plateau in the momentum decreases, and the flutter frequency
    increases.
    The change in the revival periods and the plateau can be qualitatively interpreted as an increase in the momentum
    transfer to the background gas, while the flutter follows the progression described in
    Equation~\eqref{eqn:flutter_equation}, formed from the argument presented in~\protect\cite{Mathy2012,Knap2014}.
  }
  \label{fig:changing_gamma}
\end{figure}

With the previous justification, we can access a wide range of system parameters, and view how the impurity's momentum
evolution changes within this extended parameter space.
We reiterate that within our model we have three physical parameters: the system size $N$, the initial momentum of the
impurity $Q$, and the dimensionless interaction strength $\gamma$.
The difference in the impurity's momentum evolution when changing each of these parameters can be seen in
Figures \ref{fig:changing_syssize}, \ref{fig:changing_initmom} and \ref{fig:changing_gamma} respectively.
While there are many changes throughout the parameter space, there are consistently three main features of the momentum
evolution: the regular revivals, a period of non-zero relatively constant momentum, and the small scale oscillations in
this region (dubbed ``quantum flutter'' in previous work~\cite{Mathy2012}).
These features were described in reference~\cite{Mathy2012}, and the same work thoroughly discussed both the phenomenology of the
plateau and quantum flutter, and gave an argument for the physical cause of the features.
Though previous work has thoroughly investigated two of these features, we will discuss each of them in turn over the following
sections to provide a full description of the system as a setting for future chapters.
The next section will discuss the momentum revivals, being the least investigated feature of the system, the
plateau and flutter are discussed after in Sections~\ref{sec:inf_time}, and \ref{sec:flutter}.

\begin{figure}[ht!]
  \centering
  \includegraphics[width=\linewidth]{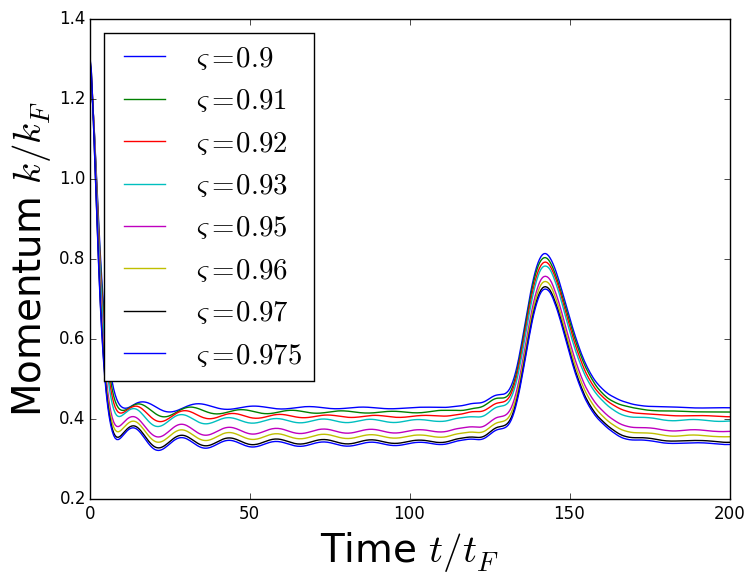}
  \caption{\label{fig:changing_overlap_unnormalised}
    The change in $\expval{P_{\downarrow}(t)}$ over $\varsigma$ with fixed $N = 99$, $\gamma = 3$ and $Q = \frac{4}{3}k_F$.
    Here a larger system size than previous is used, as the features which differ with changing overlap are sometimes
    obscured by finite size effects.
    As $\varsigma$ increases, the revival period is constant, the flutter frequency increases, and the plateau decreases.
    The flutter frequency increases with increasing overlap, but reaches a constant value at a $\varsigma$ of about $0.95$,
    while the plateau tends to some value, but has not saturated in the $\varsigma$ range shown.
  }
\end{figure}

\begin{figure}[h!]
  \centering
  \includegraphics[width=\linewidth]{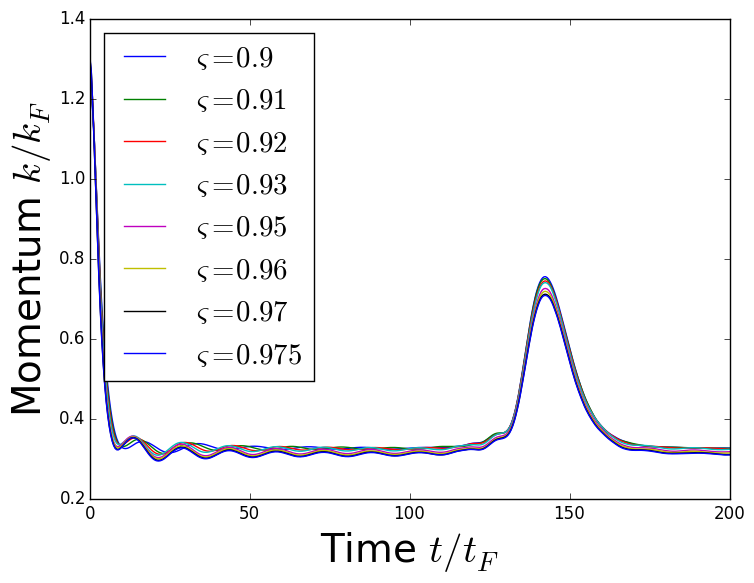}
  \caption{\label{fig:changing_overlap}
    The change in $\widetilde{\expval{P_\downarrow(t)}}$ with $\varsigma$ with fixed $N = 99$, $\gamma = 3$ and $Q = \frac{4}{3}k_F$.
    Of the two features that change with $\varsigma$, the plateau change can be almost factored out with normalisation.
  }
\end{figure}

Within these sections, we also show the change in $\expval{P_{\downarrow}(t)}$ over the non-physical parameter $\varsigma$ (of which the general shape is
shown in Figures \ref{fig:changing_overlap_unnormalised} and \ref{fig:changing_overlap}) where relevant.
While the change in the momentum evolution with $\varsigma$ is not a physical property of the system, but a property of the solution
we are using for this model, it is still important to know how $\expval{P_{\downarrow}(t)}$ changes with $\varsigma$, both in
order to gain insight into how the solution behaves, and because it is not always possible to reach large enough $\varsigma$
for a strong limit on the maximum error.
Knowing how $\expval{P_{\downarrow}(t)}$ changes with $\varsigma$ means we can investigate larger systems with an
understanding of what errors we are letting into our results.

As can be seen in Figures~\ref{fig:changing_overlap_unnormalised}, and~\ref{fig:changing_overlap}, the revival period is independent of the last $0.1$ in the $\varsigma$, though both
the flutter and plateau undergo changes.
While some of the change in the plateau can be normalised out by using $\widetilde{\expval{P_\downarrow(t)}}$ in place of
$\expval{P_{\downarrow}(t)}$ as described in Section~\ref{sec:justification}, we have no way to convert the flutter of an
under-saturated $\varsigma$ to what would occur with perfect saturation.

\clearpage
\section{Momentum Revivals}
\label{sec:revivals}

Of the three main features in $\expval{P_{\downarrow}(t)}$, the momentum plateau and the flutter are features present in the
thermodynamic limit of $N \to \infty$ with constant $\nicefrac{N}{L}$, while the revivals in the momentum are finite size effects
which can only be seen in systems small enough for the given parameters and time range.

While these revivals would not be present in a macroscopic scale gas, they're relevant in experiment, which often use
gases of the same order of magnitude $N$ as we can numerically probe~\cite{Kinoshita2005,Fertig2005,Palzer2009,Trotzky2011}.
The momentum revivals can be problematic, as they can mask the evolution of the other momentum features for smaller system
sizes, but they are an interesting feature themselves which have not yet been thoroughly investigated.

From a semi-classical argument, we can attribute the cause of the momentum revivals to the finite size of the system, with the
momentum packet put into the background gas by the impurity moving through the gas until it reaches and excites the impurity
again.
Were there hard boundaries in the system, the time this effect should appear would be influenced by the initial position of
the impurity; however in our model, the periodic boundary conditions give us translational invariance, which means the time
period of the momentum revivals is only determined by the physical parameters of the system that we have previously
discussed.  If we take the revival period as
\begin{equation}
  \label{eqn:revival_estimate}
  t_{rev} \approx \frac{L}{2 p_{relative}}
\end{equation}
where $t_{rev}$ is the revival period, $L$ is the length of the system, and $p_{relative} = Q - 2p$ (for $p$ representing the
value of the plateau seen) is the momentum in the packet put into the background gas relative to that of the impurity's
plateau, we can see this predicts a linear change in revival period with increasing system size, which matches what we
see in Figure~\ref{fig:changing_syssize}.
\begin{figure}[ht!]
  \centering
  \includegraphics[width=\linewidth]{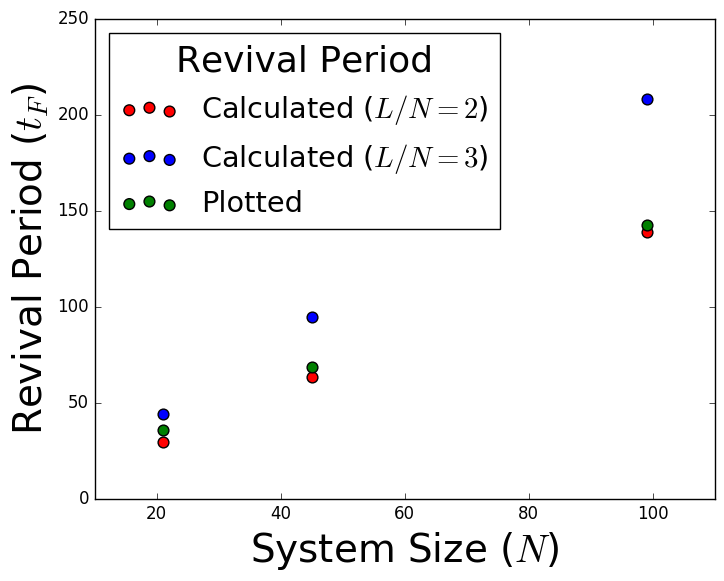}
  \caption{\label{fig:revivals_vs_syssize}
    The progression of the revival period $t_{rev}$ with changing system size $N$ increases linearly, which is in good
    agreement with Eqn~\eqref{eqn:revival_estimate}, and shows how the revivals are a finite size effect, diverging as
    $N \to \infty$.
    This plot was created with $Q = \nicefrac{4 k_{F}}{3}, \gamma = 3$, but the particulars of how $\gamma$ is set to its
    value are important for the prediction of Equation~\eqref{eqn:revival_estimate}.
    We find empirically, that fixing $\nicefrac{L}{N} = 2, g = 3$ gives the best predictions for
    Eqn~\eqref{eqn:revival_estimate}, but for $\nicefrac{L}{N} = 3, g = 2$ (as an example) the estimations are further off.
  }
\end{figure}
A similar increase in revival period and qualitative argument for the increase with system size was presented in
\cite{Mathy2012sup}, but was not fully explored.

\begin{figure}[ht!]
  \centering
  \includegraphics[width=\linewidth]{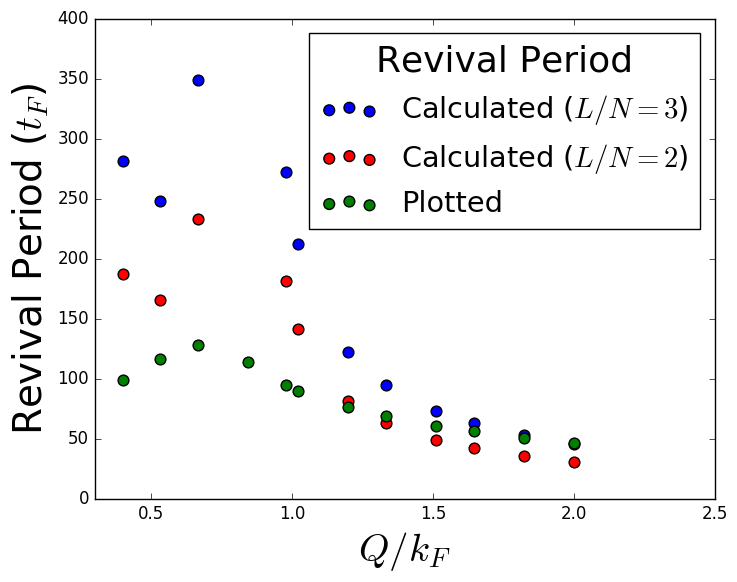}
  \caption{\label{fig:revivals_vs_initmom}
    The change with $Q$ of the revival period $t_{rev}$ progresses in a non-trivial manner, mostly following the progression
    of the momentum plateau (to be shown in Section~\ref{sec:inf_time}, see Figure~\ref{fig:infmom_vs_initmom}).
    For a low initial momentum $Q < k_{F}$, the approximation in Eqn~\eqref{eqn:revival_estimate} is very poor, failing to
    even qualitatively reproduce the progression, but as the initial momentum becomes greater than the Fermi momentum it
    gives a better prediction.
    Note the estimated revival periods for $Q = \frac{38}{45}k_{F}$ are not shown as they are greater than $1000t_{F}$, once again
    demonstrating this estimate is not useful for a low initial momentum.
    As with Figures~\ref{fig:revivals_vs_syssize} and~\ref{fig:revivals_vs_gamma}, the choice of how to set $\gamma$ is
    important for this prediction.
    The current plot was created with $\nicefrac{L}{N} = 2$ for the red points, and $\nicefrac{L}{N} = 3$ for the blue ones.
    While in Figures~\ref{fig:revivals_vs_syssize} and ~\ref{fig:revivals_vs_gamma}, the choice of $\nicefrac{L}{N} = 2$ has
    been shown to be the most accurate for $Q = \frac{4}{3}k_{F}$, this plot demonstrates a dependence of the optimum
    choice on the initial momentum $Q$, though the exact relation is currently unknown.
  }
\end{figure}

When this equation is applied to describe the change in the revivals over initial momentum $Q$ and the interaction strength
$\gamma$ it qualitatively matches what we see, the change in momentum passed to the background gas contains most of the
non-linearity of these changes.
Though it qualitatively reproduces changes for most parameters this prediction is far from perfect, completely neglecting the
fact that $\gamma$ can change with both $g$ and $L$, and failing to even qualitatively predict the progression for a low initial
momentum $Q > k_{F}$ (see Figure~\ref{fig:revivals_vs_initmom}).
\begin{figure}[ht!]
  \centering
  \includegraphics[width=\linewidth]{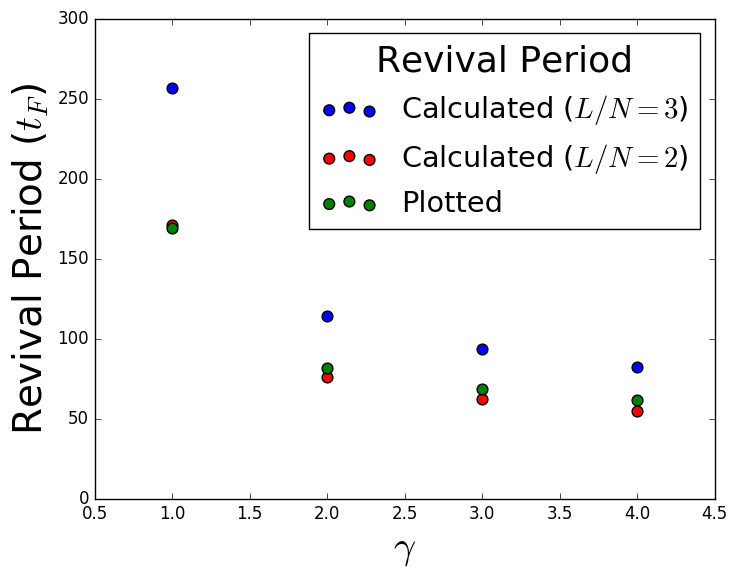}
  \caption{\label{fig:revivals_vs_gamma}
    The change in revival period $t_{rev}$ with respect to $\gamma$, shows that Eqn~\eqref{eqn:revival_estimate} qualitatively
    reproduces the progression of the momentum revivals.
    It should be noted that the revival period plotted for $\gamma = 1$ is highly suspect as the revival is not nearly
    as clear as other points, having a relative peak spanning $100t_{F}$, we chose the highest point of this peak, which was
    near its end.
    Like in Fig~\ref{fig:revivals_vs_syssize} and \ref{fig:revivals_vs_initmom} the way $\gamma$ is set has a strong
    influence on the accuracy of Eqn~\eqref{eqn:revival_estimate}, with two example data sets shown, we fix $\nicefrac{L}{N}$,
    at $2$ for the red points, and $3$ for the blue ones.
  }
\end{figure}
In Figures \ref{fig:revivals_vs_initmom}, \ref{fig:revivals_vs_syssize}, and \ref{fig:revivals_vs_gamma} we show how the
choice of $\nicefrac{L}{N}$ and $g$ for a fixed $\gamma$ affects the prediction of Eqn~\eqref{eqn:revival_estimate}.
The progression with $Q$ indicates Eqn~\eqref{eqn:revival_estimate} should have some dependence on the initial momentum,
possibly defining the ratio of $g$ to $L$ with which $\gamma$ is formed.

The fact this prediction is dependent on the ratio of $g$ to $L$, while the actual momentum is only dependent on their
product shows the limitations of the simple interpretation that leads to Eqn~\eqref{eqn:revival_estimate}.
Nevertheless the equation remains useful for systems with high initial momentum to show the qualitative progression when this
particular fault is sidestepped.

\begin{figure}[ht!]
  \centering
  \begin{subfigure}{0.9\linewidth}
    \includegraphics[width=\linewidth]{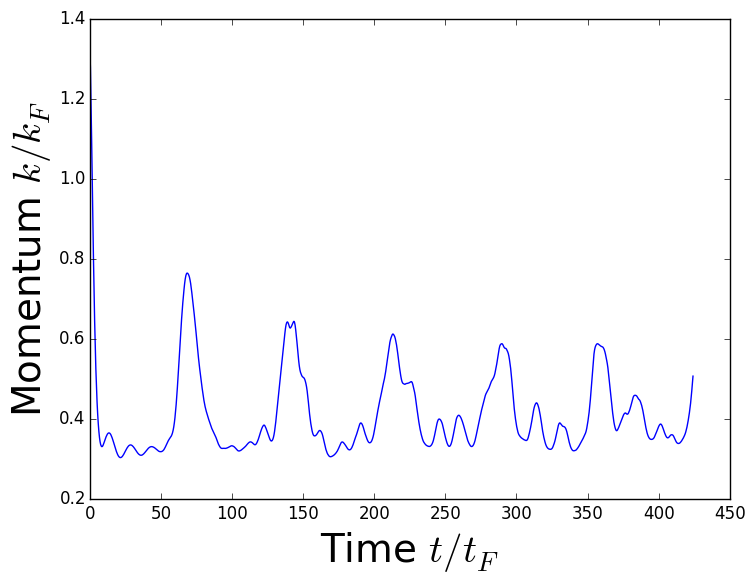}
    \caption{}
    \label{fig:medium_term_momentum_evolution}
  \end{subfigure}
  \begin{subfigure}{0.9\linewidth}
    \includegraphics[width=\linewidth]{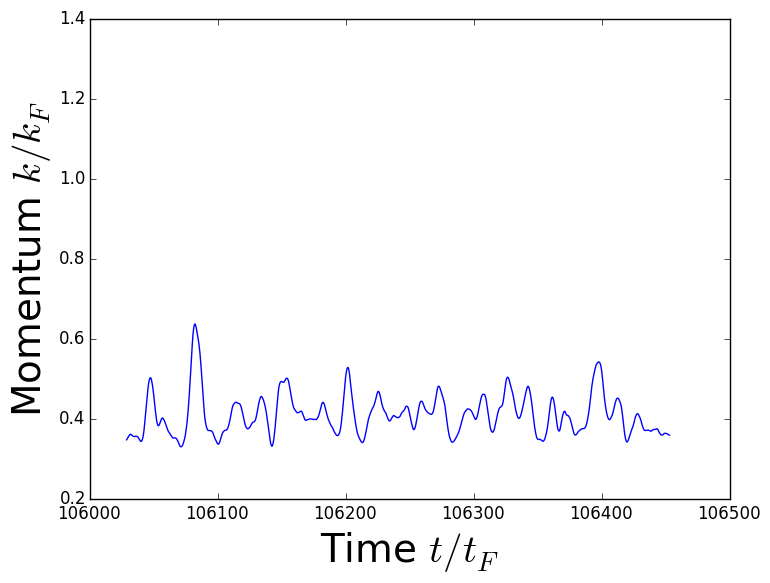}
    \caption{}
    \label{fig:long_term_momentum_evolution}
  \end{subfigure}
  \caption{
    \subref{fig:medium_term_momentum_evolution}) A plot of $\expval{P_{\downarrow}(t)}$ for many revivals shows how the momentum
    revivals initially decohere with increasing time, but the plot of very long $t \gg t_{rev}$ in
    \subref{fig:long_term_momentum_evolution}) shows that despite this, the momentum at very long term values is not
    completely stable.
    }
  \label{fig:long_term_evolution}
\end{figure}

Figure~\ref{fig:medium_term_momentum_evolution} shows a longer term evolution of the momentum, demonstrating an initial
decoherence of the revivals with increasing time.
Despite this apparent progression in the short term, a plot of $\expval{P_{\downarrow}(t)}$ for $t \gg t_{rev}$ shows no
point where they have been fully dispelled (see Fig~\ref{fig:long_term_evolution}), with some ranges of $t$ still showing quite
strong revivals.
From this information, we can see that for a finite system, the impurity never reaches a fully thermalised state, only ever
reaching the plateau before finite size effects set in.

As the period of the revivals is constant after a very low value of $\varsigma$ as seen in
figure~\ref{fig:changing_overlap}, we can infer the major contribution to this feature comes from states with among the
greatest $\abs{\braket{FS|f_Q}}$, as will be demonstrated alongside a thorough exploration into attributing eigenstate pairs
to each feature in $\expval{P_{\downarrow}(t)}$ in Chapter~\ref{sec:subsets}.

\section{Infinite Time}
\label{sec:inf_time}

The fact the impurity's saturation momentum is non-zero is an interesting phenomenon.
While a non-zero infinite time momentum inevitably draws comparisons to a superfluid, it must be stressed that typical
superfluidity, like the Bose Einstein condensate, does not in general survive the transition to one
dimension~\cite{Buchler2001,Sykes2009,Cherny2012}, and what aspects do cross over are strongly dependent on the particulars of
the system~\cite{Astrakharchik2004,Brand2005,Cherny2009,Ovchinnikov2010}.
A non-zero saturation momentum has previously been predicted via ballistic transport, through an argument based on
the dynamical conductivity in a system very similar system to this~\cite{Castella1995,Zotos2002}, but the current feature is
different.
The discovery of non-zero infinite time momentum in the current system~\cite{Mathy2012} has been of immediate interest,
receiving further investigation and generalisation in
references~\cite{Knap2014,Burovski2014,Gamayun2014,Gamayun2014b,Lychkovskiy2014}.
This section provides results corroborating what was seen in reference~\cite{Mathy2012} and also discusses how the plateau
found and the theoretical infinite time value $\expval{P_{\downarrow}(\infty)}$ relate to each other over a wide range of
changing system parameters.

\begin{figure}[ht!]
  \centering
  \includegraphics[width=\linewidth]{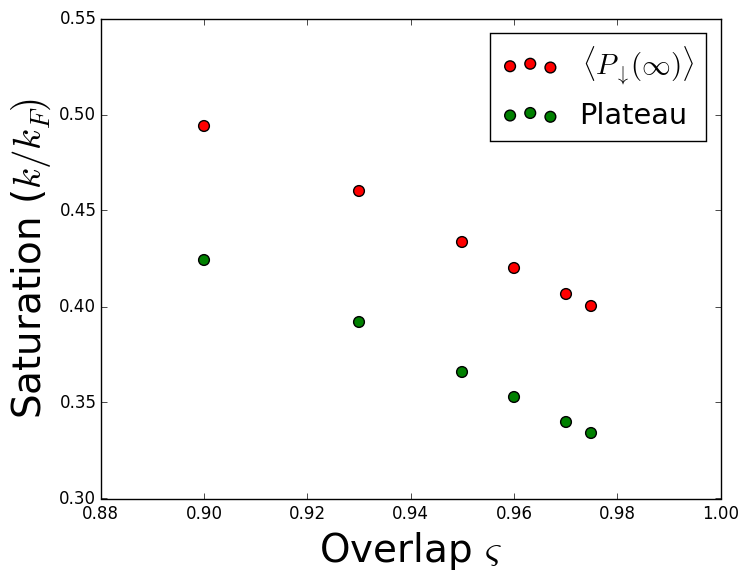}
  \caption{\label{fig:infmom_vs_overlap}
    Both saturation measures have a near linear dependence on the value of $\varsigma$ reached in the calculation,
    which can be normalised out to find the limit that would be reached for $N_{s} = \infty$.
    The gradient of this progression changes with different parameters, but once the overlap is large enough,
    $\varsigma \gtrsim 0.95$ then the linearity has always been seen to exist.
    This graph has been plotted for $N=45$, $\gamma = 3$, and $Q = \frac{4}{3}k_{F}$ and does not normalise the results
    by $\varsigma$.
  }
\end{figure}

We describe the change in both the apparent saturation momentum of the impurity, given by the plateau in
$\expval{P_{\downarrow}(t)}$ before the momentum revival takes place, and the theoretical saturation momentum of the
impurity from summing the time-independent contributions in Eqn~\eqref{eqn:momentum_against_time}.
The time-independent contributions are all elements in the sum where $\ket{f_{Q}} = \ket{f_{Q}^{'}}$, as these state pairs
are the only ones where the difference in energy is identically $0$, and hence the exponential in
Eqn~\eqref{eqn:momentum_against_time} is constant over all time $t$.
Particular care must be taken when measuring the plateau, as this is the feature of $\expval{P_{\downarrow}(t)}$ that most
strongly depends on the value of $\varsigma$ reached for the simulation as can be seen in Figures~\ref{fig:changing_overlap}
and \ref{fig:infmom_vs_overlap}, to mitigate this, we reach $\varsigma \ge 0.99$ where feasible, and for this section
we always normalise by $\varsigma$ achieved, in the manner described in Section~\ref{sec:justification} unless otherwise
specified.

\begin{figure}[ht!]
  \centering
  \includegraphics[width=\linewidth]{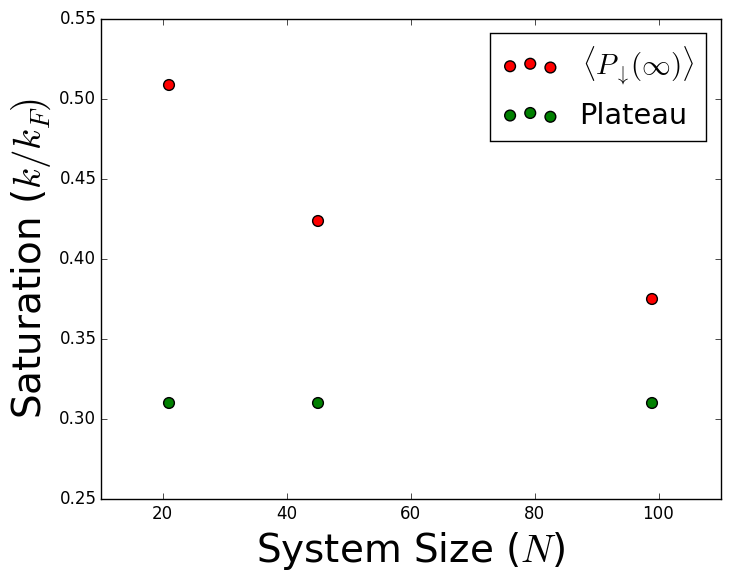}
  \caption{\label{fig:infmom_vs_syssize}
    Changing time-independent and momentum plateau values with system size for constant $\gamma = 3$ and
    $Q = \frac{4}{3}k_{F}$.
    While the plateau value in the momentum momentum stays constant with system size, the theoretical value is initially much
    greater than the plateau, and decreases towards it with a power law relation as $N \to \infty$.
    Hence, while the momentum of the impurity is not the same as its theoretical thermalised value, this is a finite size
    effect, and disappears as the system moves into the thermal regime.
    This fact could be used to obtain an approximate value for the saturation momentum in a thermalised system by finding the
    plateau of a much smaller system, using less computational resources than otherwise, but measuring the plateau is
    intrinsically imprecise due to the flutter around it, so this is only useful as an approximation.
  }
\end{figure}

In an infinite system, the plateau would have the same value as $\expval{P_{\downarrow}(\infty)}$, but we find there is a
difference, which decreases with the size of the system $N$.
This difference decreases with increasing system size with a power law relation
as can be seen in Fig~\ref{fig:infmom_vs_syssize}, so the values would be equal in the thermodynamic limit as might be
expected.
From these differences, we have two different, yet reasonable, values for the infinite time momentum for those system
parameters our program can access, which means wherever we inspect this value we have two options to choose from.
Where there is a significant difference between the two values, we will mention both in the remaining text.

\begin{figure}[ht!]
  \centering
  \includegraphics[width=\linewidth]{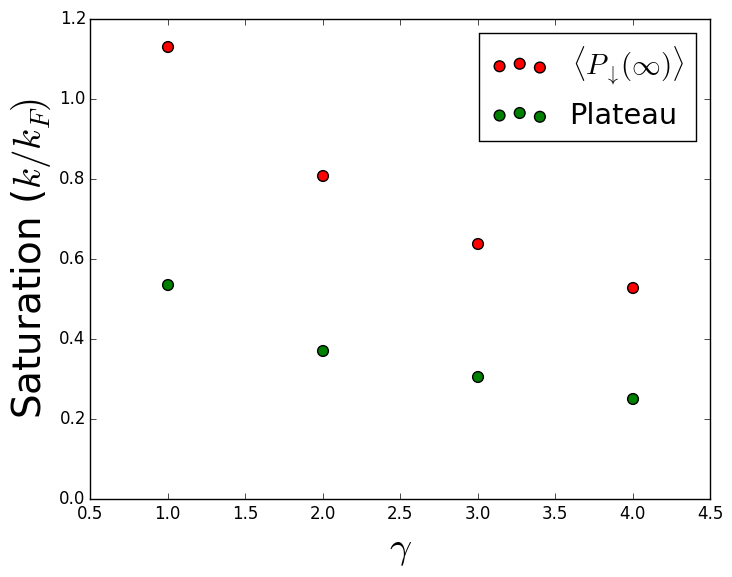}
  \caption{\label{fig:infmom_vs_gamma}
    How the saturation values are modified by interaction strength $\gamma$.
    As $\gamma$ diverges, the saturation values both converge to a non-zero value, and to each other.
    The progression of the theoretical infinite time value was discussed in reference~\protect\cite{Mathy2012sup}, and the current
    plot shows the same progression.
    While the plateau is independent of the way $\gamma$ is chosen, $\expval{P_{\downarrow}(\infty)}$, like Eqn~\eqref{eqn:revival_estimate} in
    Section~\ref{sec:revivals}, does depend on how $\gamma$ is formed, with the current plot formed for $\nicefrac{L}{N} = 2$.
    For $\nicefrac{L}{N} = 3$ the values of $\expval{P_{\downarrow}(\infty)}$ follow a progression of the same shape, but
    between those shown in this figure.
  }
\end{figure}

\begin{figure}[ht!]
  \centering
  \includegraphics[width=\linewidth]{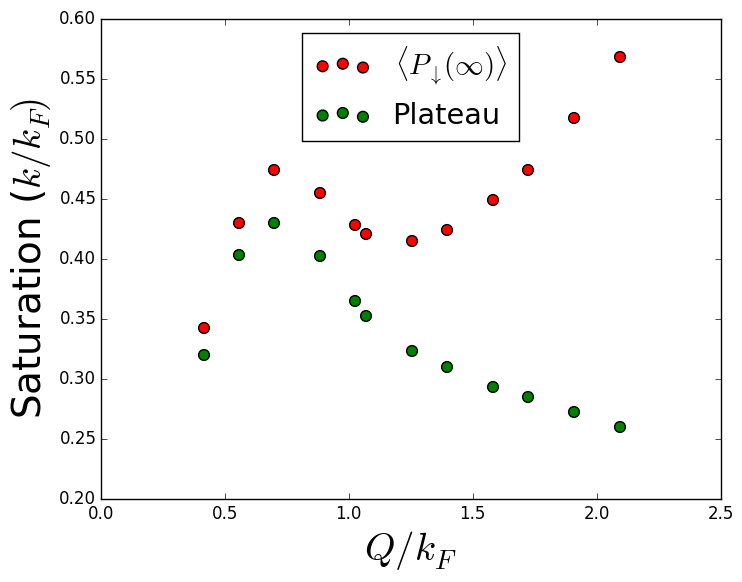}
  \caption{\label{fig:infmom_vs_initmom}
    The change with $Q$ of both $\expval{P_{\downarrow}(\infty)}$ and the momentum plateau.
    Once the initial momentum is above $k_F$, the plateau decreases with increasing $Q$, while $\expval{P_{\downarrow}(\infty)}$ increases.
    There is a maximum in the momentum plateau for an initial momentum some point below $k_F$ as seen in~\protect\cite{Mathy2012}.
    The difference between $\expval{P_{\downarrow}(\infty)}$ and the momentum plateau increases with initial momentum, showing how the interplay between
    states becomes more important for the impurity's momentum as the initial momentum goes above $k_{F}$.
    The progression of the plateau matches what is seen in~\cite{Mathy2012sup}, the progression of
    $\expval{P_{\downarrow}(\infty)}$ was not mentioned there.
    This plot was created fixing $\gamma = 3$ and $N = 45$.
  }
\end{figure}

As noted in Section~\ref{sec:revivals}, the saturation value of the impurity's momentum changes with both $Q$ and $\gamma$
(see Figures~\ref{fig:changing_initmom}, \ref{fig:changing_gamma}, \ref{fig:infmom_vs_gamma}, and
\ref{fig:infmom_vs_initmom}), for both progressions the relation of the plateau matches what is described
in Reference~\cite{Mathy2012sup}, but while the progression of $\expval{P_{\downarrow}(\infty)}$ is qualitatively the same
with changing $\gamma$, it is noticeably different with the increase of $N$ and the change in $Q$.
As $Q$ increases, the progression of $\expval{P_{\downarrow}(\infty)}$ matches that of the plateau for $Q < k_{F}$, but is
very different once $Q > k_{F}$, where $\expval{P_{\downarrow}(\infty)}$ increases with increasing initial momentum, and the
plateau decreases (Figure~\ref{fig:infmom_vs_initmom}).
This is another demonstration of how the behaviour of the system is different when $Q$ is above the Fermi momentum.
While these two momentum saturation values can be drastically different for a chosen system size, as seen in
Figure~\ref{fig:infmom_vs_syssize}, they do tend to each other in the thermodynamic limit.

The manner in which the two values converge, the plateau staying constant as $\expval{P_{\downarrow}(\infty)}$ decreases to meet it, is an interesting
characteristic, as it implies $\expval{P_{\downarrow}(\infty)}$ for an infinite system will eventually reach the consistent
value of the plateau, and hence can be found by looking at the plateau for a smaller system.
This method can only provide an approximate value for $\expval{P_{\downarrow}(\infty)}$, as which point in the quantum
flutter corresponds to the limit of $\expval{P_{\downarrow}(\infty)}$ is not well defined, making selection of the plateau by
eye intrinsically imprecise.

\clearpage
\section{Quantum Flutter}
\label{sec:flutter}

Like the non-zero infinite time value, an oscillation in the momentum of an impurity is an effect that has been
discussed in other one-dimensional systems, usually from Bloch oscillations in periodic structures~\cite{Morsch2001,Zotos2010}.
References~\cite{Gangardt2009,Schecter2012,Schecter2012a} predicted via quantum hydrodynamics arguments based on the impurity's
dispersion relation, that application of a constant external force to an impurity would create Bloch oscillations in a 1D gas
without a periodic potential, though recently the range of parameters for which this result is applicable to a
Tonks-Girardeau gas has been under discussion~\cite{Gamayun2014,Schecter2014,Gamayun2014a}.

In contrast to Bloch oscillations, quantum flutter is present in a system with no external potential acting on the impurity.
Rather than an external potential, it has so far been attributed to the superposition of plasmon and magnon states with the
impurity at momentum $Q \approx k_F$ having lost any excess momentum to the background gas~\cite{Mathy2012,Knap2014}.
This physical argument results in an exact equation for the frequency of the flutter in this model~\cite{MishaEquation}.
\begin{equation}
  \label{eqn:flutter_equation}
  \omega_{flutter} = 2k_{F}^{2}( \frac{1}{2} -
  \frac{\gamma^{2}(\frac{2 \pi}{\gamma} +
    \arctan(\frac{2 \pi}{\gamma}) +
    \frac{4 \pi^{2} \arctan(\frac{2 \pi}{\gamma})}{\gamma^{2}})}
       {4 \pi^{3}})
\end{equation}

\begin{figure}[ht!]
  \centering
  \includegraphics[width=\linewidth]{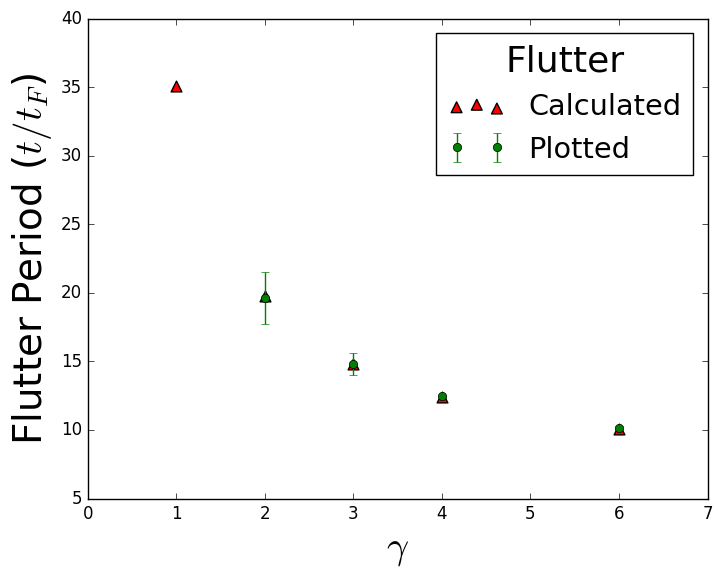}
  \caption{\label{fig:flutter_freq_vs_gamma}
    The flutter frequency we see matches what is predicted from Eqn~\eqref{eqn:flutter_equation}, to a high accuracy, the
    error bars on the period seen show the maximum and minimum value measured for the flutter, which come from finite
    size effects obscuring the flutter.
    Note there is no ``plotted'' point for $\gamma = 1$ as the oscillation was completely obscured by the revival (as can be
    seen in Figure~\ref{fig:changing_gamma}).
    Despite these difficulties, the flutter period we see closely follows the prediction from~\protect\cite{MishaEquation}.
  }
\end{figure}

Notably, the frequency is only dependent on one of the physical parameters we can change, so there should be no change with
system size and with initial momentum.
We show quantitative agreement with Eqn~\eqref{eqn:flutter_equation} to within the accuracy of our measurements in
Fig~\ref{fig:flutter_freq_vs_gamma}, and see the predicted independence on system size and initial momentum.

The dependence of the flutter on the non-physical parameter $\varsigma$ is between the dependence of the revivals in
Section~\ref{sec:revivals}, and the momentum saturation measures in Section~\ref{sec:inf_time}.
While the revivals reach a stable value for a relatively small $\varsigma$, and the saturation momentum requires the
$\varsigma$ to be very high (without using the normalisation from Section~\ref{sec:inf_time}), the frequency and amplitude of
the flutter both increase until they reach saturation at a $\varsigma$ value that is between those required for the other two
features (see Fig~\ref{fig:changing_overlap}).
As the flutter saturates before the values of the plateau and $\expval{P_{\downarrow}(\infty)}$, this implies there is a
subset of state pairs that are the cause of the feature.
From the fact the $\varsigma$ required for the flutter to saturate is much higher than what is required to see the momentum revivals,
we might guess the eigenstate pairs that cause of the flutter have a lower overlap than those determining the revivals.
This hypothesis shall be explored in detail in Chapter~\ref{sec:subsets}.

\section{Conclusion}
\label{sec:observables_conclusion}

The features of the momentum evolution for our system are discussed and plotted for a range of system parameters.
Agreement is found for all statements in references~\cite{Mathy2012,Knap2014}, and a further discussion has been made on both the
momentum revivals and the change with the non-physical parameter $\varsigma$.
We push the limits of our program to system sizes of $N = 99$ (larger than seen before~\cite{Mathy2012,Knap2014}), showing
the quantum flutter and momentum plateau without interference from the revivals.
At this system size, we are feasibly restricted to $\varsigma \approx 0.97$ which is not large enough to provide confidence
in our results by itself.
For further confidence, we present an observation based justification in Section~\ref{sec:justification}, where we show that
for the systems investigated, the frequency of the momentum revivals reaches a stable point at a very low value of
$\varsigma$, while the flutter requires greater, but still achievable $\varsigma$ to stabilise.
Furthermore, while the plateau in the momentum requires a very large $\varsigma$ to stabilise, the consistency of how it changes
allows a normalisation for a reasonably accurate prediction of the plateau for systems with a low $\varsigma$.

The time between revivals in the momentum of the impurity is shown to be qualitatively predicted by a semi-classical argument
based on the momentum passed to the background gas and the size of the system, and though the revivals initially disperse, there
is no time where they have completely gone away.

\chapter{Eigenstates Responsible For Momentum Features}
\label{sec:subsets}

\section{Introduction}
\label{sec:subgroups_introduction}

In the previous chapter, we discussed how the impurity's momentum evolution changes with different system parameters, both
physical and non-physical.
It was seen that the non-physical parameter $\varsigma$ changed the three different momentum features in different ways.
The momentum revivals were determined at a low $\varsigma$, without any noticeable change for $\varsigma > 0.9$, the quantum
flutter was determined at a higher value, only settling at $\varsigma > 0.95$, while the position of the momentum plateau did
not show any signs of saturation while $\varsigma$ progressed to its asymptotic value.
This chapter presents an attribution of state pair subsets to $\expval{P_{\downarrow}(t)}$ features, demonstrating how
different eigenstate subsets can be described, and showing which subsets cause which features of the momentum evolution.
Through this attribution of subsets to features, we will see why the different $\expval{P_{\downarrow}(t)}$ features reach
stability at different values of $\varsigma$.

For this chapter we use a large system ($N=99$) for all graphs, keeping the initial momentum constant at
$Q = \frac{3}{2} k_{F}$, and the interaction strength constant at $\gamma = 3$.
A system size this large shows the patterns we will discuss much more clearly, as the number of states contained is limited
by the size of the system in all subgroups we identify.
Despite the varying clarity, all patterns shown in this chapter were seen across the range of parameters we can access with
our program, with the exception of low initial momentum $Q < k_{F}$ where some broke down, again showing the qualitative
difference between systems with large and small initial momentum.
We hence limit all discussions in this chapter to systems where the impurity has been injected with initial momentum greater
than the Fermi momentum, which is where the quantum flutter has been predicted.

\section{The Pseudo Sea}
\label{sec:pseudo_sea}

We now describe a concept called the \emph{pseudo Fermi sea} that we will use throughout the rest of the current work to
categorise eigenstates.
The concept comes from a representation of the Bethe roots, related to the Bethe momenta of Equation~\eqref{eqn:Full_BA_eqns}
by
\begin{equation}
z_{i} = \frac{L}{2} k_{i}.
\end{equation}
These roots can be represented as
\begin{equation}
  \label{eqn:bethe_rapidities}
  z_{i} = \pi n_{i} - \delta_{i}, \qquad  i = 1, 2, \ldots, N+1
\end{equation}
where $n_{i}$ are a unique set of integers, and $\delta_{i}$ are bound within $0$ and $-\pi$.
Using this representation, the Bethe eigenstates are uniquely determined by the $N$ integers $n_{i}$, and as the energy of a
state is determined by the sum of the squares of $z_{i}$, the ground state has the integers
$n_{i} = \{-(N+1) / 2, \ldots, (N-1) / 2\}$.
\begin{figure}[ht]
  \centering
  \includegraphics[width=\linewidth]{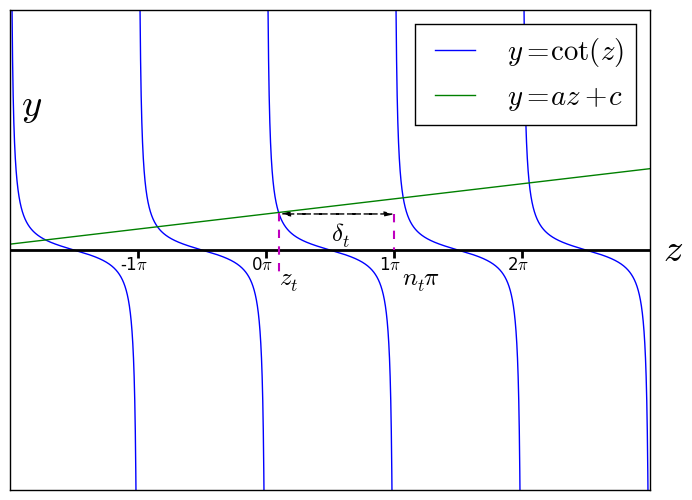}
  \caption{\label{fig:bethe_roots_solutions}
    The graphical solutions of the Bethe root equations in Eqn~\eqref{eqn:Full_BA_eqns} shows the validity of the
    representation for the Bethe roots given in Equation~\eqref{eqn:bethe_rapidities}.
    Each root can be distinguished by the range it is in, and as the gradient of the green line here is fixed by the
    physical parameters of the system, then the set of ranges in which roots are found uniquely determine the roots
    themselves.
  }
\end{figure}

An analogy can be drawn between this set of integers and the Fermi sea, as you cannot have two identical integers
in the set $n_{i}$, and the ground state's set fills all the lowest magnitude integer values.
We hence refer to this ground set of integers as the \emph{pseudo Fermi sea}.
Following this analogy, any excited state must have a number of pseudo particle/hole pairs, where a \emph{pseudo hole} is
defined as an integer in the pseudo sea but absent in $n_{i}$, and a \emph{pseudo particle} is an integer in $n_{i}$, but
absent in the pseudo sea.
This analogy and terminology is not new~\cite{Caux2009,Mathy2012sup}, but provides some very fitting terms to define the
eigenstate pair patterns that make up the bulk of this chapter.

\section{Eigenstate Families}
\label{sec:state_families}

\begin{figure}[ht!]
  \centering
  \includegraphics[width=\linewidth]{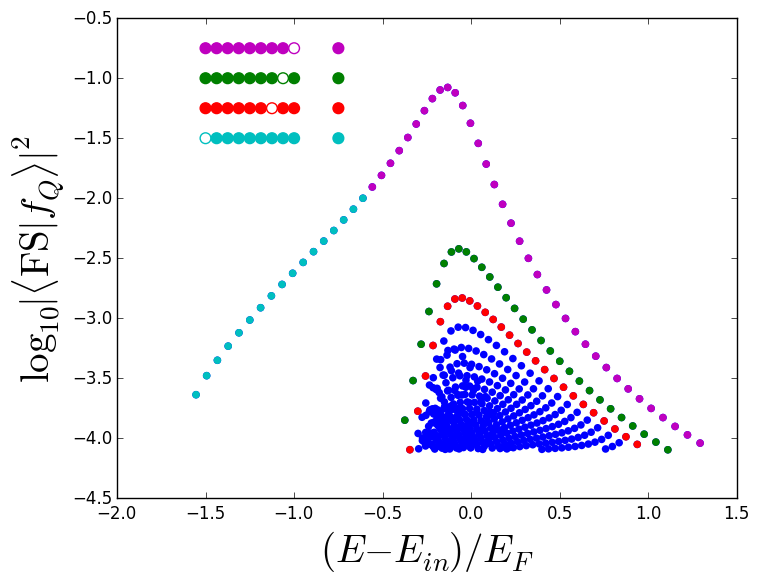}
  \caption{\label{fig:zhenya_families}
    The plot of eigenstate energy against importance shows some distinct branches.
    These branches are comprised of parametric families defined by the pseudo hole each related state shares.
    For all branches other than the main one (at the top of the figure), the branch is composed of a single parametric
    family, where all related states have the same pseudo hole.
    The main branch consists of two families, one where the pseudo hole is on the positive edge of the pseudo sea,
    $n_{hole} = \nicefrac{(N - 1)}{2}$, and one where the pseudo hole is on the negative edge,
    $n_{hole} = \nicefrac{-(N + 1)}{2}$.
    Each successive branch consists of states from a single family, whose pseudo hole is further inside the pseudo sea the
    less the average $\abs{\braket{FS|f_{Q}}}^{2}$, as schematically shown in the top left hand corner.
    %
    The most important states shown here have a single pseudo excitation, and as a single parametric family is followed from
    left to right, the pseudo particle is increased by one for each element, the energy of each state being the sum of the
    squares of Bethe roots $z_{i}$ where $ 0 \leq n_{i} - z_{i} \leq 1$.
  }
\end{figure}

The plot of energy against $\log_{10}(\abs{\braket{FS|f_{Q}}}^{2})$ of each eigenstate shown in Figure~\ref{fig:zhenya_families}
shows some clear branches, with a few having a much greater contribution to $\varsigma$ than others~\cite{Burovski2014}.
In Figure~\ref{fig:zhenya_families}, (which shows the same type of plot as seen in~\cite{Burovski2014}) each branch consists
of two parametric families, consisting of states sharing a pseudo hole.
When the pseudo hole of an eigenstate is $\nicefrac{(N-1)}{2}$, it is in the branch with the greatest average overlap.
If the pseudo hole is $\nicefrac{(N-1)}{2} - 1$ then the eigenstate is in the second most important branch, and this pattern
continues for all positive pseudo holes.
The other family in each branch is given by those states with a matching \emph{negative} pseudo hole, so states with a pseudo
hole of $\nicefrac{-(N + 1)}{2}$ are in the same branch as those with a pseudo hole of $\nicefrac{(N-1)}{2}$, and similarly for
the two families of pseudo holes $\nicefrac{(N-1)}{2} - 1$ and $\nicefrac{-(N + 1)}{2} + 1$.
Eigenstates with negative pseudo holes tend to have a much lower overlap than those with positive pseudo holes, with all bar
one family of these states below the cut off in $\abs{\braket{FS|f_{Q}}}$ used for Figure~\ref{fig:zhenya_families}.
The exceptional family consists of those states with a pseudo hole from the negative edge of the pseudo sea, at
$\nicefrac{-(N + 1)}{2}$, sharing its branch with the $\nicefrac{(N-1)}{2}$ family, which is the greatest branch in
Figure~\ref{fig:zhenya_families}.
For example, if the pseudo sea spans the integers $\{-50, -49, ..., 0, ..., 48, 49\}$ then the main branch evident
in Fig~\ref{fig:zhenya_families} would consist of all states whose pseudo hole is either $49$ or $-50$, which corresponds to
two parametric families, one with the pseudo hole $49$, another with the hole $-50$.
Note that this graph demonstrates the individual states which contribute most to $\varsigma$ have a single pseudo excitation.
This is a particular case of a ``rule of thumb'' in the literature~\cite{Caux2009,Mathy2012sup} where the contribution to $\varsigma$
from states with a small number of particle/hole pairs is dominant.
These branches and families have been noted before, and it has been shown that in the asymptotic limit of
$\gamma^{2} \log N \to 0$ and $\gamma^{2}N \to \infty$ the states from just the main family saturate
$\varsigma$~\cite{Burovski2014}.

As we know a subset of states can determine the momentum of the impurity at infinite time, and there is a strong pattern in
the description of these states within the pseudo particle/hole terminology, a natural question to ask is whether there is a
subset within the transitions between these states, and description of that subset, that determine features in the time
evolution of the momentum.
We shall show how there are in fact two separate (though related) subsets of transitions between eigenstates that together
describe the momentum features of the system.
One of these subsets describes the overall shape of $\expval{P_{\downarrow}(t)}$, including the plateau at non-zero momentum
and the revivals in momentum, while the other describes the flutter that occurs on top of the plateau.
It is the separation of these subsets, and particulars of the state pairs in each, that cause the saturation of these
different features to happen at different values of $\varsigma$, as seen in Chapter~\ref{sec:observables}.

\section{General Shape}
\label{sec:general_shape}

When deconstructing the individual contributions to the momentum of a system, it is natural to investigate the Fourier
transform.
The particulars of the method we are using make this a trivial task, as it is the Fourier transform we start with, and the
momentum is calculated from there (see Equation~\eqref{eqn:momentum_against_time}).
This also means that each point in the Fourier transform corresponds to a transition between a specific pair of eigenstates
which may then be inspected for any pattern in the pseudo particle/hole pairs that describe them.

\begin{figure}[ht!]
  \centering
  \includegraphics[width=\linewidth]{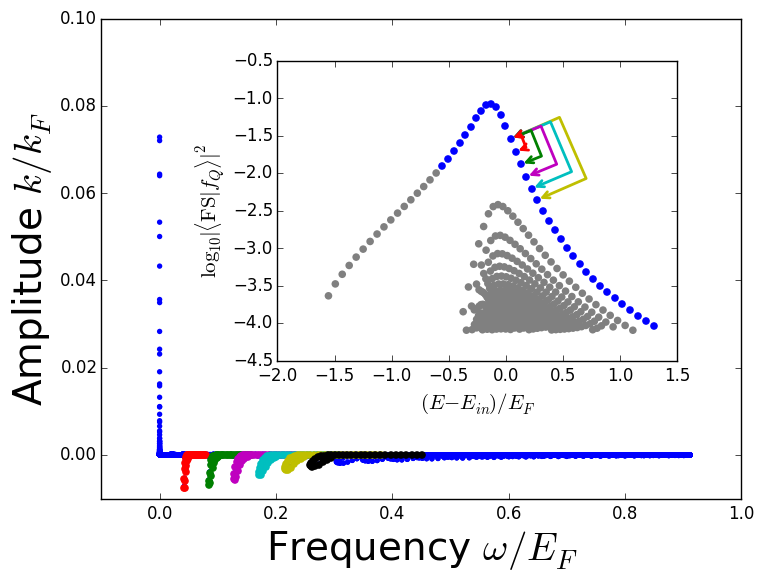}
  \caption{\label{fig:fourier_subgroups}
    The Fourier transform of the impurity's momentum with inset showing which state pairs cause each small peak.
    The main plot shows the Fourier transform of the impurity's momentum against time.
    Each point here is a contribution from a single state pair to the total $\expval{P_{\downarrow}(t)}$.
    There are two strong features, the large peak at $\omega = 0$ and the set of negative amplitude peaks around each
    integer multiple of the revival frequency.
    Inset shows all states with a single pseudo excitation on the same axis as Figure~\ref{fig:zhenya_families} (more clearly
    showing the branches discussed previously), the coloured arrows show example state pairs for some of the coloured peaks
    in the main plot.
    In all state pairs from the coloured peaks, both states are in the main family (those not coloured grey), with a pseudo
    hole of $\nicefrac{(N-1)}{2}$, so the difference between state pairs is only in the difference between the pseudo particles
    of each state in the pair.
    Transitions which give the contributions in the first negative peak of the Fourier transform are between states whose
    pseudo particles differ by one, while transitions causing the second peak are between states whose pseudo
    particles differ by two.
    This pattern continues for all the different peaks.
  }
\end{figure}

A typical Fourier transform of our impurity's momentum is shown in Fig~\ref{fig:fourier_subgroups}, which has two
obvious features to the casual observer.
The first is the strong peak at $\omega = 0$ that comes from all contributions in Eqn~\eqref{eqn:momentum_against_time} where
$\ket{f_{Q}} = \ket{f_{Q}^{'}}$, and determines $\expval{P_{\downarrow}(\infty)}$ by contributing a time-independent shift in
the impurity's momentum.
The other is the series of negative amplitude peaks that occur at integer multiples of the revival frequency, with
decreasing amplitude as the multiple increases.
As the states in the peak at $\omega = 0$ have been discussed in other works~\cite{Burovski2014}, and we have presented our
own observations on these states in Section~\ref{sec:inf_time}, we now discuss the set of negative amplitude peaks and the
state pairs that make them up.

Each point on the Fourier transform comes from a pair of eigenstates.
Upon inspection of the state pairs that form these peaks, a simple pattern in the pseudo particle/hole representation can be
found.
All states, in all state pairs of these notable peaks, come from the main family in Figure~\ref{fig:zhenya_families}, so they
all share with each other $N$ integers in the set $n_{i}$ and differ in the pseudo particle they have from their excitation.
Furthermore, each peak is formed by taking all possible pairs from this subset subject to the constraint that the pseudo
particles of the two states are a fixed number apart, where this fixed number corresponds to the integer multiple of the
revival frequency that the current peak will contribute.
For example, the first peak, coloured red in Fig~\ref{fig:fourier_subgroups}, consists of all state pairs where both
states come from the main family (i.e. whose pseudo hole is $\nicefrac{(N-1)}{2}$), and where the pseudo particles of the two
states differ by one.
Hence the pair of sets $\{n_{i}^{1}, n_{i}^{2}\}, \quad i = 1, 2, \ldots, N+1$ that enumerate each individual contribution to
the momentum in Eqn~\eqref{eqn:momentum_against_time} and Fig~\ref{fig:fourier_subgroups}, are of the form
$\{n_{i}^{base} \cup \{p\}, n_{i}^{base} \cup \{p + n\}\}$ where $p$ is some integer outside the pseudo sea, $n_{i}^{base}$
is a shared set of integers, and $n$ is an integer defining which peak this pair is in.
This relation is diagrammatically shown in the inset of Figure~\ref{fig:fourier_subgroups}.

\begin{figure}[ht!]
  \centering
  \includegraphics[width=\linewidth]{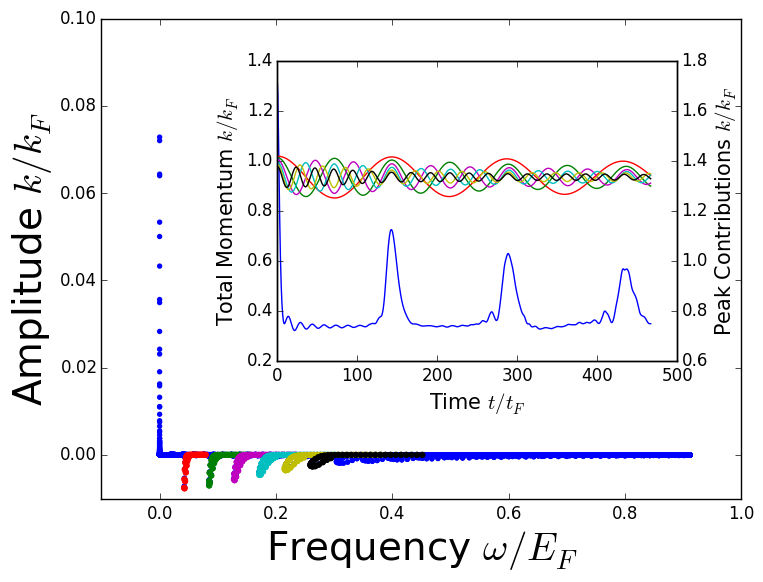}
  \caption{\label{fig:fourier_contributions}
    The contribution to $\expval{P_{\downarrow}(t)}$ from each negative amplitude peak in the Fourier transform.
    The main plot shows the Fourier transform of the momentum, highlighting each negative amplitude peak, while
    the inset compares the contribution of each of these peaks to the total momentum evolution of the impurity.
    The colour of each contribution in the inset correlates with the corresponding colour of the peak in the Fourier
    transform, and these contributions are plotted on an axis of the same scale, but shifted for clarity.
    Each peak adds a wave almost harmonic to the revival frequency, and Figure~\ref{fig:main_vs_total} shows that their
    superposition describes both the plateau and the revivals.
    Figure~\ref{fig:main_vs_total} shows this superposition of all peaks describes the general shape of the total
    $\expval{P_{\downarrow}(t)}$, but does not describe the flutter.
    Figure~\ref{fig:main_normalised_vs_total} demonstrates the contribution from these peaks to the plateau value of the
    total momentum is proportional to the $\varsigma$ value reached when just accounting for the eigenstates whose
    transitions form these peaks.
  }
\end{figure}

An initial analysis of the effect these peaks in the Fourier transform have can be done by viewing the contribution to
$\expval{P_{\downarrow}(t)}$ that each peak makes in turn.
These contributions can be seen in Fig~\ref{fig:fourier_contributions} and show that each peak combines to add a remarkably
smooth wave of period close to some integer multiple of the revival period.
Though the contributions have very good alignment for the first few revivals, it is apparent that they become misaligned as
time increases.

\begin{figure}[ht!]
  \centering
  \includegraphics[width=\linewidth]{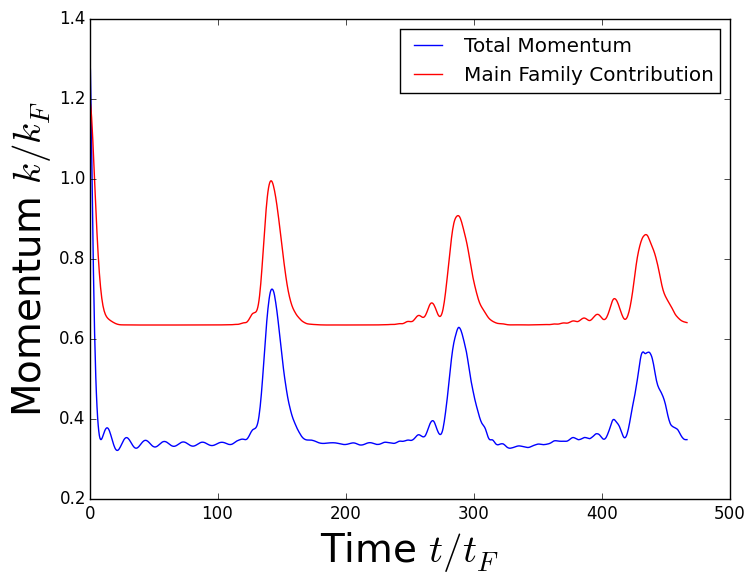}
  \caption{\label{fig:main_vs_total}
    Comparison of $\expval{P_{\downarrow}(t)}$ to the contribution from all transitions between states in the main parametric
    family, which is equivalent to the contribution from all peaks seen in the Fourier transform.
    This comparison shows how this limited number of transitions describes the majority of features in
    $\expval{P_{\downarrow}(t)}$, their contribution provides the majority of the revival amplitude, and there is a non-zero
    plateau in the momentum.
    There are notable differences though: the plateau is not at the same momentum as the plateau of the total momentum, and
    there is no quantum flutter around it.
  }
\end{figure}

Combined, the state pairs in all of these peaks plus each diagonal element from the main family (which make up the majority of
the peak at $\omega = 0$), constitute all combinations of states selected only from the main family.
This contribution is what is plotted in Figure~\ref{fig:main_vs_total}, reproducing the general shape of
$\expval{P_{\downarrow}(t)}$, but not the flutter or the exact momentum of the plateau.

Because this subset of pairs forms all transitions between a subset of states, we can apply the normalisation from
Equation~\eqref{eqn:renormalisation} to see what momentum would occur were $\varsigma$ saturated by the main family alone.
When this is done, we see that the normalisation does not account for the difference in the momentum plateau, with a notable
gap between the normalised plateau and that of the total momentum, showing that this main family does not contribute
proportionally to the momentum plateau of the impurity.
While the main family does not accurately represent the total, the main branch in the graph of
Figure~\ref{fig:zhenya_families} consists of two parametric families, one from either edge of the pseudo sea.
When both these families are accounted for, i.e. the entire branch is taken into account, the momentum plateau is much better
approximated, matching to within the variation from quantum flutter.

\begin{figure}[ht!]
  \centering
  \includegraphics[width=\linewidth]{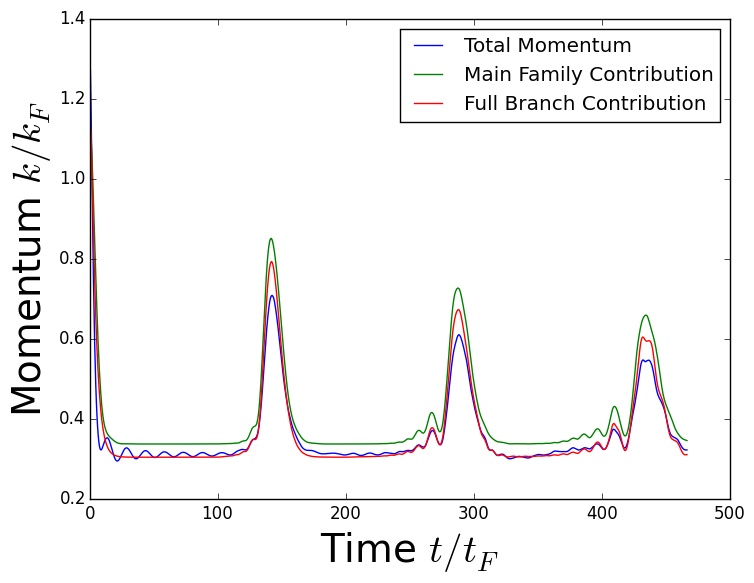}
  \caption{\label{fig:main_normalised_vs_total}
    Comparison between the normalised contributions of the main parametric family, main branch in
    Figure~\ref{fig:zhenya_families}, and all states calculated for a typical system.
    The normalised branch reproduces the plateau of the entire $\expval{P_{\downarrow}(t)}$ quite well, while the
    main parametric family does not. Both the normalised contributions from the main family and main branch have a greater
    revival amplitude than the actual $\expval{P_{\downarrow}(t)}$, demonstrating that they contribute relatively more to the
    momentum revivals than other states.
  }
\end{figure}

\begin{figure}[ht!]
  \centering
  \includegraphics[width=\linewidth]{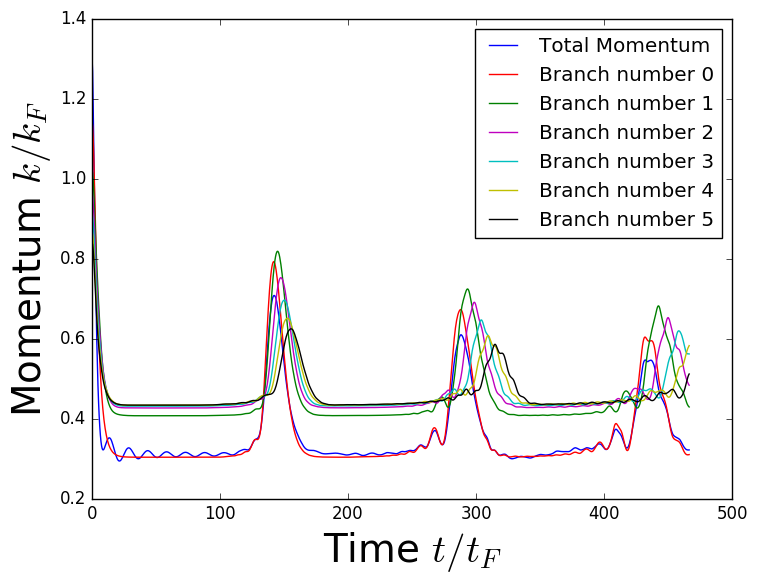}
  \caption{\label{fig:normalised_branches_vs_total}
    Comparison between the normalised contributions of each branch from Figure~\ref{fig:zhenya_families} and the total
    momentum contribution.
    Each contributes a similar shape to the momentum, though without the normalisation of each branch the shape of most would
    not be visible here.
    The gradual increase in the revival period can be attributed to the fact that the less important branches seen in
    Figure~\ref{fig:zhenya_families} have points closer together on the energy axis.
  }
\end{figure}

In both normalised contributions, the revivals have a greater amplitude than the total $\expval{P_{\downarrow}(t)}$, which
implies that the main family/branch contributes relatively more to the momentum revivals than other states.
The fact the revivals are determined mainly by the transitions between states in the main branch, and hence states that have
a large $\abs{\braket{FS|f_{Q}}}^{2}$, explains why the revivals are determined from a low $\varsigma$ onward, as was seen
in Section~\ref{sec:revivals}.
These states are the first to be accounted for, which means the momentum revivals have been found at a very early stage in
the saturation of $\varsigma$.
The effects of this normalisation, and hence the representative nature of the main branch, can be seen in
Figure~\ref{fig:main_normalised_vs_total}.
We talk of how ``representative'' a momentum contribution from a set of states is to refer to how well it recreates the total
momentum once normalised using Equation~\eqref{eqn:renormalisation}.
The representative nature of the main branch contribution to the momentum plateau is a useful feature, allowing one to find
the value of the momentum plateau only accounting for small subset of states whose description is known beforehand, meaning
the calculation time required would be less than the square root of what it would be otherwise.
Unfortunately, this feature is limited in scope.
While it exists for all system parameters we can probe with this program, in the thermodynamic limit of $N \to \infty$ with
constant $\frac{N}{L}$, the infinite time contribution of the main family tends to $0$ while the total
$\expval{P_{\downarrow}(\infty)}$ does not (see Section~\ref{sec:pinf_vs_p1st_thermalised}).
It is of interest to note that the contributions from all the parametric families and branches seen in
Figure~\ref{fig:zhenya_families} are similar, reproducing the general shape of $\expval{P_{\downarrow}(t)}$, however, the
amplitude of each contribution decreases with the importance of the states, and the resulting contributions are less
representative of the total.
The contribution of each branch normalised by Eqn~\eqref{eqn:renormalisation} is compared to the total momentum in
Figure~\ref{fig:normalised_branches_vs_total}.
While the general shape stays constant, there are notable differences.
The period of the revivals is slightly larger for lesser families, the normalised amplitude of the revivals decreases, and
the momentum plateau of the normalised contribution is further from the total.
This shows that these families with lower $\abs{\braket{FS|f_{Q}}}^{2}$ not only contribute less overall that the main one,
but they also contribute less proportional to their contribution to the $\varsigma$ saturation.
I.e. when normalised by Equation~\eqref{eqn:renormalisation} their contribution to the total momentum is still less than the
similarly normalised contribution of the main family.
When combined, the contributions from all intra-branch pairs in the parametric families create a better
approximation of the momentum plateau than the main branch alone, but the normalisation applied previously cannot be applied
again as this contribution does not come from all transitions within a subset of states.

\begin{figure}[ht!]
  \centering
  \includegraphics[width=\linewidth]{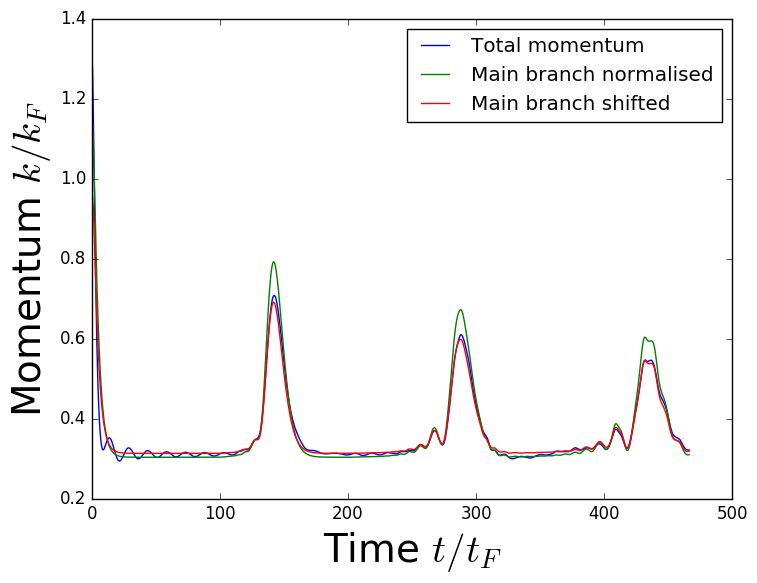}
  \caption{\label{fig:shifted_approximation}
    Comparison of the approximations gotten from normalising the main branch contribution via
    Equation~\eqref{eqn:renormalisation} to that of adding the time-independent contribution of other states to the full
    contribution of the main branch.
    Accounting for the time-independent contributions of all states provides a better approximation around the momentum
    revivals.
  }
\end{figure}

The good results obtained by normalising the contribution of states in the main branch of Figure~\ref{fig:zhenya_families}
using Equation~\eqref{eqn:renormalisation} neglect the contribution coming from the majority of states.
Using the numerical observation that the value of $\matrixel{f_{Q}}{P_{\uparrow}}{f_{Q}^{'}}$ for a given state $\ket{f_{Q}}$
is greater when $\ket{f_{Q}^{'}} = \ket{f_{Q}}$ than otherwise, a better approximation of the total momentum can be obtained.
This approximation is made by taking the full contribution of the same subgroup of transitions as before, but rather that
normalising this contribution using Equation~\eqref{eqn:renormalisation}, other states are accounted for using just their
time-independent contribution.
While accounting for these extra contributions is much more time-intenstive, it produces a better approximation of the total
momentum evolution, especially near the momentum revivals.
This can be seen in Figure~\ref{fig:shifted_approximation}, which compares the two approximations with the total, and again
in Figure~\ref{fig:singly_excited_shifted} where the same type of approximation is used.
This alternate approximation provides a better estimate of the momentum plateau at lower initial momenta $Q$, as shown in
Figure~\ref{fig:compare_infmom_approximations}, where various plateau approximations are presented for a range of different
$Q$.

\section{Flutter}
\label{sec:flutter_groups}

The last section showed that the overall shape of $\expval{P_{\downarrow}(t)}$ comes from all transitions where both states
come from the same branch of Figure~\ref{fig:zhenya_families}.
This section will demonstrate that the oscillation dubbed ``quantum flutter'' is a feature resulting from pairs of states
where each is in a different branch.
While the Fourier transform highlighted the states most culpable for the general shape, we found no feature around the
quantum flutter frequency in any relation we looked at.
This is because the eigenstates from the branches discussed in the previous section dominated the structure of any plots
taking into account all eigenstates.

\begin{figure}[ht!]
  \centering
  \includegraphics[width=\linewidth]{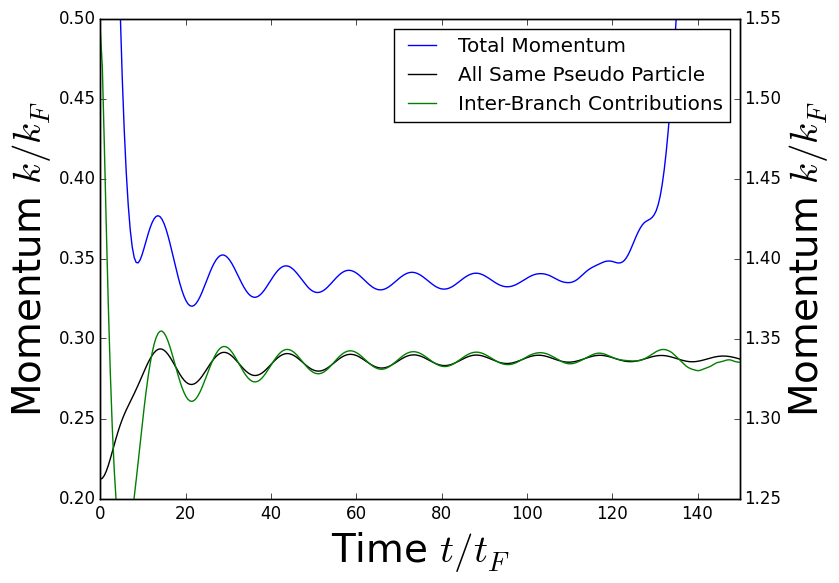}
  \caption{\label{fig:all_inter_branch}
    Plot comparing the quantum flutter in the total momentum evolution with the contributions to the momentum from transitions
    between state branches, and contributions from transitions between states in different branches, but sharing a pseudo
    particle as highlighted in Figure~\ref{fig:flutter_states}.
    The total quantum flutter is reproduced by all inter-branch transitions, and the restricted set of transitions between
    states sharing a pseudo particle reproduces the frequency, and most of the amplitude of the flutter.
  }
\end{figure}

It can be reasoned, given that eigenstates with a single pseudo excitation are the greatest contributors to
$\expval{P_{\downarrow}(t)}$ and transitions between states in the same branch from Figure~\ref{fig:zhenya_families} describe
the general shape of the impurity's momentum without the flutter, that the flutter may come from transitions between states
in different branches of Figure~\ref{fig:zhenya_families}.
This conjecture is borne out in Figure~\ref{fig:all_inter_branch}, where the contribution of the inter-branch transitions can
be seen to match the flutter of the total $\expval{P_{\downarrow}(t)}$, barely contributing any overall shift of the impurity
from the initial momentum of $\frac{4}{3}k_{F}$.
The fact that flutter comes from these inter-branch transitions can be used to explain the observation that the quantum
flutter saturates at a $\varsigma$ of about 0.95.
The phenomenon of quantum flutter requires including many of the parametric families from Fig~\ref{fig:zhenya_families}, but
does not depend on any states that have more than a single pseudo excitation.
This means that a large number of states must be accounted for to describe the flutter, but these are states that are
accounted for first by the sampling algorithm of our code.

\begin{figure}[ht!]
  \centering
  \includegraphics[width=\linewidth]{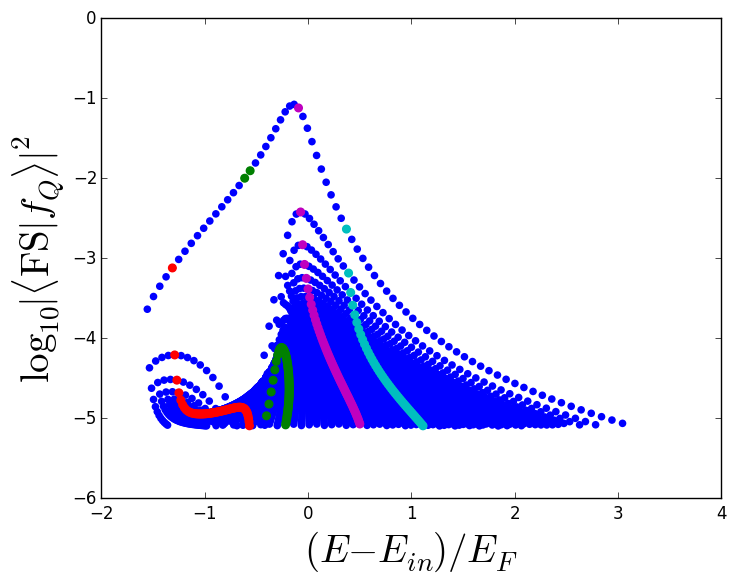}
  \caption{\label{fig:flutter_states}
    All singly excited states with a $\abs{\braket{FS|f_{Q}}}^{2}$ above a certain threshold.
    Some parametric families different to the type discussed before are highlighted in colours other than dark blue.
    In these parametric families, related states share a pseudo particle and have different pseudo holes.
    To avoid confusion, we will not refer to these sets as parametric families in the text.
    Intra-family transitions from these families create the major contribution to the quantum flutter in
    $\expval{P_{\downarrow}(t)}$.
    While one family of the type described in Figure~\ref{fig:zhenya_families} clearly contributes more to
    $\expval{P_{\downarrow}(t)}$ than others, the contributions of the families shown here are relatively similar, and we
    cannot isolate a single one as providing the main contribution to the flutter.
    %
  }
\end{figure}

While this identification of the contribution to the flutter as the inter-branch transitions between singly excited states
does cut the computational resources required to investigate this feature, this is still a large subset compared to that
which we found was representative of the general shape in Section~\ref{sec:general_shape}.
We can find a much smaller group that still reproduces the flutter frequency using an analogy to the parametric families
discussed in Section~\ref{sec:state_families}, whose intra-branch transitions were shown to determine the general shape of
the impurity's momentum evolution in Section~\ref{sec:general_shape}.
Those representative families were defined by the set of states that share their only pseudo hole, and we find that
transitions between states which share their only pseudo particle (and have different pseudo holes) determine the flutter
around the momentum plateau.
Hence, the transitions which cause the majority of the flutter are between pairs of states that can be represented as
$\left \{\big(n^{ground} \setminus \{h\}\big) \cup \{p\}, \big(n^{ground} \setminus \{k\}\big) \cup \{p\}\right \}$, where
$n^{ground}$ is the set of integers defining the ground state, $h$ and $k$ are integers defining the pseudo hole missing in
each state, and $p$ is the integer defining the pseudo particle that both states share.
In this representation, all three of $h$, $k$, and $p$ can change while the state pair is still in the set of contributing
transitions.
Examples of these sets are highlighted in Figure~\ref{fig:flutter_states}, which shows the same plot as
Fig~\ref{fig:zhenya_families}, but filters out states with more than one pseudo particle/hole excitation, and is cut off at a
lower $\abs{\braket{FS|f_{Q}}}^{2}$.
While we could normalise the contribution from one of these sets, the combination of all transitions from each of these sets
is not normalisable, hence we show the unnormalised contribution to $\expval{P_{\downarrow}(t)}$ from intra-set
transitions in black, and compare it to the unnormalised total $\expval{P_{\downarrow}(t)}$ in
Figure~\ref{fig:all_inter_branch}.
The contribution from those state pairs whose pseudo particles differ by one, two, or more has the same oscillation as shown
in Figure~\ref{fig:flutter_states}, but with an ever decreasing amplitude as the pseudo particles are further apart.

\begin{figure}[H]
  \centering
  \begin{subfigure}[b]{0.9\linewidth}
    \includegraphics[width=\linewidth]{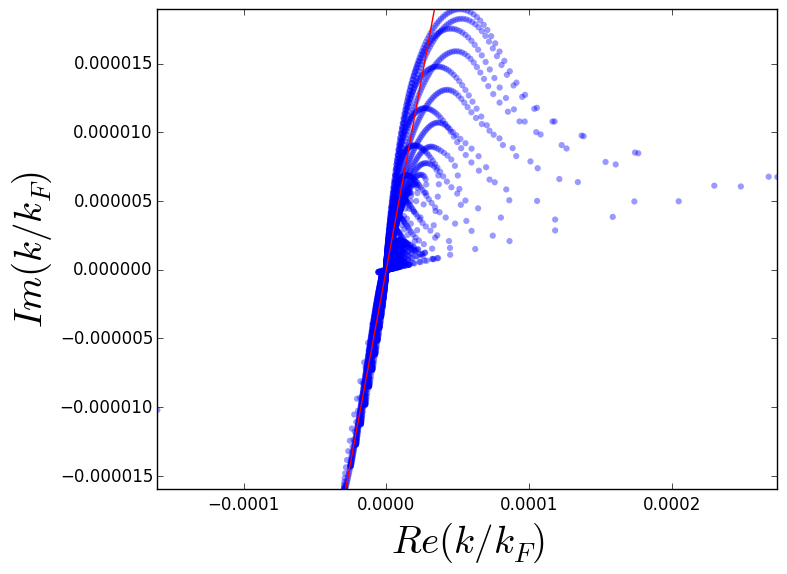}
    \caption{}
    \label{fig:polar_plot_highmom}
  \end{subfigure}
  \begin{subfigure}[b]{0.9\linewidth}
    \includegraphics[width=\linewidth]{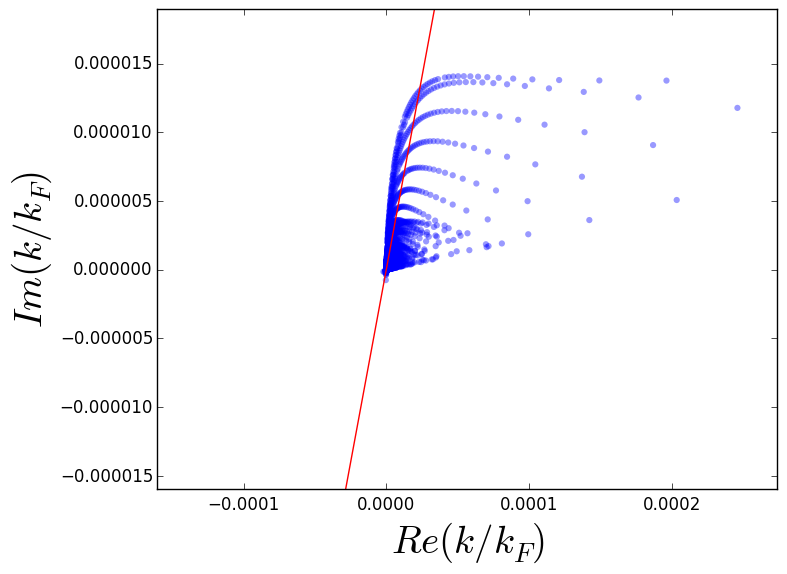}
    \caption{}
    \label{fig:polar_plot_lowmom}
  \end{subfigure}
  \caption{
    The plot of the imaginary and real parts of the terms in Eqn~\eqref{eqn:momentum_against_time} for those state pairs found to be
    major flutter contributors.
    \subref{fig:polar_plot_highmom}) While the contributions with the greatest amplitude are far from the flutter
    frequency, there is a strong peak around that frequency coming from state pairs where one state is excited from the
    negative edge of the pseudo sea.
    \subref{fig:polar_plot_lowmom}) For a small initial momentum $Q < k_{F}$, the structure of the Bethe Ansatz means states excited from the negative edge
    with a positive particle can't exist, so the branches are lost.
    In both plots, the red radial line denotes the flutter frequency.
  }
  \label{fig:polar_plots}
\end{figure}

When looking for patterns in the individual contributions before, the Fourier transform highlighted some contributions which
could be used to describe the general shape of $\expval{P_{\downarrow}(t)}$.
Using the same technique on those contributions identified as the main contributions to the flutter is not as helpful, as
the largest contributions in the subset we have identified are not at the flutter frequency.
A more effective visualisation of these state pairs is to plot the real and imaginary parts of each term on the RHS of
Eqn~\eqref{eqn:momentum_against_time} that comes from state pairs sharing a pseudo particle for a fixed time $t = 1$.
This plot is effectively a polar plot, with the radius of each point determined by the amplitude of the term, and the
angle anticlockwise from the positive real axis determined by the term's frequency.
Two examples of such plots are shown in Figure~\ref{fig:polar_plots}, one from a system with a high initial momentum
$Q > k_{F}$ and one with $Q < k_{F}$.
Note that in Figure~\ref{fig:polar_plots} we have limited the contributions to those with positive frequency for clarity in
the graphs.
This is done without loss of information, as each contribution is symmetric with respect to the order of the eigenstates,
having equal amplitude and opposite sign frequency.
These plots show another form of branch, and again, each branch consists of all entries in some parametric family.
Each family of state pairs in these plots share an eigenstate, and since in every contribution shown, the states share
their pseudo particle, moving along the family only changes the pseudo hole of the transitions second state.
The greatest amplitude families come from the greatest amplitude individual states, and all have a frequency that is less
than $\nicefrac{\pi}{4}$, so the visible states with positive imaginary and real parts all have positive amplitude and those
with negative imaginary and real parts have negative amplitude.

\begin{figure}[ht!]
  \centering
  \includegraphics[width=\linewidth]{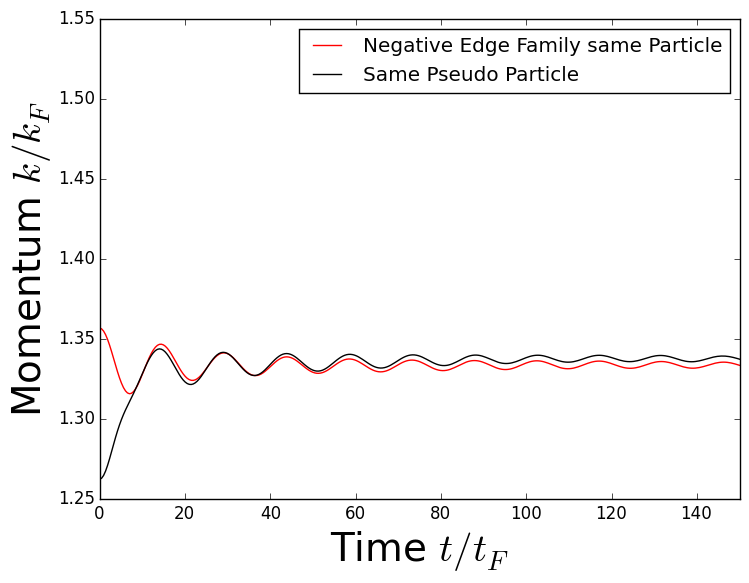}
  \caption{\label{fig:negative_edge_flutter_states}
    Comparison of the contribution from all state pairs sharing a pseudo particle, to the further subset of these state pairs
    where one of the states has its pseudo hole as the negative edge of the pseudo sea.
    The oscillation that relates to the flutter is the same between these two subsets, sharing both the amplitude and
    frequency.
    This shows how the state pairs contributing to the flutter in the subset of state pairs who share a pseudo particle all
    have one state from the negative edge of the pseudo sea.
  }
\end{figure}

In Figure~\ref{fig:polar_plot_highmom}, we see how, despite the greatest amplitude contributions coming from states with a
much smaller frequency than the flutter, there is a large, negative amplitude peak around the flutter frequency.
This peak consists of those families whose fixed state has a pseudo hole on the negative edge of the pseudo sea,
i.e. the fixed state defining the new branch is in the negative edge parametric family of type described in
Section~\ref{sec:state_families}.
Using the insight given to us by Figure~\ref{fig:polar_plots} where the greatest amplitude states with a frequency near that
of the flutter observed come from an even smaller subset of those state pairs identified previously, we plot the sum of the
contributions from transitions in this peak.
Figure~\ref{fig:negative_edge_flutter_states} shows just taking those state pairs where one of the states has a pseudo hole
on the negative edge of the Fermi sea doesn't actually decrease the flutter contribution further from that obtained when
limiting the state pairs to those sharing a pseudo particle.
With this final filter we have found a subset of state pair contributions that describe the flutter, at a lower
amplitude to the total, yet require accounting for a much smaller number of contributions in
Equation~\eqref{eqn:momentum_against_time}.

It should be stressed that though this small subset of transitions reproduces the flutter frequency, there are other
contributors within the larger subset of inter-branch transitions that increase the amplitude of said oscillation.
When the initial momentum is only slightly greater than the Fermi momentum, these states contribute more to the amplitude
than otherwise, and though the transitions to and from states on the negative edge between others with the same pseudo
particle still reproduce the frequency, the proportion of the flutter observed is less.

\begin{figure}[ht!]
  \centering
  \includegraphics[width=\linewidth]{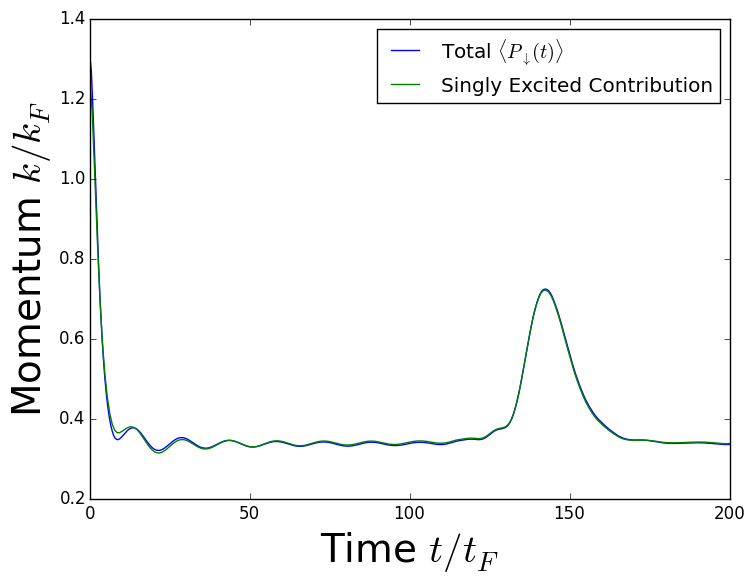}
  \caption{Comparison of the unnormalised total momentum evolution of the impurity to the momentum evolution that
    comes from assuming the only time-dependent contributions come from transitions between states with a single pseudo
    excitation.
    While the plateau of the approximation is not as level as the total, the majority of all features are shown, with both
    the flutter and revivals presenting good approximation in both amplitude and frequency.}
  \label{fig:singly_excited_shifted}
\end{figure}

In the previous sections we have shown that the main contribution to both time-dependent features of the system come from
transitions between states which only have a single pseudo excitation.
In order to stress this point, Figure~\ref{fig:singly_excited_shifted} compares the total momentum evolution of the system to
what occurs when assuming the time-dependent contribution of all states with more than a single pseudo excitation is $0$.
This is done by taking all transitions between singly excited states, and adding to that the time-independent contribution of
all other states.

\begin{figure}[ht!]
  \centering
  \includegraphics[width=\linewidth]{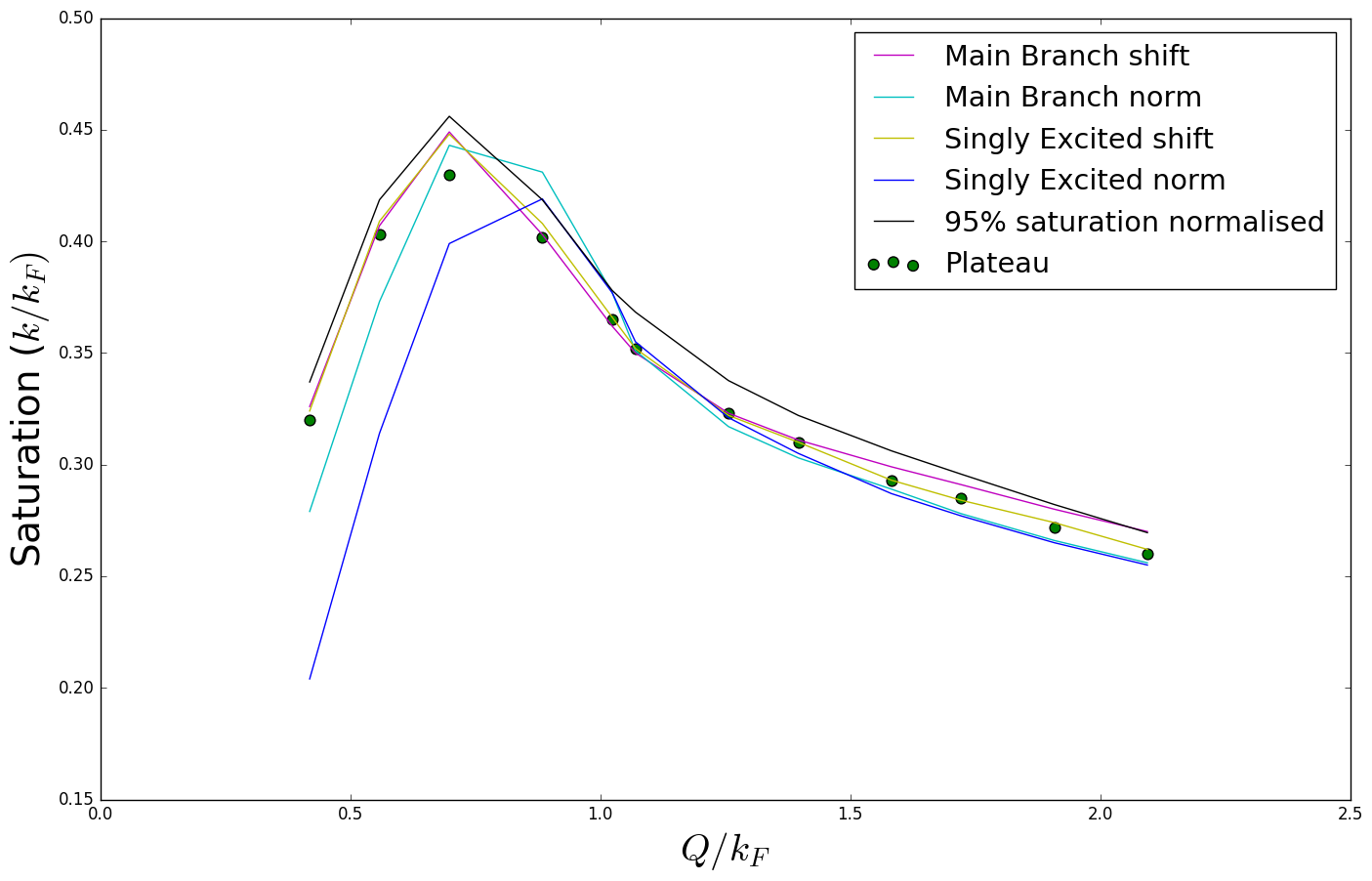}
  \caption{\label{fig:compare_infmom_approximations}
    The plateau values of different approximations and their progression with the initial momentum $Q$.
    We can see that the approximation from normalising a subgroup defined via pseudo excitation patterns becomes
    progressively less accurate as the initial momentum of the impurity decreases.
    On the other hand, the approximations from taking account of the time-independent contributions of all states still
    provides a reasonable plateau value.
    Moreover, normalising the contribution found from only obtaining a saturation of $\varsigma = 0.95$ also provides a
    reasonable plateau over all initial momenta $Q$.
  }
\end{figure}

The approximations presented in this section are generally only useful in the regime where $Q > k_{F}$.
The shape of the momentum evolution and the plateau obtained using the normalisation approach become unacceptably inaccurate
when $Q < k_{F}$.
However, the plateau approximation from discarding the time-dependent contribution of states outside the two main subgroups
identified here is still reasonably accurate, as demonstrated in Figure~\ref{fig:compare_infmom_approximations}.

\clearpage
\section{Conclusion}
\label{sec:subgroup_conclusion}

While reference~\cite{Burovski2014} showed that in the limit $\gamma^{2} \log{N} \to 0$ and $\gamma^{2} N \to \infty$, states
in the main family saturate the sum rule, we demonstrate that for those system parameters reachable by our program,
transitions within these states describe the overall shape of the impurities momentum.
More generally, we show transitions between states in a branch contribute to the general shape of the momentum evolution and
transitions between states in different branches contribute to the flutter, with their combination describing the
non-negligible contributions to the time dependent motion of the impurity.
I.e., the set of contributions from intra-branch transitions, where a branch is defined by the pseudo hole of an excitation,
determines the overall $\expval{P_{\downarrow}(t)}$ shape, while the contributions from inter-branch transitions determine
the flutter.

Within those transitions causing the flutter, the main contribution comes from transitions between state pairs where one state
has its pseudo hole on the negative edge of the pseudo sea, and both states have the same pseudo particle.
Similarly, the main contribution to the general shape can be found from those state pairs where both states are in the main
family.
A final decomposition can be made, as the revival frequency is seen from contributions between states in the main
family which differ by a nearest-neighbour displacement of one pseudo particle.

These observations can be used to approximate any individual feature of the momentum evolution with a much smaller
computational expense than has previously been possible, paving the way to more research on the physics of the system without
as much focus on how said physics must be calculated.
From the approximately ${N^{2} \choose 2}$ state pairs that come from all states with a single pseudo excitation, we have
identified two distinct subsets of approximately $N^{2}$ states which can be used to find each of the different features of
the impurity's momentum evolution.

\begin{figure}[ht!]
  \centering
  \includegraphics[width=\linewidth]{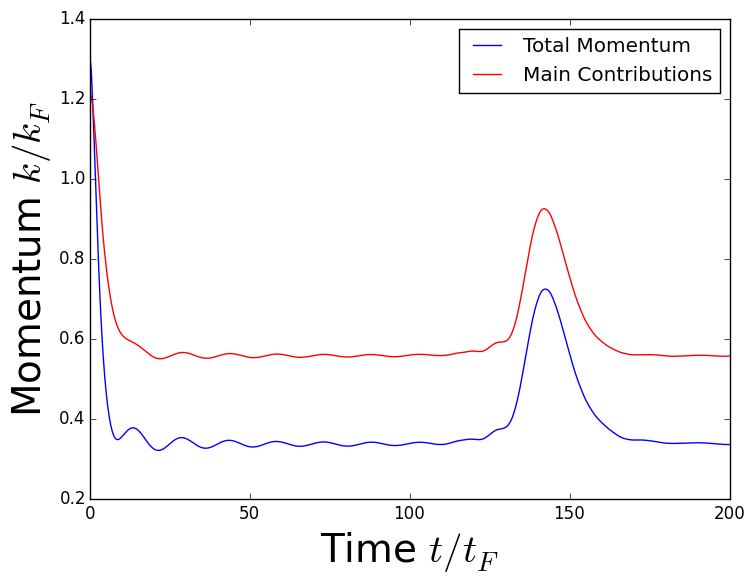}
  \caption{\label{fig:total_approximation}
    Comparison of $\expval{P_{\downarrow}(t)}$ to the sum over transitions between the two main state pair contributors
    identified.
    The inclusion of all transitions between states in the main branch matches the overall shape of the momentum evolution,
    while the inclusion of those transitions between states excited from the negative edge of the pseudo sea and those which
    share a pseudo particle with them, add flutter to the plot.
    The approximate flutter does not reproduce the same amplitude as the total, and the value of the momentum plateau is not
    matched, but the frequency of the revivals and flutter are accurately reproduced.
    In this graph, the total plot required calculating 400000000 state pairs, while the approximate value required
    calculating only 11865 state pairs, a computational speedup of about 4 orders of magnitude.
  }
\end{figure}

\chapter{Other Investigations}
\label{sec:other}

\section{Thermalisation}
\label{sec:thermalisation}

The question of if and how a one-dimensional system thermalises from its initially excited state is interesting and
open~\cite{Kollar2008,Faribault2009,Polkovnikov2011,Banuls2011,Polkovnikov2011,Caux2012}.
Experimental data has shown that such systems do not always relax into a thermalised state~\cite{Kinoshita2006,Gring2012}, and
theoretical work on the subject has shown there can be a non-thermal steady state that a system can relax
into~\cite{Cazalilla2006,Rigol2007,Manmana2007,Kollath2007,Kollar2008,Eckstein2008,Mossel2012,Nessi2013}, while other work
has inspected the locality of this state and how looking at a wider scope affects conclusions~\cite{Gangardt2008,Cramer2008}.
Within this area the effects of dimensionality and whether a system is closed or integrable seem to be
strong~\cite{Rigol2007,Manmana2007,Barthel2008,Rigol2009,Mossel2010,Rigol2014}

Recently it has been argued~\cite{Caux2013} that in the thermodynamic limit, the expectation value of an operator
$\mathcal{O}$ which is local in space can be found using the projection of the ground state onto a single eigenstate of the
system $\Phi_{s}$ (see Eqn~\eqref{eqn:thermalised_timedep}).
Also that as $t \to \infty$ expectation values of observables in the system can be found from the expectation of that single
state~\cite{Caux2013} as shown in Eqn~\eqref{eqn:thermalised_inftime}
\begin{equation}
  \label{eqn:thermalised_timedep}
  \lim_{N \to \infty} \expval{\mathcal{O}(t)} =
  \lim_{N \to \infty} \bigg[\frac{\matrixel{\Psi}{\mathcal{O}(t)}{\Phi_{s}}}{2 \braket{\Psi|\Phi_{s}}}
  + \Phi_{s} \leftrightarrow \Psi\bigg]
\end{equation}
\begin{equation}
  \label{eqn:thermalised_inftime}
  \lim_{t \to \infty} \lim_{N \to \infty} \expval{\mathcal{O}(t)} =
  \lim_{N \to \infty} \frac{\expval{\mathcal{O}}{\Phi_{s}}}{\braket{\Phi_{s}|\Phi_{s}}}.
\end{equation}
Here the item $\Phi_{s} \leftrightarrow \Psi$ denotes the previous term in the equation with terms $\Phi_{s}$ and $\Psi$
swapped.

The current system, being integrable, one-dimensional, and evolving from an initial state far from equilibrium, is a good
example of those that have been of such interest.
Though Eqn~\eqref{eqn:thermalised_inftime} was derived under assumptions that don't hold in our system, we can make some
numerical observations about whether it may hold for the momentum operator here, providing complementary information in a
similar system.
The first steps of such an analysis have been made, and our limited results are presented below.

\begin{figure}[h]
  \centering
  \includegraphics[width=\linewidth]{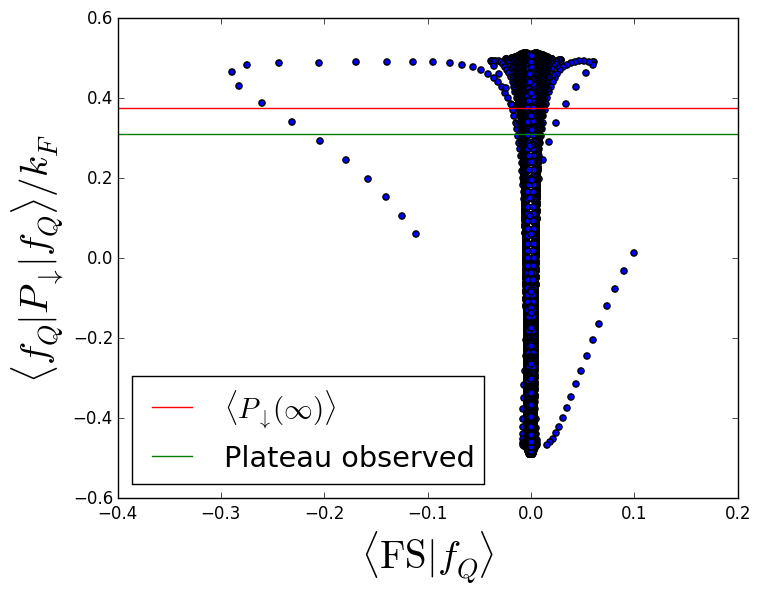}
  \caption{\label{fig:thermalisation_scatter_graph}
    Distribution of each state's momentum contribution, and the weight that contribution has.
    This plot is limited to states that are singly excited for clarity, as the structure in states with extra excitations
    can't be seen at this scale.
    The red line shows the time-independent contribution of all states, and the green line denotes the plateau seen in the
    momentum.
    While there are states that have the momentum which would occur at infinite time (whichever of the two definitions we
    use), there is no obvious feature in the distribution around this point.
  }
\end{figure}

While no claim has been made of the relative import of $\Phi_{s}$ one might guess that it is a state with a large
contribution to the momentum of the impurity, so looking only at those states with a single pseudo excitation seems a
reasonable starting point.
Plotting $\expval{P_{\uparrow}}{\Phi_{s}}$ against $\abs{\braket{FS|f_Q}}$ for each such eigenstate found in the system
gives the graph shown in Fig~\ref{fig:thermalisation_scatter_graph}.
While there clearly are states whose momenta are near the two values for infinite time momentum we have (the plateau on our
plots and the theoretical $\expval{P_{\downarrow}(\infty)}$), none have a particularly notable $\abs{\braket{FS|f_Q}}$, and
there is no obvious feature leading to some state we can take as an initial guess for $\Phi{s}$.

\begin{figure}[h]
  \centering
  \includegraphics[width=\linewidth]{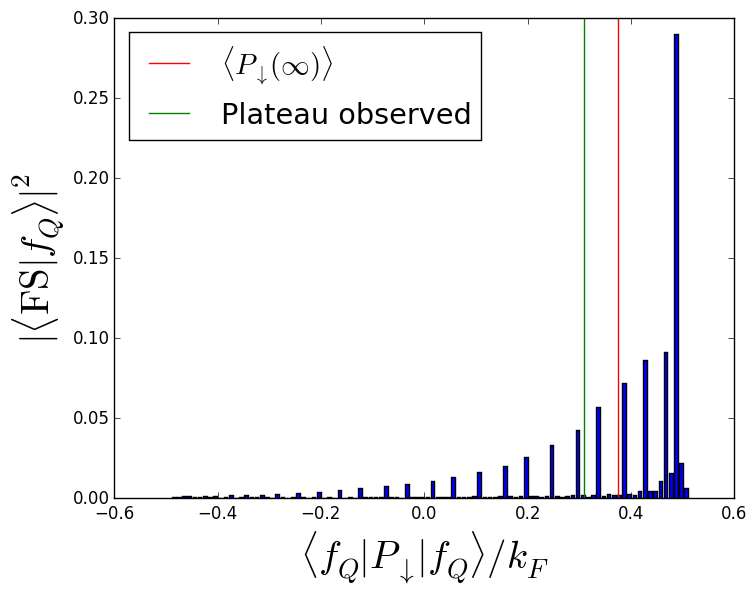}
  \caption{\label{fig:thermalisation_histogram_sum}
    Contribution to infinite time momentum from singly excited states.
    Each bar is the sum of the $\abs{\braket{FS|f_Q}}^{2}$ for the states in that region, the red vertical line denotes the
    infinite time contributions from all states, and the green line denotes the position of the plateau seen when the total
    momentum against time is plotted.
    While the contributions increase markedly near the upper bound on the momentum, the peak is slightly beforehand.
    There are two apparent contributions: one from the progression of the most important family, and one from others, but
    both have a peak at the same point.
  }
\end{figure}

Though these plots don't provide any conclusive data on the thermalisation hypothesis, they do give some more information
about the structure of the singly excited eigenstates.
We can see the states in this subset are bounded in momentum, and those states where the impurity has a positive
momentum (i.e. the impurity is travelling in the same direction as it was initially going) have a greater
$\abs{\braket{FS|f_Q}}^{2}$ on average than those where it is negative.
This asymmetry can be seen in Figure~\ref{fig:thermalisation_histogram_sum}, which shows a much stronger directionality than
is obvious in Figure~\ref{fig:thermalisation_scatter_graph}, Figure~\ref{fig:thermalisation_histogram_sum} shows the import
of each range of momentum, plotting a histogram showing the value of the equation
\begin{equation}
  \label{eqn:histogram_sum}
  \sigma = \sum_{f_{Q}} \expval{P_{\uparrow}}{f_{Q}} |\braket{FS|f_{Q}}|^{2}
\end{equation}
for each bucket, where the sum is over those $\ket{f_{Q}}$ whose momentum is within the bucket's range.
Not only does this figure show a very strong directionality, but it also demonstrates two separate progressions in $\sigma$:
one that follows the contribution of the main branch, and another from all states in different branches of
Fig~\ref{fig:zhenya_families}.

\begin{figure}[h]
  \centering
  \includegraphics[width=\linewidth]{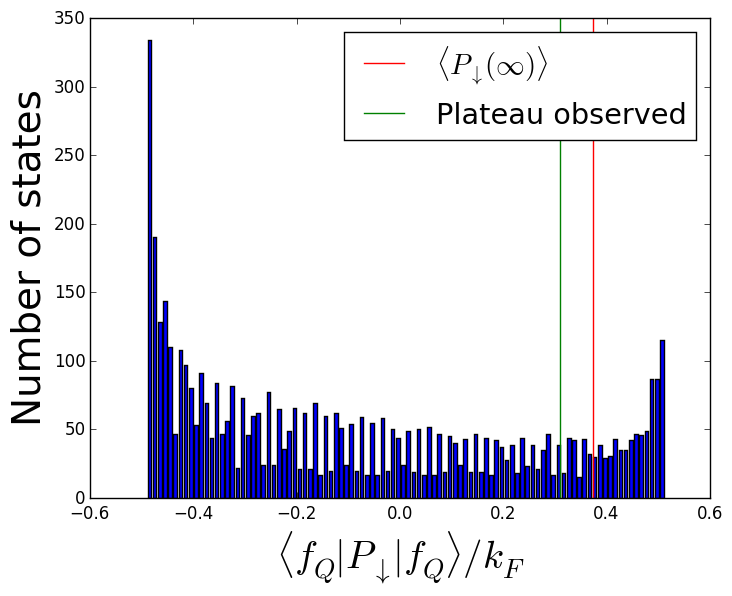}
  \caption{\label{fig:thermalisation_count_histogram}
    Distribution of the singly excited states across their expectation values of the impurity's momentum.
    Each bar represents the number of states within that region of momentum, the red vertical line denotes the infinite
    time contributions, and the green line denotes the position of the plateau when the total momentum
    against time is plotted.
    The momentum of these states is bounded in both directions, and there are more states near these bounds than elsewhere.
    While the number of states with negative momentum is greater than those with positive momentum, the
    $\abs{\braket{FS|f_Q}}^{2}$ weights mean the contribution from positive momentum states is much greater
    (see Fig~\ref{fig:thermalisation_histogram_sum}).
  }
\end{figure}

While there is a much stronger positive contribution to the impurity's momentum seen by accounting for the importance of each
state there are actually more eigenstates where the impurity has a negative momentum than otherwise.
We show this in Figure~\ref{fig:thermalisation_count_histogram}, which plots the number of states over the same buckets as
used for Figure~\ref{fig:thermalisation_histogram_sum}.
It can be seen in this figure that while there are peaks in the number of states at both bounds in the momentum, the number
of states near the negative bound is much larger than at the positive one.

\section{Asymptotic $\expval{P_{\downarrow}(\infty)}$}
\label{sec:pinf_vs_p1st_thermalised}

\begin{figure}
  \centering
  \includegraphics[width=\linewidth]{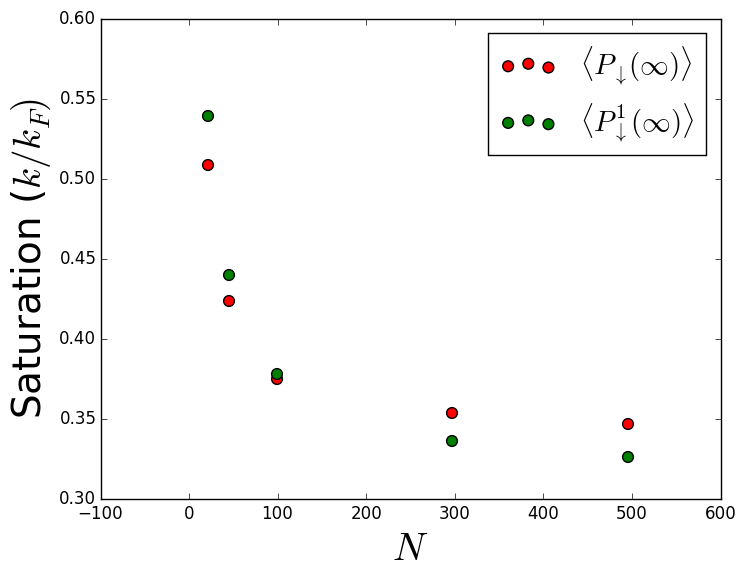}
  \caption{\label{fig:pinf_and_p1st_vs_syssize}
    How the normalised time-independent contributions to the momentum from the main family compare with the total as the system
    size changes.
    For those systems which we can plot the time evolution of the momentum (with $N < 100$), the infinite time contributions
    of the main branch are almost representative of the total, and their representative nature increases with increasing
    system size.
    For larger systems, outside of this calculable range, the infinite time contribution of the main branch decreases further,
    away from the total.
    An analytical analysis shows that this progression continues, and as the system size diverges, the main branch's
    contribution tends to zero~\protect\cite{Sasha_told_me_so}.
  }
\end{figure}

The discussion presented in section~\ref{sec:general_shape} on the representative nature of the contribution from the main
parametric branch (seen in Fig~\ref{fig:zhenya_families}) is necessarily restricted to the range of parameters our code can
reach.
Naturally, the question of whether this behaviour persists to the thermal regime of $N \to \infty$ has been raised.
As the computational expense of calculating the full $\expval{P_{\downarrow}(t)}$ for a large system is prohibitively expensive
for system sizes greater than $N \approx 99$, we looked at how representative the infinite time contribution of the main
branch $\expval{P_{\downarrow}^{1}(\infty)}$ is of the total value as the system grows larger.
While the normalised values of the main branch seem to tend towards $\expval{P_{\downarrow}(\infty)}$ for the range of system
sizes we can calculate the momentum evolution for, as we progress into larger systems, $\expval{P_{\downarrow}^{1}(\infty)}$
continues to decrease, as shown in Figure~\ref{fig:pinf_and_p1st_vs_syssize}.
An analytical approach made by Oleksandr Gamayun~\cite{Sasha_told_me_so} has shown this is how they behave as the system
moves into the thermal regime $N \to \infty$, with the contribution of the main branch continually
decreasing, and eventually disappearing $\expval{P_{\downarrow}^{1}(\infty)} \to 0$.

\section{Conclusion}
\label{sec:thermalisation_conclusion}

While we have no conclusive results on the thermalisation hypothesis in our system, the code we have should provide an
adequate platform for research into the area.
Initial plots show that if some representative eigenstate $\Phi_{s}$ exists, it does not have a standout value of
$\braket{FS|f_Q}$, and the distribution of states does not show any standout feature near where it should be.
The plots created in order to find $\Phi_{s}$ shed some light on the infinite time momentum of the impurity, while there are
actually more states with a negative momentum, the impurity's directionality comes from the strong asymmetry in the
contribution to $\varsigma$.

The discovery in Section~\ref{sec:general_shape}, that the main family is representative of the total momentum contribution
without the flutter, should only be relevant to finite systems, as its infinite time contribution does not stay
representative.
As $N \to \infty$, then $P_{1}(\infty) \to 0$, but this progression is slow and can be discounted for the systems
investigable by our program.

\chapter{Concluding Remarks}
\label{sec:concluding_remarks}

\section{Results}
\label{sec:conclusion_results}
To conclude, this work has discussed the momentum evolution of an impurity quenched into a one dimensional Tonks-Girardeau
liquid.
We find agreement with the statements made in reference~\cite{Mathy2012} on the momentum evolution of the impurity,
reproducing all progressions in the plateau and quantum flutter.
Section~\ref{sec:revivals} presented progressions in the momentum revivals that come from finite size effects, and show that
for a large initial momentum they can be qualitatively described using a classical argument based on the momentum imparted to
the background gas by the impurity.
Though useful, this argument cannot be complete as it fails to describe the equivalency in the change to
$\expval{P_{\downarrow}(t)}$ that comes from modifying $\gamma$ via either the density of the background gas or the
interaction strength between the background gas and the impurity.
The relationship between the momentum plateau and the theoretical infinite time value obtained from time averaging
$\expval{P_{\downarrow}(\infty)}$ was explored in Section~\ref{sec:inf_time}, and it was shown that while the momentum
plateau is constant with changing system size, $\expval{P_{\downarrow}(\infty)}$ starts out much higher and decreases towards
the value of the plateau, the difference decreasing with a power law relation.

Chapter~\ref{sec:subsets} described patterns observed in the eigenstates and eigenstate pair contributions to the momentum of
the impurity, and used them to explain why different features of the momentum evolution saturate at a different accuracy as
measured by $\varsigma$.
We find that the overall shape of the evolution is determined by eigenstate transitions within the same branch of
Figure~\ref{fig:zhenya_families}, with those in the main branch contributing the most, though the contribution from this
branch becomes less representative with large systems.
These contributions can be normalised using Equation~\eqref{eqn:renormalisation}, to account for the difference in $\varsigma$
between subsets of states used and the total.
This shows that the states in the main branch are representative of the total in setting the momentum plateau, but contribute
more to the momentum revivals than their $\varsigma$ would suggest.
The quantum flutter which is the main topic of references~\cite{Mathy2012,Knap2014} was shown to come from inter-branch
transitions, which demonstrates that the contribution from those states with more than one pseudo excitation is negligible.
While the entire flutter requires all inter-branch transitions, the frequency at about a quarter of the amplitude can be
obtained from just looking at those transitions between states excited from the negative edge of the pseudo sea and those in
other branches, under the condition that both states share a pseudo particle.
This can provide an intuitive explanation for how the numerics describes the loss of flutter when the initial momentum drops
below the Fermi momentum, as the structure of the Bethe Ansatz inhibit any states excited from the negative edge when
$Q < k_{F}$.

These patterns can explain the stability of each feature once $\varsigma$ has passed a given point.
Figure~\ref{fig:zhenya_families} shows that states from the main branch have a much greater contribution to $\varsigma$ than
others, and hence are counted first.
Similarly, states with more than one pseudo excitation are accounted for much later when stepping up $\varsigma$.
Because transitions between states in the main branch define the general shape of the momentum evolution, and more concretely,
define the frequency of the momentum revivals, these features stabilise much earlier than the exact value of the momentum
plateau and the flutter around it.
The inter-branch transitions that define the frequency of the flutter require the lesser contributing branches to be
accounted for, this naturally results in a greater value of $\varsigma$ before the feature has been fully described.
However as all eigenstates of the system contribute to the momentum plateau, that value shows no saturation at $\varsigma < 1$.

Finally, the distribution of the singly excited eigenstates in the momentum was shown in Chapter~\ref{sec:other},
demonstrating no significant pattern around the thermal value of the impurity's momentum.
In the same chapter, the contribution to $\expval{P_{\downarrow}(\infty)}$ from the main branch of
Figure~\ref{fig:zhenya_families} is shown to decrease relative to the total $\expval{P_{\downarrow}(\infty)}$ as the system
size increases.
While this doesn't necessarily mean the momentum plateau from the normalised main branch contribution decreases relative to
the total plateau, it has been found elsewhere that this is the case~\cite{Sasha_told_me_so}, so the representative nature of
the main families contribution to the momentum plateau found in Section~\ref{sec:general_shape} is only valid in finite
systems.

\section{Limitations and Further Work}
\label{sec:conclusion_limitations}

The current work only directly applies to the integrable case in the Tonks-Girardeau regime, and much of the results are on
the structure of the Bethe Ansatz solution to the system, without a known physical interpretation.
Despite this limitation it is hoped that these results can allow others to probe this regime with much less computational
expense, from which more physical results can be found.
The ubiquitous nature of the Bethe Ansatz in integrable one-dimensional systems also lends credibility to a hope that such
patterns may occur in different models, both on a lattice and in the continuum.

An alternate direction of further study might be into the case of an attractive potential between the impurity and background
gas.
While requiring changing the code which solves the Bethe Ansatz to account for complex Bethe roots~\cite{McGuire1966} this
route should not require changing the code which finds the impurity's momentum.

Overall we have presented novel research in this system, writing code and discovering relations that can aid any research of
others in this area.

\begin{appendices}
\chapter{Code Details}
\label{sec:code_details}

\tikzset{Segmentation lines/.style={gray,very thin,dashed}}

\section{Introduction}
\label{sec:code_introduction}
As the central tool used throughout this thesis, the code written for calculating off-diagonal elements in the matrix of the
background gas momentum operator deserves some discussion, and is provided here.
While our method of finding eigenstates of our system is novel, it is beyond the scope of this thesis, being written by
Evgeni Burovski and not the current author.
Instead this chapter discusses the process of finding the time dependent momentum contributions from a given set of
eigenstates.
We will neglect the details of encoding those equations presented in Section~\ref{sec:m&m_model} and discuss the challenges
faced when scaling to large systems, how they have been circumvented and what trade-offs have been made.
Hence this chapter contains no information on the physics or maths of the problem, focusing solely on the implementation
details of this research.

There are two main discussions in this chapter, 1) How to efficiently spread the work required over multiple processes, and
2) the benefits and disadvantages of storing different data structures, though there is no clear separation between them, as
different methods often make different compromises between these values.

Throughout this chapter we will refer to two stages, the \emph{calculation stage} and the \emph{analysis stage}, with the
assumption that the majority of the calculation is done on a larger machine, such as a computing cluster, and the analysis
done on a much smaller personal computer.
Note all judgement calls on when a data structure was too large, or took too much computation for the analysis stage, were
made to account for a personal computer with 2GB of RAM and one hyperthreaded $2.20GHz$ processor.
The most notable decision was whether calculating $\expval{P_{\downarrow}(t)}$ from the amplitudes and frequencies in the RHS
of Eqn~\eqref{eqn:momentum_against_time} should be part of the calculation stage or if it could be done during analysis.

In order to simplify the discussion, we shall compare and contrast the approaches to two different systems: one small
system (e.g. $N = 15$) accounting for $N_{s} = 400$ states, and one large system (e.g. $N = 45$) accounting for
$N_{s} = 20000$ states.
To give the reader an idea of how the time required to find $\expval{P_{\downarrow}(t)}$ from the eigenstates of the system
changes with increasing system size $N$ a description of the calculation bottlenecks is required.
The bottleneck in calculating each individual amplitude from a pair of states comes from the calculation of the singular
value decomposition of an $N \times N$ matrix.
This scales with $N^{3}$, while the number of amplitudes that must be calculated in this manner scales with $N_{s}^{2}$.
The number of states $N_{s}$ here has a non-trivial but strongly increasing relationship with $N$ and this relationship is
shown in Figure~\ref{fig:states_overlap_systemsize}.

\section{Data Structures}
\label{sec:data_structures}

When working on a small system, worries about computational expense and memory usages are much lower that otherwise.
Hence most design decisions were made based on the ease of analysis once all terms on the RHS of
Eqn~\eqref{eqn:momentum_against_time} have been found.
The data required for the analysis made in this work are the amplitude, frequency, and pair of eigenstates for each term on
the RHS of Eqn~\eqref{eqn:momentum_against_time}, along with the system parameters (both physical and non-physical) described
in Chapter~\ref{sec:observables}.
Note however, that the form of a stored eigenstate changes throughout the process of reading from a stored cache, calculating
amplitudes, and storing with amplitudes and frequencies for a given contribution.
In order to save calculation, many values which must be calculated once for each eigenstate and used to find the amplitude of
each transition to or from that state are cached in the eigenstate structures during the calculation step.
However, in analysis, the only identification required for a state are the pseudo excitations that create it (see
Section~\ref{sec:pseudo_sea}).
This section describes the form of data structures written to disk in the calculation step for use in analysis.

\begin{figure}
  \begin{subfigure}[b]{0.3\textwidth}
    \centering
    \begin{tikzpicture}[scale=1.3]
      \draw[Segmentation lines] (0,3.5) -- +(3,0);
      \node (freq) at (1.5,4) [anchor=south] {Frequencies};

      \draw[Segmentation lines] (0,2) -- +(3,0);
      \node (amps) at (1.5,2.5) [anchor=south] {Amplitudes};

      \draw[Segmentation lines] (0,0.5) -- +(3,0);
      \node (pairs) at (1.5,1) [anchor=south] {State Pairs};

      \draw[Segmentation lines] (2,0) -- +(0,0.5);
      \node (parameters) at (0,0.25)
            [rectangle,minimum width=2,minimum height=0.5,align=center,anchor=west] {Parameters};
      \node (saturation) at (2.5,0) [rectangle,anchor=south] {$\varsigma$};

      \draw (0,0) rectangle (3,5);
    \end{tikzpicture}
    \caption{
      Small system.
    }
    \label{fig:small_data_structures}
  \end{subfigure}
  \hfill
  \begin{subfigure}[b]{0.6\textwidth}
    \centering
    \begin{tikzpicture}[x=1cm,y=1cm]
      \node[above right,rectangle,draw,minimum height=1cm,minimum width=3cm] at (0,0) {Total $\expval{P_{\downarrow}(t)}$};

      \draw (0,1.8) rectangle +(2.6,5);
      \node[anchor=west] (cache) at (3.5,7.4) {Original Eigenstate Store};
      \draw [->,thick] (3.5,7.2) to [bend right=20] (1.25,5.5);

      \draw (3.4,0) rectangle +(5.4,5.4);
      \node[font=\small,anchor=west,text width=4cm,align=center] at (4.4,3.4) {Increasing\\$E_{f}$};
      \node[font=\small,anchor=west,text width=4cm,align=center] at (3.2,2)   {Increasing\\$E_{f^{'}}$};

      \draw [->] (4.6,4.2) -- (8.2,4.2);
      \draw [->] (4.2,4.0) -- (4.2,0.4);

      \draw [->,thick] (5.5,6) to [bend right=20] (4.4,5);
      \node[anchor=south] at (6.4,6) {Amplitude Matrix};

    \end{tikzpicture}
    \caption{
      Large system.
    }
    \label{fig:large_system_data_structures}
  \end{subfigure}
  \caption{
    Schematics of data structures serialised for both large and small systems.
    For small systems a hash table containing all data required for analysis was directly serialised to disk and unserialised
    when required.
    This had the advantage of simplicity, and would not easily result in un-synchronised data, as the calculation stage
    created the data in one step, and no modifications are needed in the analysis stage.
    Frequencies, amplitudes, and eigenstate pairs are kept ordered by increasing energy, and can hence be matched
    accordingly, this means only information required for choosing interesting states is needed in the state pairs.
    When the system is large, storing all data in RAM at the same time in the analysis stage is impossible, so a different
    method of reading in data was devised (see Figure~\ref{fig:multiprocessing_schematic}).
    Said method centred around calculating the position in the stored amplitude matrix where each interesting eigenstate
    pair will be, and only reading the data stored in those positions.
    As this new method of reading those amplitudes required for partial contributions took time and required the calculation
    of each eigenstates' energies, storing the frequencies associated with each amplitude became redundant, and was removed for
    storage space concerns.
  }
  \label{fig:both_data_structures}
\end{figure}
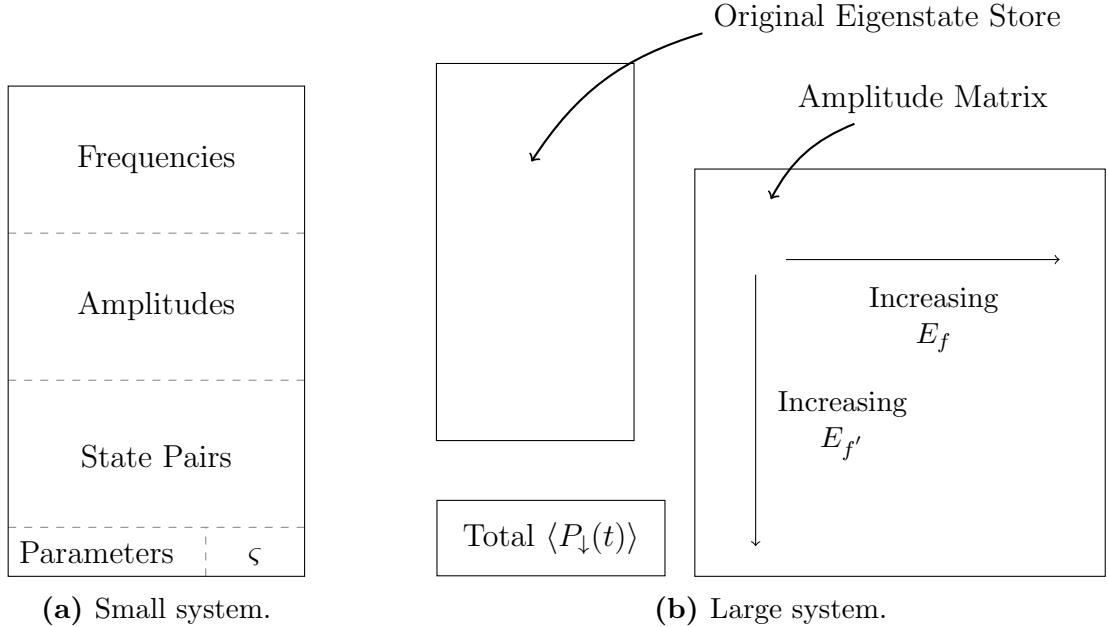

In a system near our ``small'' example, $N_{s}$ is low enough that all these values can be simultaneously stored in RAM.
Because of this ability, data structures for these systems are designed with the primary objective of being easy to read and
manipulate during analysis, storing all values required in a single file, with no recalculation required (see
Figure~\ref{fig:small_data_structures}).
From this data $\expval{P_{\downarrow}(t)}$ for any period of time is easily created, and any subsets can be found by
filtering contributions by the relevant eigenstate pairs, providing good flexibility in analysis within reasonable time
frames.
For a larger system size, and the correspondingly larger number of states used, such an approach is no longer viable, as the
calculation time of $\expval{P_{\downarrow}(t)}$ from the elements of the Fourier transform, and the space required to store
all individual contributions both become prohibitively expensive.
The consequences of these restrictions are twofold: first, the process of finding the total $\expval{P_{\downarrow}(t)}$ from
all contributions must be moved into the calculation stage, and second, a method of only reading in those contributions
required for a subset analysis must be implemented.

Implementing the first of these restrictions is a simple manner, simply choosing a range of time values to plot before
calculation, and saving the momentum and time values for each of the points requested, on disk for analysis.
Moving the calculation of $\expval{P_{\downarrow}(t)}$ from the analysis step to the calculation one is not as much of a
hindrance as one might initially suppose as the calculation of the total $\expval{P_{\downarrow}(t)}$ is rarely done more
than once.

The latter of these two consequences is implemented by storing the amplitudes on disk in a dense two dimensional array, with
each row and column corresponding to the eigenstates which create the amplitude.
As this matrix is made in order, with the eigenstate energy for each column/row increasing with increasing element index, the
amplitude for a given pair of eigenstates can be obtained by calculating the corresponding position in the array and only
reading the value in that position.
A decision to forgo storing the frequencies of each state pair was made, as the recalculation of this value is trivial, and
faster than reading a value from disk, especially after the energies for each eigenstate have been found when finding
corresponding positions in the amplitude matrix.
We also decided to keep checkpoint files for eigenstates and amplitudes separate, rather than joining them together.
This helped in parsing the data structures, reusing existing code to read eigenstates from their original structure, but
created a danger which the user must be aware of.
The two files may get out of sync if eigenstates are added to the cache which would have been accounted for in the matrix of
amplitudes.
Keeping these two sets of values separate also meant that spreading the work of calculating amplitudes across multiple
processors could be kept simple, memory mapping our matrix onto a file as will be described in
Section~\ref{sec:multiprocessing}.
The data structures saved to disk when in this large system regime, and hence those that define the majority of our codes
structure, are shown in Figure~\ref{fig:large_system_data_structures}.
These consist of a packed two-dimensional array of amplitudes, ordered by the energy of each state in the transition, an SQLite
database storing the eigenstates of the system (in the same format as delivered by the eigenstate calculation), and a numpy
named array containing the impurity's momentum and time over a predefined time period.

\section{Multiple Processors}
\label{sec:multiprocessing}
\begin{figure}[h]
  \begin{tikzpicture}[node distance=3mm and 3mm]

    \draw[gray,ultra thin,step=0.5cm] (0,0) grid (5,5);

    \node at (0,0) [anchor=south west,rectangle,thick,draw,minimum width=5cm,minimum height=5cm] (ampmat) {};
    \node[anchor=south west,rectangle,minimum width=5cm,minimum height=2cm] (p3data) at (0,0) {};
    \node[anchor=south,rectangle,minimum width=5cm,minimum height=1.5cm] (p2data) at (p3data.north) {};
    \node[anchor=south,rectangle,minimum width=5cm,minimum height=1.5cm] (p1data) at (p2data.north) {};
    \foreach \x in {2,3.5} {
    \draw [thick] (0,\x) -- (5,\x);
    \draw [dashed] (-1.3,\x) -- (-0.3,\x);
    }
    \node[fill=white] at (p2data) {Units of Work};

    \node[rectangle,draw,minimum width=5cm,minimum height=1cm] (states)
         [above=of ampmat] {$\ket{f}$};
    \node (stateslabel) [above=of states] {Increasing $E_{f}$};
    \ExtractCoordinate{states.north west}
    \ExtractotherCoordinate{stateslabel.west}
    \draw (\XCoord,\YotherCoord) -- (stateslabel.west);

    \ExtractCoordinate{states.north east}
    \ExtractotherCoordinate{stateslabel.east}
    \draw [->] (stateslabel.east) -- (\XCoord,\YotherCoord);

    \node[rectangle,draw,minimum width=1cm,minimum height=5cm] (pstates)
         [left=of ampmat] {$\ket{f^{'}}$};
    \node (pstateslabel) [left=of pstates] {\rotatebox{90}{Increasing $E_{f^{'}}$}};
    \ExtractCoordinate{pstates.north west}
    \ExtractotherCoordinate{pstateslabel.north}
    \draw (\XotherCoord,\YCoord) -- (pstateslabel.north);

    \ExtractCoordinate{pstates.south east}
    \ExtractotherCoordinate{pstateslabel.south}
    \draw [->] (pstateslabel.south) -- (\XotherCoord,\YCoord);

    \def\stepdistance{1.3cm}
    \node (calc1) at ([xshift=\stepdistance]p1data.east) [anchor=west,rectangle,draw] {Calculate};
    \node (calc2) at ([xshift=\stepdistance]p2data.east) [anchor=west,rectangle,draw] {Calculate};
    \node (calc3) at ([xshift=\stepdistance]p3data.east) [anchor=west,rectangle,draw] {Calculate};

    \draw [->] (p1data) -- (calc1);
    \draw [->] (p2data) -- (calc2);
    \draw [->] (p3data) -- (calc3);

    \node (summation) at ([xshift=2*\stepdistance]calc2.east) [anchor=west,rounded rectangle,draw] {+};

    \draw [->] (calc1.east) to [bend left=20]  node[sloped,above=1pt] {$\expval{P_{\uparrow}(t)}$} (summation);
    \draw [->] (calc2) --                      node[sloped,above=1pt] {$\expval{P_{\uparrow}(t)}$} (summation);
    \draw [->] (calc3.east) to [bend right=20] node[sloped,below=1pt] {$\expval{P_{\uparrow}(t)}$} (summation);

    \draw [->] (summation.east) -- node[above=1pt] {Total  $\expval{P_{\downarrow}(t)}$} ([xshift=2*\stepdistance]summation.east);

  \end{tikzpicture}
  \caption{
    Schematic of how work is spread over multiple processors, in this example $3$ processors are used.
    The distribution of work across multiple processors is done in a simple manner; each processor takes a block of the
    amplitude matrix and calculates the contribution to the RHS of Equation~\eqref{eqn:momentum_against_time} for a set of time
    points.
    Each of these contributions is then summed, and taken from the total momentum of the system $Q$ to find the momentum of
    the impurity over the range in time calculated over.
  }
  \label{fig:multiprocessing_schematic}
\end{figure}
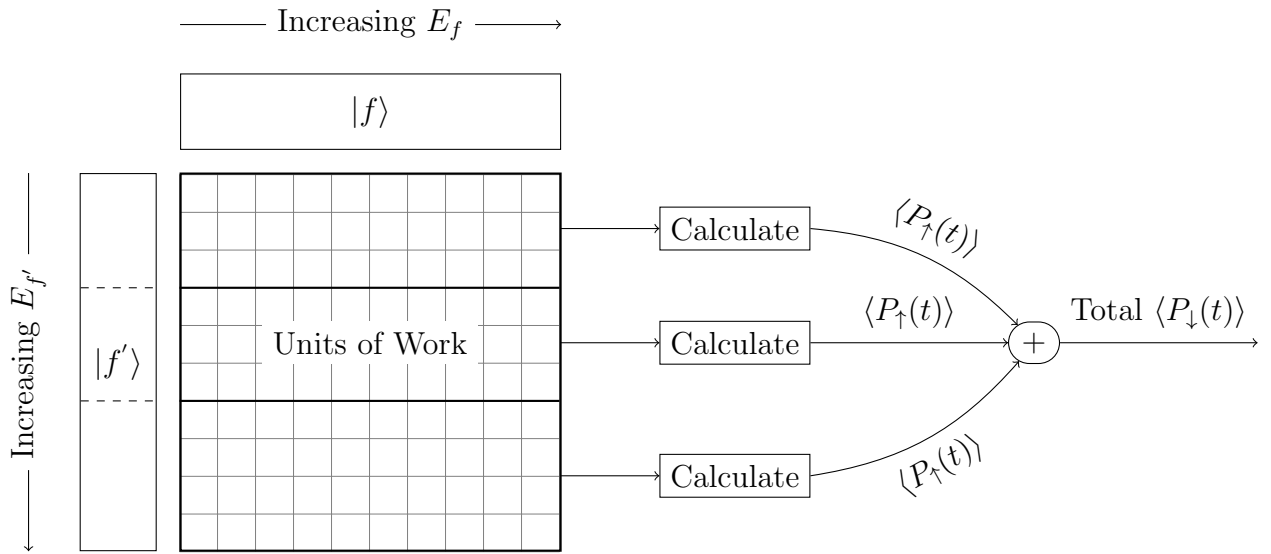

With a large system, the need to spread work over more processors becomes much greater, as the amount of computation
increases dramatically (the manner in which it increases is discussed at the start of this Chapter).
Luckily, the form of the equations solved means the bottleneck in the computation can be written in an embarrassingly parallel
manner.
Each process takes a different set of eigenstate pairs to work with, and works independently to find the background gas'
momentum contribution from that set of transitions.
They then sum these contributions with the other processes before subtracting the total from the total momentum to obtain the
momentum of the impurity.
This method, schematically shown in Figure~\ref{fig:multiprocessing_schematic}, lends itself very well to the packed data
structure of amplitudes we store on disk (see Section~\ref{sec:data_structures}).
Each process can memory map a chunk of the file as its assigned part of the matrix, and when this chunk has been flushed to
disk the amplitudes are saved in their assigned positions.
Assigning the work in this manner means the number of processes used when the amplitude matrix was initially created is
completely opaque to the user, keeping the data structures general, and allowing a user to reuse the amplitude matrix over a
different number of processes without manual intervention.

The lower level spreading of processes has been implemented in two different ways, one uses the built in Python
multiprocessing module, and the other uses mpi4py~\cite{mpi4py}.
Dual implementations are maintained as the default multiprocessing module in Python can not spread load over more than
one node in a cluster, but we see less of an overhead when using it to manage different processes on a single node.

We observe a near-linear scaling from 1 to 64 processes used, calculating $\expval{P_{\downarrow}(t)}$ for $1000$ points in
time of our large system, greater numbers of processes have not been investigated as the wait for job scheduling becomes a
limiting factor.

\section{Improvements}
\label{sec:code_improvements}

\begin{figure}[h]
  \centering
  \begin{tikzpicture}[node distance=3mm and 3mm]
    \node at (0,0) [anchor=south west,rectangle,thick,draw,minimum width=6cm,minimum height=6cm] (ampmat) {};

    \node[rectangle,draw,minimum width=6cm,minimum height=1cm] (states)
         [above=of ampmat] {$\ket{f}$};
    \node (stateslabel) [above=of states] {Decreasing $\abs{\braket{FS|f}}^{2}$};
    \ExtractCoordinate{states.north west}
    \ExtractotherCoordinate{stateslabel.west}
    \draw (\XCoord,\YotherCoord) -- (stateslabel.west);

    \ExtractCoordinate{states.north east}
    \ExtractotherCoordinate{stateslabel.east}
    \draw [->] (stateslabel.east) -- (\XCoord,\YotherCoord);

    \node[rectangle,draw,minimum width=1cm,minimum height=6cm] (pstates)
         [left=of ampmat] {$\ket{f^{'}}$};
         \node (pstateslabel) [left=of pstates] {\rotatebox{90}{Decreasing $\abs{\braket{FS|f^{'}}}^{2}$}};
    \ExtractCoordinate{pstates.north west}
    \ExtractotherCoordinate{pstateslabel.north}
    \draw (\XotherCoord,\YCoord) -- (pstateslabel.north);

    \ExtractCoordinate{pstates.south east}
    \ExtractotherCoordinate{pstateslabel.south}
    \draw [->] (pstateslabel.south) -- (\XotherCoord,\YCoord);

    \node at (ampmat.north west)
      [anchor=north west,align=center,rectangle,draw,minimum width=1.5cm,minimum height=1.5cm]
      (minor1) {$\varsigma = 0.9$};
    \node at (ampmat.north west)
      [anchor=north west,align=center,rectangle,draw,minimum width=3.5cm,minimum height=3.5cm]
      (minor2) {};
    \node at (minor1.south east)
      [anchor=north west,align=center,rectangle,minimum width=1.5cm,minimum height=1.5cm]
      {$\varsigma = 0.99$};
    \node at (minor2.south east)
      [anchor=north west,align=center,rectangle,minimum width=2.5cm,minimum height=2.5cm]
      {$\varsigma = 0.999$};

    \foreach \x in {1.5cm,3.5cm} {

      \draw [dashed] ([yshift=-\x]pstates.north west) -- ([yshift=-\x]pstates.north east);
      \draw [dashed] ([xshift=\x]states.north west) -- ([xshift=\x]states.south west);
    }

  \end{tikzpicture}
  \caption{Alternate order of amplitude matrix.
    Currently amplitudes of contributions are stored ordered by the energy of the eigenstates in each transition shown in
    Figure~\ref{fig:multiprocessing_schematic}.
    If the matrix were ordered by $\abs{\braket{FS|f}}^{2}$ instead, then finding the amplitudes for a smaller subset of
    eigenstates, taken in this order, would be much simpler.
    This would mean finding the contribution from a smaller set of eigenstates could be done without recalculation of their
    amplitudes.
  }
  \label{fig:overlap_ordered_amplitudes}
\end{figure}
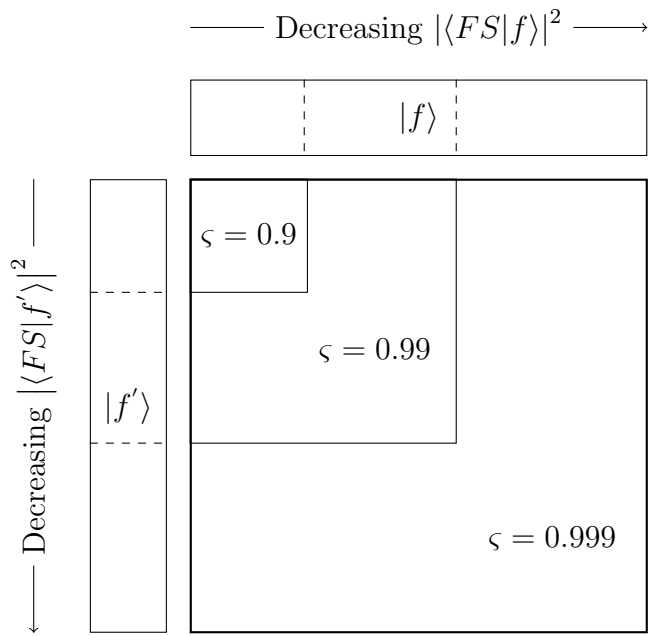

While the current formulation of this code has been proved useful and reliable when working through this project, there are a
few improvements which should have noticeable benefits, yet have not been made due to time constraints.
The simplest of these is to save the packed matrix of amplitudes ordered by eigenstate overlap
$\abs{\braket{FS|f_Q}}^{2}$ rather than energy.
This different ordering would help when inspecting how the system changes with differing $\varsigma$ saturation.
When a set of amplitudes has been calculated for a saturation of e.g. $\varsigma = 0.999$, the higher overlap eigenstates
from this set that provide a saturation of $\varsigma = 0.9$ could be read from the original data file in the simple manner
shown in Figure~\ref{fig:overlap_ordered_amplitudes}.
Currently, calculating the contribution from a smaller set requires saving another matrix of amplitudes in a separate file
that simply contains a subset of the information in the original: this results in needless duplication.
Storing the amplitude matrix ordered by energy has no inherent benefit, but the amount of programmatic complexity in order to
realise the benefits of an alternate order has so far delayed this change.

Another possible yet unimplemented feature is the filtering of eigenstate pairs in the initial computational run.
The trade-off on using this hypothetical feature would be sacrificing future flexibility in analysis for a shorter initial
computational run.
Thus certain contributions would not be calculated initially on the assumption that they will never be needed in the analysis
stage.
While the results presented in Chapter~\ref{sec:subsets} suggest that such restrictions can be made when looking at specific
features of the momentum evolution, the focus of this work never moved to using these statements, and this feature was hence
not implemented.

One final compromise which a future researcher may wish to reverse has been made to sacrifice close to a factor of $2$ in
program speed for code simplicity.
As the amplitude matrix is symmetric, the direction of transition not affecting the value, there is no reason to find every
possible transition as is currently done.
Finding the amplitudes of all off-diagonal transitions in a given order and doubling the contribution would decrease both
the required RAM and CPU time dramatically.
An equivalent optimisation is already implemented in the code for subset analysis, but it is not accounted for when
calculating the total amplitudes.
The decision has been done solely for the sake of code clarity and speed of development, and as such is a strong candidate
for change in the future.
Furthermore, in obtaining this factor of $2$ speedup, it is highly likely that a quirk in implementation can be removed,
reducing the RAM required by yet another factor of $3$.
The details of this are discussed in the code comments and are based around the use of the \texttt{numpy.frompyfunc}
command.

Overall the structure of the current program makes what the author believes to be reasonable compromises between flexibility
of analysis and computational resources, scaling well with multiple processors yet still allowing for easy selection of
transition subsets by the researcher during analysis.
There are some obvious possible improvements which the author sincerely hopes are made, and anyone wishing to use this code
is encouraged to contact the author with any questions.

\end{appendices}

\printbibliography

\end{document}